\documentclass[twocolumn,showpacs,prb,longbibliography]{revtex4-2}

\usepackage{times}
\usepackage{amsmath}
\usepackage{amssymb}
\usepackage{graphicx}
\usepackage[unicode]{hyperref}
\hypersetup{
   unicode=true,          %
   plainpages=false,
   colorlinks=true,       
   citecolor=blue,        
   urlcolor=blue
}
\usepackage[caption=false,position=top,singlelinecheck=off,justification=raggedright]{subfig}
\usepackage{color}
\usepackage{pst-node}

\begin{document}

\title{Bridging closed and dissipative discrete time crystals in spin systems with infinite-range interactions}

\author{Jayson G. Cosme}
\email[]{jcosme@nip.upd.edu.ph}
\affiliation{National Institute of Physics, University of the Philippines, Diliman, Quezon City 1101, Philippines}

\author{Jim Skulte}
\affiliation{Zentrum f\"ur Optische Quantentechnologien and Institut f\"ur Laser-Physik, Universit\"at Hamburg, 22761 Hamburg, Germany}
\affiliation{The Hamburg Center for Ultrafast Imaging, Luruper Chaussee 149, 22761 Hamburg, Germany}

\author{Ludwig Mathey}
\affiliation{Zentrum f\"ur Optische Quantentechnologien and Institut f\"ur Laser-Physik, Universit\"at Hamburg, 22761 Hamburg, Germany}
\affiliation{The Hamburg Center for Ultrafast Imaging, Luruper Chaussee 149, 22761 Hamburg, Germany}

\date{\today}
\begin{abstract}

We elucidate the role that the dissipation in a bosonic channel plays in the prevalence and stability of time crystals (TCs) in a periodically driven spin-boson system described by the Dicke model. Here, the bosons are represented by photons, and they mediate the infinite-range interactions between the spin systems. For strong dissipation, we study the dynamics using an effective atom-only description and the closed Lipkin-Meshkov-Glick model. By mapping out the phase diagrams for varying dissipation strengths, ranging from zero to infinitely strong, we demonstrate that the area in the phase diagram, where a TC exists, grows with the dissipation strength but only up to an optimal point, beyond which most of the TCs become unstable. We find TCs in both closed-system and dissipative regimes, but dissipative TCs are shown to be more robust against random noise in the drive, and are only weakly affected by the choice of initial state. We present the finite-sized behaviour and the scaling of the lifetime of the TCs with respect to the number of spins and the interaction strength within a fully quantum mechanical description.
\end{abstract}
\maketitle

\section{Introduction}

A time crystal (TC) is a nonequilibrium phase of matter signified by the spontaneous breaking of time-translation symmetry \cite{Wilczek2012,Sacha2015,Khemani2019,Else2020,Sacha2020}. This characteristic behaviour manifests itself in the emergence of a periodic pattern in time distinct from the underlying temporal symmetry of the system. For example, a TC in a system described by the periodically driven Hamiltonian $H(t)=H(t+T_d)$, where $T_d$ is the driving period, will display an  observable $\hat{O}$ oscillating at a lower frequency or higher period, i.e., $\langle \hat{O}(t) \rangle = \langle \hat{O}(t+nT_d) \rangle $ with $n>1$. TCs are formed through an interplay between periodic driving, many-body interactions and possibly, dissipation. Initial predictions and subsequent realisations of TCs involve closed systems, wherein tailored interactions and strong disorder prevent heating dynamics that would otherwise destabilise a TC \cite{Else2016,Yao2017,Khemani2016,Russomanno2017,Pizzi2021NC,Barfknecht2019, Estarellas2020,Pizzi2021a,Ye2021,Zhang2017,Choi2017,Rovny2018,Kyprianidis2021,Randall2021,Munoz2022,Smits2018,Autti2018,Huang2018}. Controlled dissipation has also been demonstrated as an alternative strategy for stabilising TCs \cite{Gong2018,Zhu2019,Iemini2018,Buca2019,Gambetta2019,Sullivan2020,Skulte2021, Michal2022, Cabot2022,Vu2022,Kessler2021,Nie2023, Kessler2020,Alaeian2022,Heugel2019,Kongkhambut2022,Taheri2022}. In most of these physical systems, time-crystalline dynamics can be understood using the spin language \cite{Else2016,Yao2017,Khemani2016,Russomanno2017,Pizzi2021NC, Barfknecht2019,Estarellas2020,Pizzi2021a,Ye2021,Ye2021,Zhang2017,Choi2017,Rovny2018, Kyprianidis2021,Randall2021,Munoz2022,Gong2018,Zhu2019,Iemini2018, Buca2019,Gambetta2019,Sullivan2020,Skulte2021,Michal2022, Cabot2022,Vu2022,Kessler2021,Nie2023}.

Focusing on fully connected spin systems or, equivalently, spins with all-to-all interactions, time-crystalline phases have been studied both for closed and dissipative systems through the Lipkin-Meshkov-Glick (LMG) model and the open Dicke model (DM), respectively. Introduced in the context of nuclear physics \cite{Lipkin1965,Meshkov1965,Glick1965}, the LMG model describes $N$ fully connected spin-$\frac{1}{2}$ particles in a transverse field \cite{Vidal2004,Morrison2008}. A similar model for photon-mediated interactions is the DM \cite{Dicke1954}. The DM typifies a spin-boson system, wherein the bosons, specifically, photons in a single mode, mediate the all-to-all interactions between the spins \cite{Kirton2019,Larson2021}. The open version of the DM includes a dissipation channel via the photon decay. 
On the one hand, discrete TCs and the related subharmonic response are predicted to exist in the periodically driven closed LMG model \cite{Russomanno2017,Pizzi2021NC,Kelly2021}. We note that direct experimental observation of a TC in such an infinite-range interacting closed system remains elusive, even though existing platforms could in principle simulate the LMG model, for example, in Refs.~\cite{Baumann2010,Korenblit2012,Jurcevic2017,Monroe2021}.
On the other hand, the paradigmatic discrete TC in open systems is proposed in the driven-dissipative DM \cite{Gong2018,Zhu2019}. Using a cavity-quantum-electrodynamics (QED) platform as a quantum simulator of the open DM, indeed, a Dicke TC has been realised experimentally \cite{Kongkhambut2022}, despite the mean-field breaking terms in cavity-QED systems that compete with the infinite-range interactions necessary for emulating the DM \cite{Tuquero2022}. 

In the limit of an extremely strong photon decay rate $\kappa \to \infty$, adiabatic elimination of the rapidly evolving photon field will map the open DM onto the closed LMG model, which establishes the relation between these two fully connected models \cite{Morrison2008,Keeling2010,Larson2021}. However, it has been suggested for selected parameters that too strong dissipation could be detrimental to the stability of TCs in the open DM \cite{Gong2018,Zhu2019}, which then poses the question of how this relates to the TC phenomenology in the closed LMG model \cite{Russomanno2017,Pizzi2021NC}. As we will show later, the precise form of driving and the choice of the initial state become crucial in the closed-system limits of vanishing and infinitely strong dissipation rates. In contrast, we will demonstrate that the time-crystalline dynamics occur more ubiquitously in the dissipative regime.

\begin{figure}[!t]
\centering
\includegraphics[width=1\columnwidth]{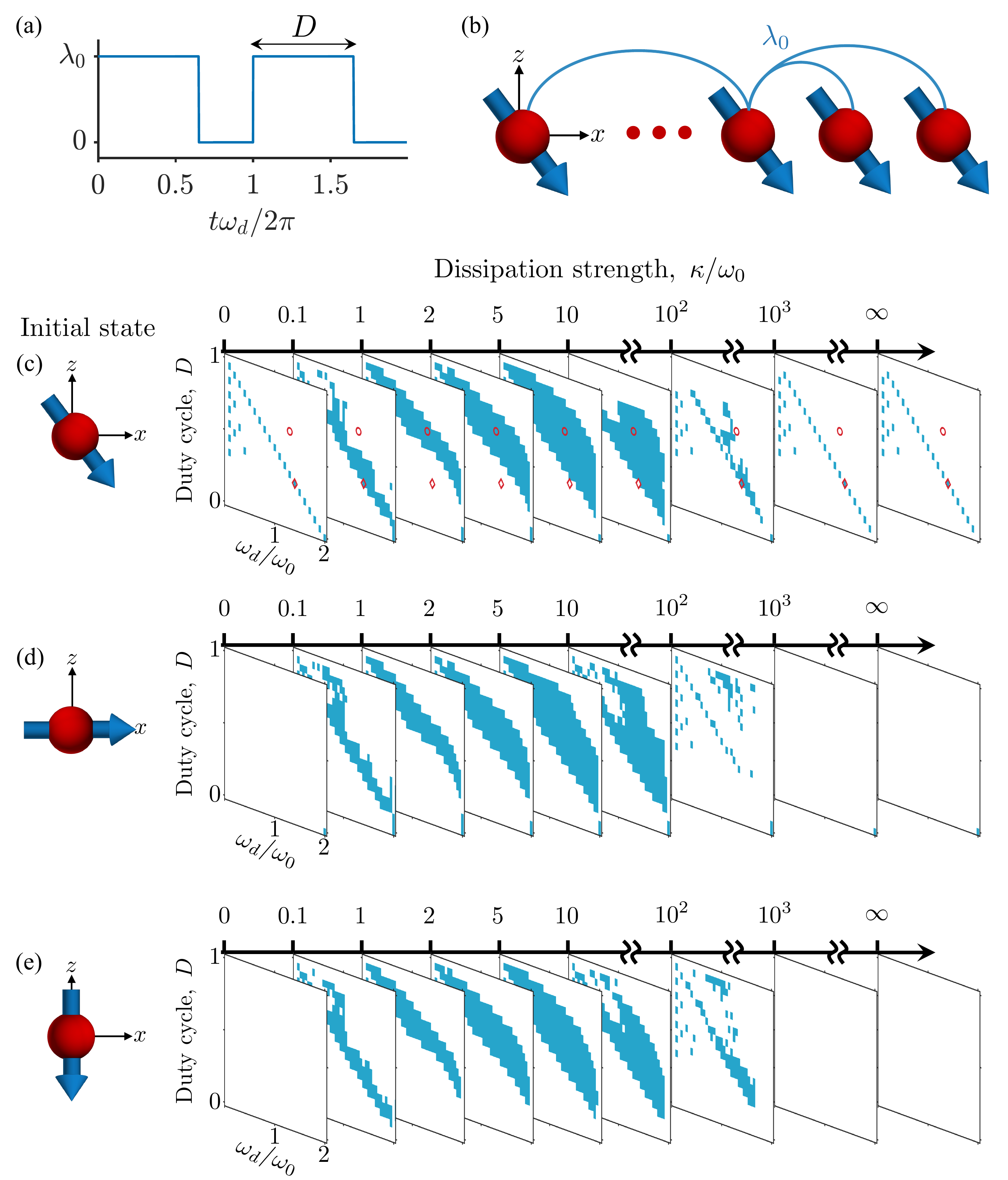}
\caption{(a) Driving protocol. The duty cycle $D$ is the duration of the bright time in one Floquet cycle. (b) During the bright times, photons mediate the all-to-all interactions between the spins. (c) and (d) Dynamical phase diagrams as a function of $D$ and driving frequency $\omega_d$ for varying dissipation strengths $\kappa$. The light-matter coupling is fixed at  $\lambda_0 = 1.1\lambda_\mathrm{cr}$, and the photon frequency is $\omega_p=\omega_0$. As depicted in the left-most panels of (c) and (d), three types of initial product states are considered: (c) one of the $\mathbb{Z}_2$-symmetry broken states, (d) spins polarised along the positive $x$ direction, and (e) spins polarised along the negative $z$ direction. The dark areas in each phase diagram denote the period-doubling   time crystalline phases. We use the Dicke model (DM) for $\kappa/\omega_0 \leq 10^3$, the atom-only DM for $10^3 \leq \kappa/\omega_0 < \infty$, and the Lipkin-Meshkov-Glick (LMG) model for $\kappa = \infty$. }
\label{fig:1} 
\end{figure}

In this paper, we present a thorough investigation of TCs in the transition from closed-system to dissipative limits, or vice versa, for spin systems with infinite-range interactions mediated by photons. By doing so, we shed light on the precise roles of dissipation and the form of driving on the emergence of TCs in infinite-range interacting systems, such as the cavity-QED setup used in the realisation of the dissipative Dicke TC \cite{Kessler2021}. To describe the system, we use the open DM for weak and intermediate dissipations, and an effective atom-only description and the LMG model for strong dissipations in which the photons are adiabatically eliminated. We consider a binary drive wherein the system periodically switches between interacting and noninteracting Hamiltonians as shown in Fig.~\ref{fig:1}(a). Mapping out the phase diagrams for a range of dissipation strengths $\kappa \in [0,\infty)$, we connect the TCs in the closed and dissipative regimes and demonstrate that the areas in the phase diagram with time crystals and time quasicrystals (TQCs) expand with increasing dissipation but only up to an optimal value, as depicted in Figs.~\ref{fig:1}(c) and \ref{fig:1}(d). We also find numerical evidence suggesting that the mechanism for generating TCs in the dissipative system is a period-doubling instability arising from a parametric resonance, and therefore we generalise the conditions first proposed in Ref.~\cite{Gong2018}. Furthermore, the TCs in the open DM are found to be more robust against random errors in the drive and are less sensitive to the choice of initial states than their counterparts in the closed-system limits, $\kappa=0$ and $\kappa \to \infty$. Nevertheless, the TCs in the closed LMG model display enhanced stability for few spins, wherein quantum effects dominate, as their lifetimes can be increased by simply increasing the interactions strength without changing the number of spins, and they have longer lifetimes than the TCs in the open DM, in general.

This paper is organised as follows. In Sec.~\ref{sec:mod}, we introduce the relevant physical models, namely, the DM, its atom-only description, and the LMG model, and the driving protocol. In Sec.~\ref{sec:mf}, we explore using mean-field theory the dynamical phase diagrams for varying dissipation strengths, and the robustness of TCs against noises in the drive and choices of initial states. In Sec.~\ref{sec:qm}, we investigate the properties of TCs for both closed-system and dissipative limits in the quantum regime of few spins. Finally, we conclude in Sec.~\ref{sec:conc}.

\section{Models and Driving Protocol}\label{sec:mod}
The Hamiltonian for the open DM is \cite{Kirton2019}
\begin{equation}\label{eq:dm}
\hat{H}/ \hbar = \omega_p \hat{a}^\dagger \hat{a} +\omega_{0}\hat{J}_z +\frac{2\lambda}{\sqrt{N}}\left(\hat{a}^\dagger+\hat{a} \right) \hat{J}_x,
\end{equation}
where $N$ is the total number of spins, $\hat{a}$ ($\hat{a}^\dagger$) is the bosonic annihilation (creation) operator for the photons, and $\hat{J}_\mu = \sum_{i=1}^N \sigma^{\mu}_i~(\mu=x,y,z)$ are the collective spin operators. The light-matter coupling strength is $\lambda$, the photon frequency is $\omega_p$, and the transition frequency of the two-level atoms represented by the spins operators is $\omega_0$.
In the presence of photon losses, the dynamics of the system can be described by the Lindblad master equation \cite{Dimer2007}
\begin{equation}\label{eq:rho}
\partial_t \hat{\rho} = -i[\hat{H}/\hbar,\hat{\rho}] + \kappa D[\hat{a}] \hat{\rho},
\end{equation}
where $D[\hat{a}]\hat{\rho} = 2 \hat{a} \hat{\rho} \hat{a}^\dagger - \left(\hat{a}^\dagger \hat{a} \hat{\rho} + \hat{\rho} \hat{a}^\dagger \hat{a}\right)$. The rate of photon emission is characterised by the photon decay rate or dissipation strength $\kappa$.

An effective atom-only description can be obtained for large but finite dissipation strength, $\infty > \kappa \gg \omega_0$ \cite{Damanet2019,Jager2022}, which, in this work, will be called the atom-only Dicke model (ADM). The ADM Hamiltonian is  \cite{Jager2022}
\begin{align}\label{eq:adm}
&\hat{H}_\mathrm{ADM}/\hbar =   \omega_{0}\hat{J}_z  -  \biggl( \frac{4\lambda^2 \omega_p}{N(\kappa^2 + \omega_p^2)} \biggr) \hat{J}^2_x \\ \nonumber
&-  \biggl( \frac{4\lambda^2 \kappa \omega_p\omega_0}{N(\kappa^2 + \omega_p^2)^2} \biggr) \{ \hat{J}_x,\hat{J}_y\} -  \biggl( \frac{2\lambda^2 \omega_0 (\omega_p^2-\kappa^2)}{N(\kappa^2 + \omega_p^2)} \biggr)\hat{J}_z 
\end{align}
In the thermodynamic limit, the ADM Hamiltonian yields the correct set of equations of motion  obtained in Ref.~\cite{Damanet2019}.

In the limit of $\kappa \to \infty$, the photonic mode can be adiabatically eliminated to obtain a Hamiltonian that depends only on the spins, equivalent to the anisotropic LMG model \cite{Lipkin1965,Meshkov1965,Glick1965,Engelhardt2013}
\begin{align}\label{eq:lmg}
\hat{H}_\mathrm{LMG}/ \hbar &=  \omega_{0}\hat{J}_z  -  \biggl( \frac{4\lambda^2 \omega_p}{N(\kappa^2 + \omega_p^2)} \biggr) \hat{J}^2_x.
\end{align}
The last term in Eq.~\eqref{eq:lmg} reveals that indeed the photons mediate the effective all-to-all interactions between the spins. In addition to the light-matter coupling strength $\lambda$, the parameters related to the photonic degree of freedom, namely, the photon frequency $\omega_p$ and dissipation rate $\kappa$, also contribute to the strength of the effective spin-spin interactions.

In the thermodynamic limit, we rescale $a = \langle \hat{a} \rangle/\sqrt{N}$ and $j_{\mu\in\{x,y,z \} } = \langle \hat{J}_\mu \rangle / N$. 
The three models described above all possess a symmetry-breaking phase transition at a critical value of the coupling strength given by \cite{Kirton2019,Damanet2019,Morrison2008,Dimer2007}
\begin{equation}
\lambda_\mathrm{cr} = \frac{1}{2}\sqrt{\frac{\omega_0}{\omega_p}(\omega_p^2+\kappa^2)}.
\end{equation}
Below the critical coupling strength, the stable phase or steady state corresponds to all the spins pointing in the negative $z$ direction, $\{j_x,j_y,j_z\} = \{0,0,-\frac{1}{2}\}$. This phase is sometimes referred to as the normal phase (NP) and for the DM. Another defining feature of the NP is the absence of photons $a=0$.
Above the critical coupling strength, the system undergoes a quantum phase transition as it spontaneously breaks the $\mathbb{Z}_2$ symmetry, $\{\hat{a},\hat{J}_x\} \to \{-\hat{a},-\hat{J}_x\} $. The steady state in the symmetry-broken phase has a spin configuration of \cite{Dimer2007}
\begin{equation}\label{eq:sr1}
\{j_x,j_y,j_z\} = \frac{1}{2}\biggl\{\pm \sqrt{1-\left(\frac{\lambda^2_\mathrm{cr}}{\lambda^2}\right)^2},0,-\frac{\lambda^2_\mathrm{cr}}{\lambda^2} \biggr\}.
\end{equation}
In the DM	, the photon mode is occupied in the symmetry broken phase, also known as the superradiant phase. The corresponding steady-state photon amplitude is
\begin{equation}\label{eq:sr2}
a = \mp  \frac{\lambda}{\omega - i\kappa} \sqrt{1- \left(\frac{\lambda^2_\mathrm{cr}}{\lambda^2}\right)^2}.
\end{equation}

We are interested in a binary Floquet drive or  bang-bang protocol wherein the interactions periodically switch according to
\begin{equation}\label{eq:lamt}
\lambda(t) = \biggl\{
\begin{matrix}
\lambda_0, & nT_d \leq t < (n+D)T_d \\
0, & (n+D)T_d \leq t < (n+1)T_d,
\end{matrix}
\end{equation}
where $n \in [0,1,2,\dots]$, $T_d$ is the driving period related to the driving frequency via $\omega_d  = 2\pi/T_d$, and $D \in [0,1]$ is a unitless quantity called the duty cycle. The duty cycle controls the duration of the dark ($\lambda=0$) and bright ($\lambda=\lambda_0$) times in a driving cycle. For $D=0$, the the light-matter coupling is always off, while for $D=1$, the light-matter coupling has a constant nonzero value $\lambda_0$ for all times. This binary driving protocol has been shown to host a period-doubling dissipative TC for $D=0.5$ \cite{Gong2018,Zhu2019}. 
We note that, for $D\to1$, this protocol is not identical to the kicking protocol considered in Ref.~\cite{Russomanno2017} because, there, the spins are flipped using a $\pi$ pulse along the $x$ direction during the kicking times, i.e., the transverse field $\omega_0$ is driven. Instead of applying a spin-flip operation, we allow the spins to rotate freely according to the coherent time evolution during the dark times, at least for the closed-system or nondissipative regimes.

\section{Mean-field results}\label{sec:mf}

We first consider the thermodynamic or mean-field limit of a large number of spins $N$. In the limit of a large number of spins, cavity-QED systems based on quantum gases  \cite{Ritsch2013,Mivehvar2021} are ideal platforms for quantum simulations since the typical number of atoms, emulating the two-level systems, reaches $N\sim 10^5$. In fact, various phenomena predicted in the DM ranging from the normal-superradiant phase transition \cite{Baumann2010,Klinder2015} to the formation of dissipative discrete TCs \cite{Kessler2021} have been observed using quantum-gas-cavity systems.

The mean-field dynamics can be obtained by solving the corresponding semiclassical equations of motion. Depending on the value of $\kappa$, we use the appropriate model, i.e., the DM for $\kappa/\omega_0 < 10^3$, the ADM for $10^3 \leq \kappa/\omega_0 < \infty$, and the LMG model for $\kappa = \infty$. The semiclassical equations of motion for the three models are presented in Appendix \ref{app:eom}. In the following, we numerically integrate the equations of motion and mainly focus on the dynamical behaviour of the expectation value of the total magnetisation along the $x$-component, $j_x$. We consider a total driving time of $t_f = 100T_d$ in accordance with the typical timescales in state-of-the-art experiments on closed and dissipative discrete TCs \cite{Zhang2017,Choi2017,Rovny2018,Randall2021,Kongkhambut2022}.

In Secs.~\ref{subsec:dp} and \ref{subsec:rob}, we choose as the initial state one of the $\mathbb{Z}_2$-symmetry-broken states amounting to all spins having a non-zero component in the positive $x$-direction, which is denoted by the upper sign solution in Eq.~\eqref{eq:sr1}. For the DM, the additional initial condition for the photon amplitude is given by Eq.~\eqref{eq:sr2}. In Sec.~\ref{subsec:isd}, we investigate other types of initial states, namely spins that are fully polarised either along the positive $x$ direction or the negative $z$ direction. 
We fix the light-matter coupling to $\lambda_0 = 1.1 \lambda_\mathrm{cr}$ and the photon frequency to $\omega_p = \omega_0$. Fixing $\lambda_0/\lambda_\mathrm{cr}$ makes the results for the LMG model independent of $\omega_p$ and $\kappa$ since the interaction strength in the LMG Hamiltonian Eq.~\eqref{eq:lmg} only depends on this ratio. In Appendix \ref{app:param}, we show similar results for other choices of $\lambda$ and $\omega_p$.

\subsection{Dynamical phases}\label{subsec:dp}

\begin{figure*}[!htpb]
\centering
\includegraphics[width=2\columnwidth]{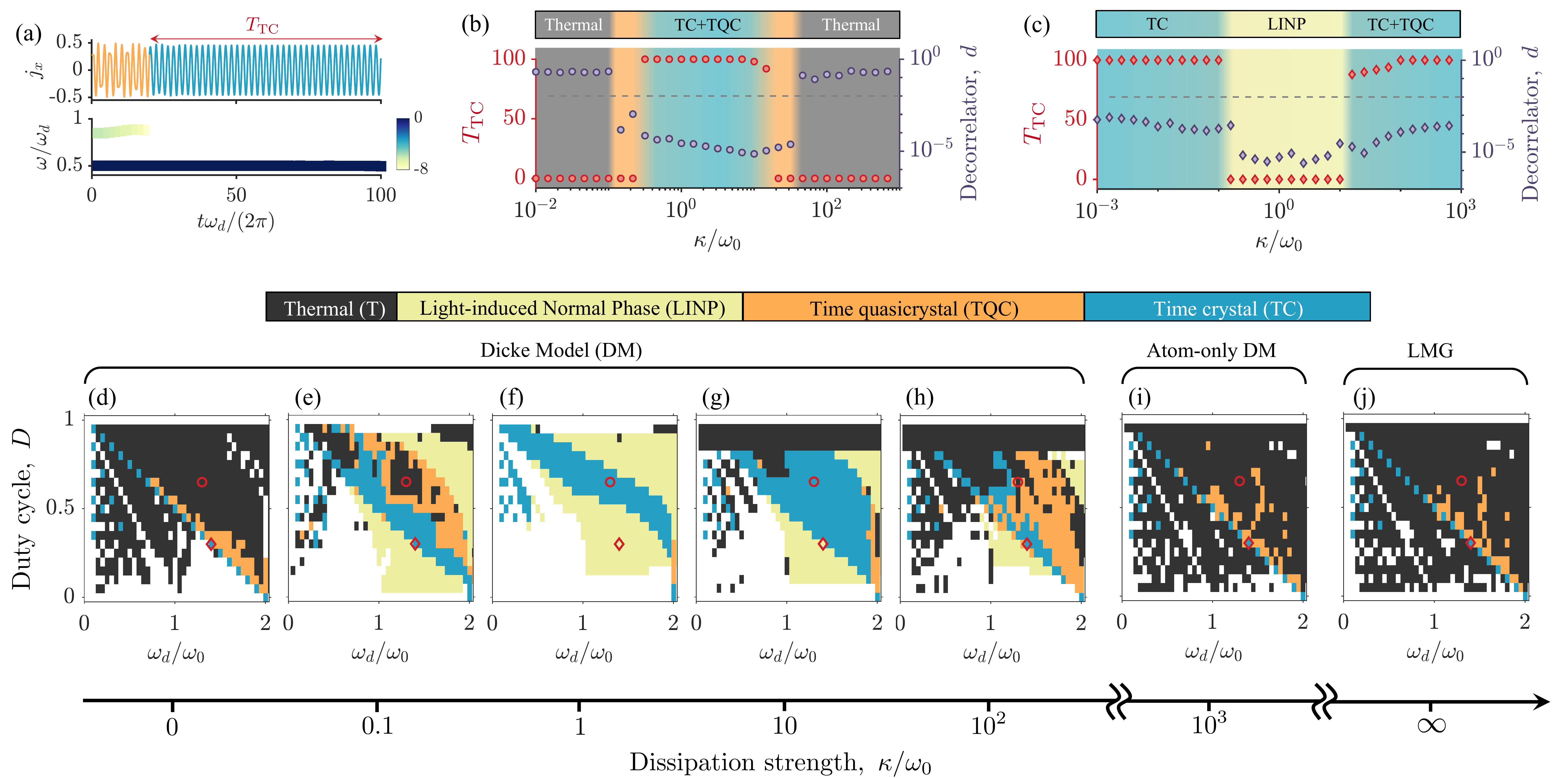}
\caption{(a) (Top panel) Exemplary dynamics of $j_x$ for $\kappa/\omega_0 = 20$ with driving parameters $\{D,\omega_d\}=\{0.3,1.4\omega_0\}$. These driving parameters are denoted as a circle in (d)-(j). The dark curve with duration $T_\mathrm{TC}$ denotes the period-doubling response, while the light curve corresponds to the transient quasi-period-doubling behaviour. (Bottom panel) Corresponding time-frequency power spectrum of $j_x$ in log scale for the two highest peaks. (b) and (c) Dependence of $T_\mathrm{TC}$ and the decorrelator $d$ on the dissipation strength $\kappa$ for driving parameters marked by (b) circles and (c) diamonds in (d)-(j). The driving parameters are (b) $\{D,\omega_d\}=\{0.65,1.3\omega_0\}$ and (c) $\{D,\omega_d\}=\{0.3,1.4\omega_0\}$. The horizontal dashed lines denote $d=0.01$, which is the threshold used to distinguish between thermal and nonthermal phases. (d)-(j) Various phases identified for different driving parameters $D$ and $\omega_d$. Each panel corresponds to a specific value of $\kappa$. Harmonic, superharmonic, and $n$-tupling dynamics are all represented in the white areas.  For the results shown here, the photon frequency and light-matter coupling strength are fixed to $\omega_p = \omega_0$ and $\lambda_0 = 1.1\lambda_\mathrm{cr}$, respectively.}
\label{fig:2} 
\end{figure*}

A generic many-body system with periodic driving, especially in the absence of dissipation, is expected to inevitably heat up and approach a featureless state \cite{Alessio2014,Bukov2015}. TCs in closed systems are particularly interesting since they are exceptions to this. To distinguish between nontrivial phases and a thermal or chaotic phase, we define the decorrelator:
\begin{equation}
d = \frac{1}{(t_f-t_i)}\sum_{t=t_i}^{t_f}   \biggl( \left | j_x(t) \right |  -  \left | j'_x(t) \right | \biggr),
\end{equation}
where $j'_x(t)$ is the dynamics of a slightly perturbed initial state relative to $j_x(t)$. Specifically, we choose $j'_x(0) = j_x(0)-0.5 \times 10^{-3}$, $j'_y(0) = 0$, and $j'_z(0) = -\sqrt{1-|j'_x(0)|^2}/2$. The decorrelator provides a measure for the distance between the time-evolving observables to probe the emergence of chaos \cite{Pizzi2019,Pizzi2021}. A large decorrelator $d \sim 1$ signifies sensitivity to initial conditions consistent with classical chaos. We obtain the decorrelator for a time window spanned by $t_i = 50T_d$ and $t_f=100T_d$. To classify thermal phases, we set a threshold of $d \geq 0.01$.

In the top panel of Fig.~\ref{fig:2}(a), we show an exemplary dynamics exhibiting both a TQC for transient times and a period-doubling TC for long times. To quantify the lifetime of the TC $T_\mathrm{TC}$, we calculate a time-frequency power spectrum according to $P(\omega,t) = {|\mathcal{F}(\omega,t)|^2}/{\sum_{i=1}^{N_f} |\mathcal{F}(\omega,t)|^2}$, where $\mathcal{F}(\omega,t)$ is the Fourier transform of $j_x$ for the time window starting at time $t$ and ending at $t_f=100T_d$.  The total number of discrete frequencies resolved by the Fourier transform is $N_f$. To demonstrate how we obtain $T_\mathrm{TC}$ using $P(\omega,t)$, we present in the bottom panel of Fig.~\ref{fig:2}(a) an example of the time-frequency power spectrum. A TQC is marked by the appearance of a secondary peak in the power spectrum in addition to the primary peak associated with the period-doubling response \cite{Pizzi2019}. We then use the appearance of a secondary peak in the power spectrum with $\ln P(\omega,t') > {-8}$ as a criterion for detecting TQC phases. That is, the lifetime of the TC phase for simulation times considered here is $T_\mathrm{TC} = 100T_d - t'$. In Fig.~\ref{fig:2}(a), we indeed find a secondary peak around $t' \approx 20T_d$ consistent with a visual inspection of the dynamics shown in the top panel. Thus, for this example, the system is in a time-quasicrystalline phase for $t<t'$, and the TC emerging for $t>t'$ has a lifetime of at least $T_\mathrm{TC} = 80T_d $.  

The lifetime $T_\mathrm{TC}$ and the decorrelator $d$ as a function of the dissipation strength $\kappa$ are shown in Figs.~\ref{fig:2}(b) and \ref{fig:2}(c), which correspond to driving parameters $\{D,\omega_d\}=\{0.65,1.3\omega_0\}$ and $\{D,\omega_d\}=\{0.3,1.4\omega_0\}$, respectively. In Fig.~\ref{fig:2}(b), the values of the decorrelator $d$ for thermal phases are several orders of magnitude larger than those for nonthermal phases. We set $T_\mathrm{TC}=0$ for thermal phases, regardless of whether a transient TQC is found for early times or a time-crystalline signal is detected for a single mean-field trajectory. 

For time-translation symmetry-breaking responses, we find the following phases: (i) pure TC, (ii) pure TQC, and (iii) mixed TC and TQC. A pure TC is characterised by having period-doubling dynamics for the entire duration of the simulation $T_\mathrm{TC}=100T_d$, as exemplified by $\kappa/\omega_0 = 1$ in Fig.~\ref{fig:2}(b) and $\kappa/\omega_0 = 10^{-3}$ in Fig.~\ref{fig:2}(c). On the other hand, a pure TQC, while insensitive to initial conditions $d<0.01$ still has $T_\mathrm{TC}=0$, since its spectrum has at least one additional subharmonic frequency peak, which in general is incommensurate with the driving frequency for the entire simulation time. An example of the dynamics and the power spectrum for a pure TQC is $\kappa/\omega_0 = 21$ shown in Appendix \ref{app:ex}. Lastly, a mixed TC and TQC phase is denoted by a transient TQC at early times and a TC at long times, as shown in Fig.~\ref{fig:2}(a), for example. We label the pure TC phase and mixed TC-and-TQC phase as simply TC for the rest of the paper since both have long-time period-doubling behaviour.

The results presented in Figs.~\ref{fig:2}(b) and \ref{fig:2}(c) highlight one of the key findings of this paper, which is the nonmonotonic behaviour in the presence and lifetime of TCs as a function of the dissipation strength. The optimal dissipation strength will strongly depend on the specific choice of driving parameters. This is illustrated by the absence of TCs for $\kappa/\omega_0 < 10^{-1}$ and $\kappa/\omega_0 > 10^2$ in Fig.~\ref{fig:2}(b) while they are present in Fig.~\ref{fig:2}(c) for the same regimes of dissipation strength . In fact, for intermediate dissipation strengths $ 10^{-1} < \kappa/\omega_0 < 10^2$, wherein TCs are seen in Fig.~\ref{fig:2}(b), the driving parameters in Fig.~\ref{fig:2}(c) push the system into a light-induced NP, which is a NP dynamically stabilised by the drive and is defined by having zero photon number despite $\lambda_0>\lambda_\mathrm{cr}$ \cite{Skulte2021} (see also  Appendix \ref{app:ex}).

In Figs.~\ref{fig:2}(d)-\ref{fig:2}(j), the dynamical phase diagrams as a function of the driving parameters are shown, wherein each panel corresponds to a particular choice of dissipation strength $\kappa$. That is, we demonstrate in Figs.~\ref{fig:2}(d)-\ref{fig:2}(j) how the dynamical phase diagram changes with the dissipation strength. In the following, we will not discuss harmonic, superharmonic, and $n$-tupling dynamics, which are all indicated by the white areas in the dynamical phase diagrams. Instead, we concentrate on the influence of dissipation on the thermal, time-crystalline, and time-quasicrystalline phases.

\subsubsection{Closed systems}
We find TC and TQC phases in the closed-system limits, namely the closed DM ($\kappa=0$), the ADM ($\kappa/\omega_0=10^3$), and the LMG model ($\kappa \to \infty$), albeit only in a relatively narrow region of the driving parameter space.
The dynamical phase diagrams for closed systems in Figs.~\ref{fig:2}(d), \ref{fig:2}(i), and \ref{fig:2}(j) share a strong similarity with each other, especially in the location of the TC phases. The qualitative agreement between the ADM and LMG phase diagrams implies the applicability of the LMG model for dynamical states, such as a TC, which is in contrast to the limitation of the LMG model in describing steady states \cite{Damanet2019,Jager2022}.

\begin{figure}[!t]
\centering
\includegraphics[width=1\columnwidth]{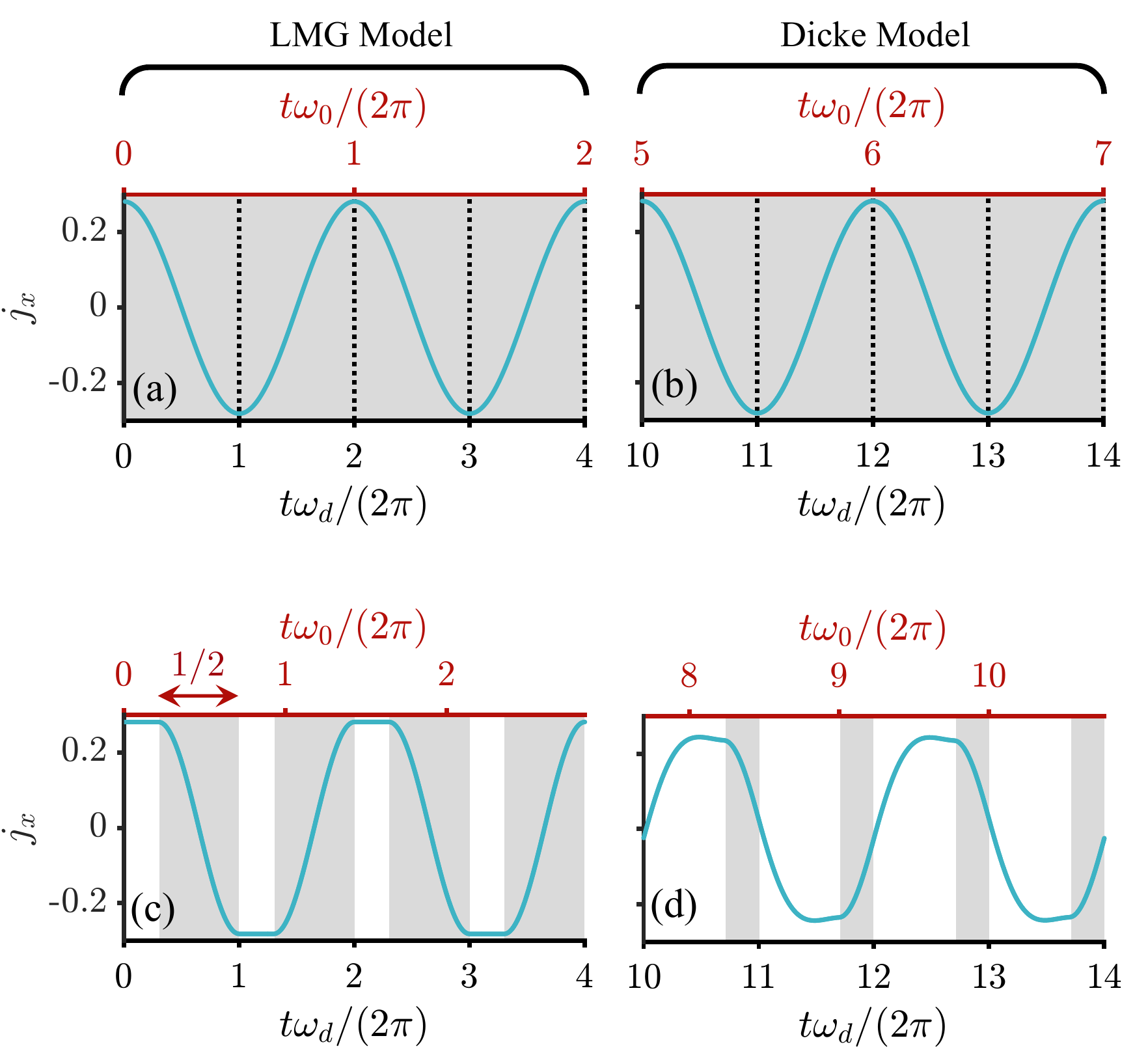}
\caption{Dynamics of $j_x$ in the (left panels) Lipkin-Meshkov-Glick (LMG) model and (right panels) Dicke model with $\kappa/\omega_0 = 1$. (a) and (b) In the absence of driving and for a sudden quench $D=0$, the spins precess around the $z$ direction due to the first term in the Hamiltonian in Eq.~\eqref{eq:lmg}. The top axis displays time in units of the precession period $T_0 = \omega_0/(2\pi)$, while the bottom axis shows time in units of the driving period $T_d = \omega_d/(2\pi)$. This apparent period doubling is trivial as the periodic driving is absent. (c) and (d) Time crystals for the periodically driven systems with parameters (c) $\{D,\omega_d\}=\{0.3,1.4\omega_0\}$ and (d) $\{D,\omega_d\}=\{0.7,1.3\omega_0\}$. The shaded areas indicate the dark time when the spin interactions are switched off.  The arrow in (c) indicates that the dark time is $t_\mathrm{dark} = T_0/2$. The remaining parameters are the same as in Fig.~\ref{fig:2}.}
\label{fig:3} 
\end{figure}

The apparent period-doubling response seen for $D=0$, as illustrated in Fig.~\ref{fig:3}(a), can be considered trivial since this simply corresponds to a sudden quench at $t=0$ from $\lambda = 1.1\lambda_\mathrm{cr}$ to $0$. Within the LMG model, this leads to a coherent dynamics of the spins precessing around the $z$-axis at a frequency $\omega_0$, i.e., a precession period of $T_0 = 2\pi/\omega_0$. For a driving frequency of $\omega_d = 2\omega_0$, such a response will seemingly appear as subharmonic even though the periodic drive is actually absent for $D=0$, as illustrated in Fig.~\ref{fig:3}(a).

Based on the location of the TC phases in the phase diagrams in Figs.~\ref{fig:2}(d), \ref{fig:2}(i), and \ref{fig:2}(j), for the closed DM, ADM, and LMG model, we identify that a period-doubling instability emerges for bang-bang protocols when the duty cycle follows
\begin{equation}\label{eq:cond}
D_\mathrm{ins} = 1-\frac{\omega_d}{2\omega_0}.
\end{equation}
The above condition appears as a line in the phase diagram and it can be analytically understood as follows. The magnetisation $j_x$ for the noninteracting limit will have the same magnitude but opposite sign as its initial value every $(n+1/2)T_0$, where $n$ is an integer. Hence, for the driven system, the dark time must be exactly half the precession period in the absence of spin interactions $t_\mathrm{dark}=T_0/2$. The instability condition Eq.~\eqref{eq:cond} precisely satisfies this:
\begin{equation}\label{eq:dark}
t_\mathrm{dark} = (1 - D_\mathrm{ins}) T_d = \frac{2\pi}{2\omega_0} = \frac{T_0}{2}.
\end{equation}
The state at times $t=(n+1/2)T_0$ is the symmetry broken partner of the initial state, which is chosen to be an eigenstate of the Hamiltonian with spin-spin interactions. As such, the states do not change during the bright times of each driving cycle, as depicted in the white areas in Fig.~\ref{fig:3}(c), which then yields the apparent period-doubling response for the bang-bang protocol. Therefore, the emergence of a period-doubling response in the absence of dissipation strongly hinges on the appropriate timing of when the interactions are switched on and off. This interplay between the internal dynamics of the spins and the timing of the external drive is also argued to be important for the $n$-tupling response in a variable-range interacting spin model with binary driving \cite{Kelly2021}.

We remark that the equivalence of the dynamics in the ADM and the LMG model for a TC is solely attributed to the specific form of the binary drive. For both models, during the bright times, the state of the system is the same initially prepared symmetry-broken phase defined by Eq.~\eqref{eq:sr1}. During the dark times, the additional terms in the ADM Hamiltonian [last two lines in Eq.~\eqref{eq:adm}] are also set to zero, which means that the resulting equations of motion are the same for both models. Thus, the spins in the ADM will simply precess in the same way as they would in the LMG model during the dark times.

In general, for an integer $m$, the period doubling arises if $t_\mathrm{dark} = (m+1/2)T_0$. The driving parameters for the isolated islands of TCs in Figs.~\ref{fig:2}(d), \ref{fig:2}(i), and \ref{fig:2}(j), and more clearly in Fig.~\ref{fig:1}(c) for $\kappa/\omega_0=\{0,10^3,\infty\}$ satisfy this general condition for the period-doubling instability. We emphasise that the arguments discussed so far hold only if the initial state is an eigenstate of the Hamiltonian during the bright times. Otherwise, the system will dephase and a clean period-doubling dynamics will not be observed. This will be discussed further in Sec.~\ref{subsec:isd}. Moreover, Eqs.~\eqref{eq:cond} and \eqref{eq:dark} only apply to binary drives, in which the system becomes noninteracting at well-defined times. For smooth sinusoidal driving, we do not find any clean period-doubling response for all relevant types of initial states in the absence of dissipation, as shown in Appendix \ref{app:contdrive}. This underscores the sensitivity to the specific driving protocol of the TC in the closed-system limit.

\subsubsection{Open systems}

We now discuss the results for the open DM with dissipation strength $0<\kappa/\omega_0 < 10^3$.
For $D=0$, the photonic and spin degrees of freedom decouple, leading to a spin dynamics equivalent to the $D=0$ case in the LMG model. The initially nonzero photon number eventually vanishes due to dissipation. The magnetisation $j_x$ oscillates at a frequency $\omega_0$ around zero, as seen in Fig.~\ref{fig:3}(b), and the apparent period doubling for $D=0$ is trivial since the periodic driving is in fact absent. 
We show an example of a dissipative TC in the DM in Fig.~\ref{fig:3}(d), in which the specific driving parameters yield bright and dark times that are both incommensurate to the precession period, $t_\mathrm{bright}  \approx 0.5385 T_0$ and $t_\mathrm{dark}  \approx 0.2308 T_0$, respectively. This demonstrates that the period-doubling instability conditions for the nondissipative limits based on Eqs.~\eqref{eq:cond} and \eqref{eq:dark} are no longer applicable, in general, when dissipation is present. 

Going from $\kappa=0$ to $\kappa/\omega_0 =0.1$ [Figs.~\ref{fig:2}(d) and \ref{fig:2}(e)] we see that, while time-crystalline phases remain along the line defined by Eq.~\eqref{eq:cond}, new TCs start to emerge in other parts of the phase diagram associated with driving parameters that would otherwise lead to thermal phases in the closed DM. Moreover, some of the thermal phases for $\kappa=0$ are converted to not only TCs but also TQCs after dissipation is introduced. 
Thus, we provide a concrete demonstration of dissipation, the photon decay, counteracting the heating induced by the periodic drive. Increasing the dissipation strength pushes the TCs away from the instability line in the closed-system limit, as seen from the change in the phase diagram from $\kappa/\omega_0 =0.1$ to $\kappa/\omega_0 =1$, see Figs.~\ref{fig:2}(e) and \ref{fig:2}(f). Further increase in the dissipation strength leads to an expansion of the area in the phase diagram with TCs, as demonstrated in Figs.~\ref{fig:1}(c) and \ref{fig:2}(d) for $\kappa/\omega_0 \in [1,10]$. 

Note, however, that the dissipation-induced enhancement of TC in the phase diagram only applies up to a certain value of $\kappa$. In Fig.~\ref{fig:1}(c), comparing the area of the time-crystalline phase in $\kappa/\omega_0 = 5$  and $\kappa/\omega_0 = 10$, we find that the TC area decreases for $\kappa/\omega_0 > 5$. While the overall shape of the area with both TC and TQC is not significantly changed from $\kappa/\omega_0=10$ to $\kappa/\omega_0=10^2$, as displayed in Figs.~\ref{fig:2}(g) and \ref{fig:2}(h), there are more TQCs in the phase diagram for $\kappa/\omega_0=10^2$ than for $\kappa/\omega_0=10$, which implies that the TCs are converted to TQCs with increasing dissipation strength. This can also be inferred from the expansion of the TQC domain as the dissipation strength increases from $\kappa/\omega_0=10$ to $\kappa/\omega_0=21$ in Appendix \ref{app:ex}.

We have seen that, for the ADM and LMG model, the TCs are restricted along the instability line Eq.~\eqref{eq:cond}. The question remains whether the phase diagrams for dissipative systems will change gradually or suddenly as $\kappa$ increases to large enough values, such that the adiabatic approximation and thus the ADM and LMG model can be applied. To address this issue, we consider even stronger dissipation strengths on the order of $\kappa/\omega_0 \sim 10^2$ while still solving the full semiclassical equations including the photon dynamics. For even stronger dissipation beyond the optimal value, we find that the dynamical phase diagram gradually develops features that resemble its closed-system counterpart, as seen in Figs.~\ref{fig:1}(c) and \ref{fig:2}(d) for $\kappa/\omega_0 = 10^2$. Comparing $\kappa/\omega_0 = 10^2$ and $\kappa/\omega_0 = \{10^3,\infty\}$, some of the TQCs, which were previously TCs for weaker dissipation, turn into thermal phases in the closed-system models. Moreover, going from  $\kappa/\omega_0 = 10$ to $\kappa/\omega_0 = 10^2$, the time-crystalline phases start to gather toward the instability line Eq.~\eqref{eq:cond} for closed systems. 

To summarise Sec.~\ref{subsec:dp}, we identify the condition for creating TCs in the closed-system limit with a periodic binary drive or bang-bang protocol. We also demonstrate that dissipation, in general, leads to the expansion of the TC and TQC areas in the phase diagram. The two limits $\kappa=0$ and $\kappa\to\infty$ are smoothly connected by the gradual change of the phase diagram as dissipation is increased.

\begin{figure}[!htpb]
\centering
\includegraphics[width=1\columnwidth]{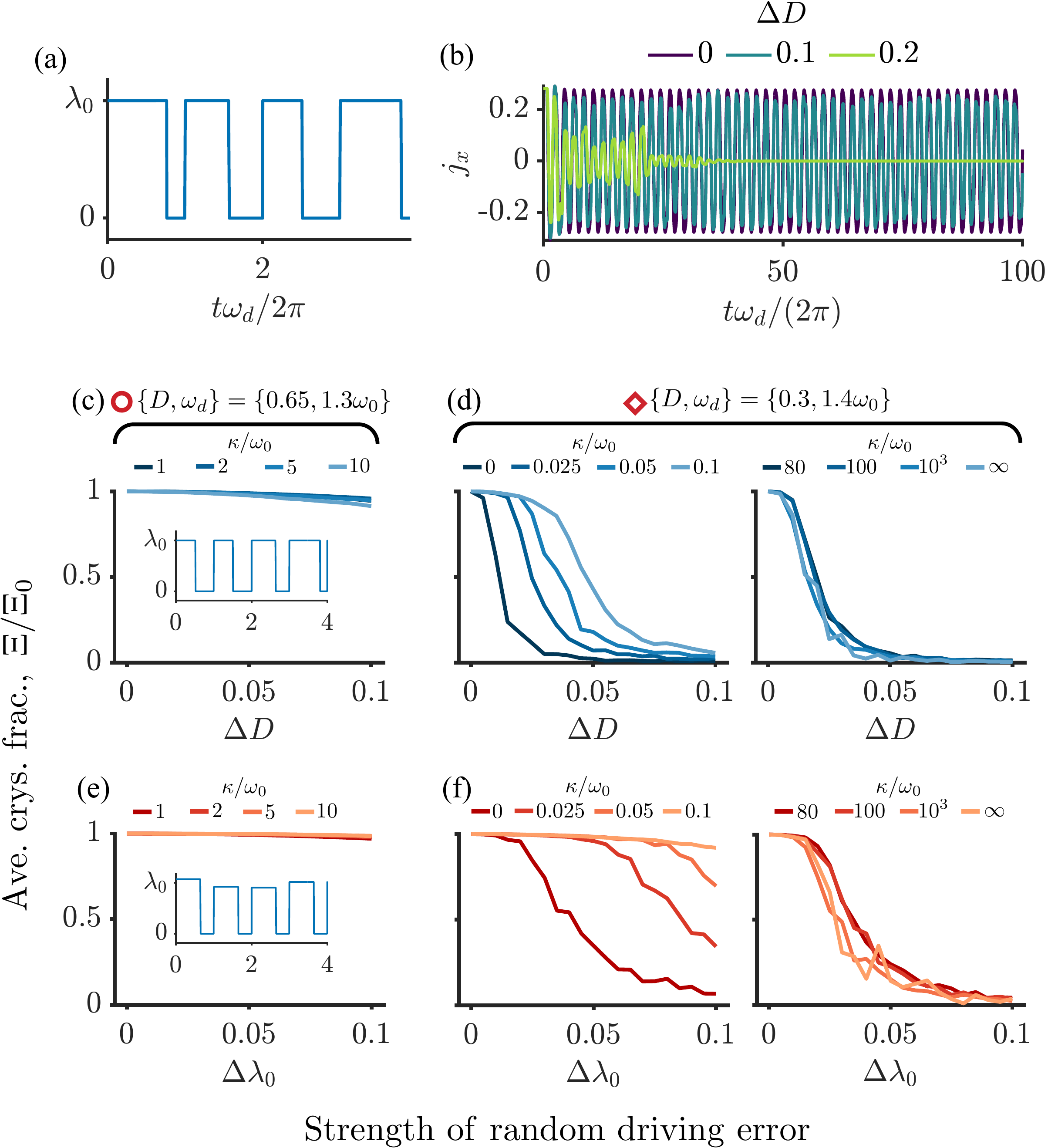}
\caption{(a) One realisation of the disordered drive. (b) Dynamics of the total magnetisation along the $x$ direction $j_x$ for different disorder strengths $\Delta D$ as indicated in the legend. The driving parameters are $\{D,\omega_d\}=\{0.65,1.3\omega_0\}$, and the dissipation strength is $\kappa/\omega_0 = 1$. The initial state is a $\mathbb{Z}_2$-symmetry broken phase for $\lambda_0 = 1.1\lambda_\mathrm{cr}$. (c) and (d) Dependence of the relative crystalline fraction  $\Xi/\Xi_0$ on the strength of the random driving error or temporal disorder $\Delta D$. (e) and (f) Like (c) and (d) but for a noisy light-matter coupling with disorder strength $\Delta \lambda$. Insets: One realisation of the disordered drive.}
\label{fig:4} 
\end{figure}

\subsection{Robustness against random driving errors}\label{subsec:rob}

We will now investigate the role of dissipation on the robustness of TCs against temporal noise. To this end, we introduce a random driving error in the duty cycle for every Floquet drive:
\begin{equation}\label{eq:dis1}
\lambda(t) = \biggl\{
\begin{matrix}
\lambda_0, & nT_d \leq t < (n+D_n)T_d \\
0, & (n+D_n)T_d \leq t < (n+1)T_d,
\end{matrix}
\end{equation}
where $D_n = D+\Delta D_n$, and $\Delta D_n$ is a random number drawn from a box distribution $\Delta D_n \in [-\Delta D,\Delta D]$. A single realisation of this disordered drive is depicted in Fig.~\ref{fig:4}(a) [see also the inset of Fig.~\ref{fig:4}(c)]. 
We also consider another kind of temporal perturbation, namely, in the light-matter coupling strength such that
\begin{equation}\label{eq:dis2}
\lambda(t) = \biggl\{
\begin{matrix}
\lambda_0+ \lambda_n, & nT_d \leq t < (n+D)T_d \\
0, & (n+D)T_d \leq t < (n+1)T_d,
\end{matrix}
\end{equation}
where $\lambda_n/\lambda_0 \in [-\Delta \lambda_0,\Delta \lambda_0]$. An example of a periodic drive with this disorder is shown in the inset of Fig.~\ref{fig:4}(e).

In the following, we use driving parameters corresponding to the circles and diamonds in Fig.~\ref{fig:2}, where TCs exist for clean driving or in the absence of temporal disorder.
We take 100 disorder realisations when calculating the dynamics of $j_x$ and the crystalline fraction $\Xi$, which we define as the average of the power spectrum of $j_x$ at $\omega_d/2$. 
\begin{figure*}[!htpb]
\centering
\includegraphics[width=1.9\columnwidth]{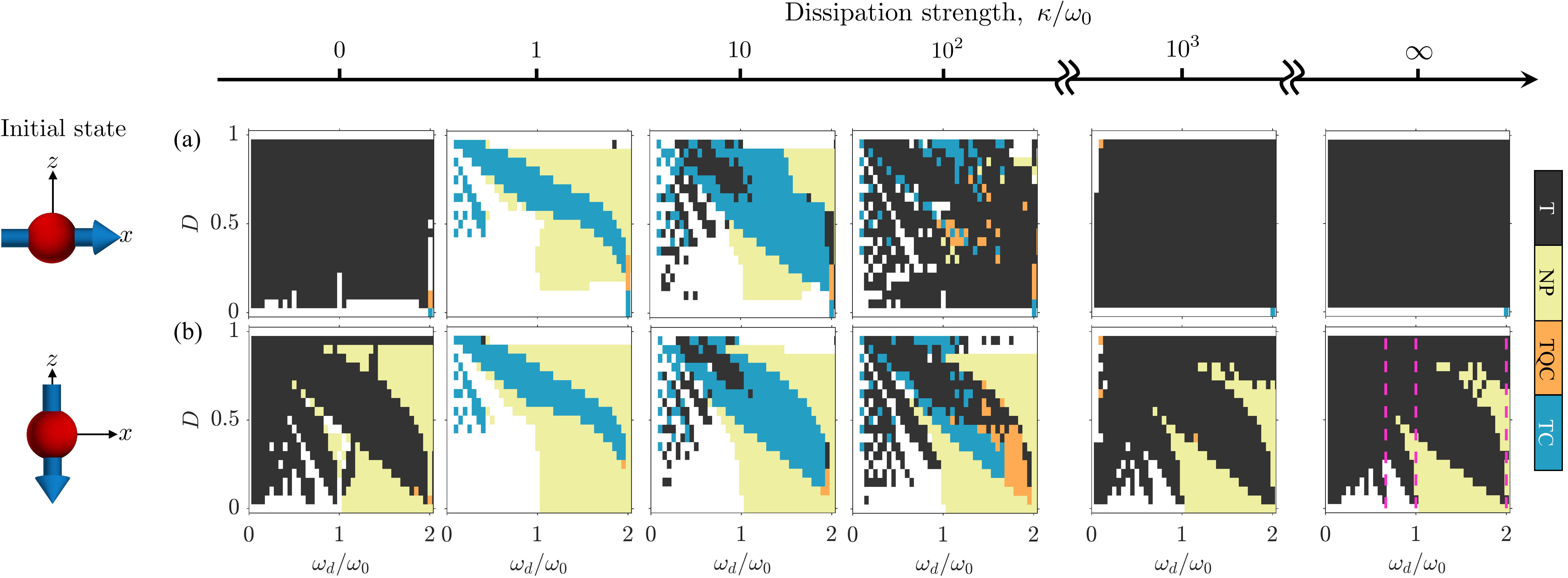}
\caption{Dynamical phase diagrams like Fig.~\ref{fig:2} but for initial fully polarised states (a) $\lvert\Rightarrow\rangle$ and (b) $\lvert\Downarrow\rangle$ as schematically illustrated in the left-most panels. The dashed vertical lines in $\kappa=\infty$ denote the parametric resonance condition $\omega_d/\omega_0 = 2/n$, where $n=\{1,2,3\}$. The system parameters are the same as in Fig.~\ref{fig:2}.}
\label{fig:5} 
\end{figure*}
We present in Fig.~\ref{fig:4}(b) the disorder-averaged dynamics of $j_x$ for a noisy duty cycle, Eq.~\eqref{eq:dis1}. As the disorder strength increases, the oscillation amplitude of $j_x$ deviates from being a constant as the TC becomes unstable. This is expressed in the reduction of the relative crystalline fraction $\Xi/\Xi_0$, where $\Xi_0$ is the crystalline fraction in the clean case, for increasing disorder strength $\Delta D$, as shown in Figs.~\ref{fig:4}(c) and \ref{fig:4}(d). Additional frequencies introduced by the noise broaden the power spectrum of $j_x$ and thereby decrease the crystalline fraction. More importantly, Figs.~\ref{fig:4}(c) and \ref{fig:4}(d) demonstrate another key finding of this paper, which is the role of dissipation in making a TC more robust against temporal noise. For the nondissipative cases $\kappa/\omega_0=\{0,10^3,\infty\}$ in Fig.~\ref{fig:4}(d), the crystalline fraction decays rapidly with $\Delta D$. In contrast, the decay is slower when dissipation is introduced, i.e., the crystalline fraction remains large over a wide range of disorder strengths. This is evident in Fig.~\ref{fig:4}(c) for intermediate dissipation strengths, wherein the crystalline fraction is found to slowly decrease with $\Delta D$. 

The dissipation-induced robustness against temporal noise can be understood as a consequence of the dissipation-induced expansion of the TC area in the phase diagram discussed in the previous subsection. In the phase diagram for $\kappa/\omega_0=10$ in Fig.~\ref{fig:2}(g), the TC corresponding to the driving parameters marked by the circle is surrounded by other period-doubling TCs, and thus, a perturbation in $D$, $\omega_d$, and $\omega_0$ will not easily push the system into a different dynamical phase. On the other hand for closed systems, we see in Figs.~\ref{fig:2}(d), \ref{fig:2}(i), and \ref{fig:2}(j), that for driving parameters marked by the diamonds, a slight variation in $D$ away from the instability condition Eq.~\eqref{eq:cond} will take the system to a different dynamical phase other than a period-doubling TC. This leads to a TC that is less robust against temporal perturbations of the driving parameters $D$ and $\omega_d$. This also explains the relatively weak robustness observed for strong dissipation in the right panel of Fig.~\ref{fig:4}(d) since the TC area is relatively small and highly fragmented for dissipation strengths of this order of magnitude, as seen  for $\kappa/\omega_0 = 10^2$ in Fig.~\ref{fig:1}(c).
In Figs.~\ref{fig:4}(e) and \ref{fig:4}(f), we observe similar findings for a drive with noisy light-matter coupling. Both dissipative and nondissipative models appear to be more robust against this type of noise, as seen from the larger plateaus in the crystalline fractions in Fig.~\ref{fig:4}(f) than those in Fig.~\ref{fig:4}(d). This can be attributed to the presence of TCs even for higher values of $\lambda_0$, as seen in Appendix \ref{app:param}.

\subsection{Initial fully polarised states}\label{subsec:isd}

For potential applications and experimental realisations, we discuss how close the initial state must be to the desired state to create a TC. 
So far, we have considered one of the symmetry-broken states as the initial state. 
In Ref.~\cite{Russomanno2017}, robustness against the choice of initial state for TCs in the kicked LMG model has been demonstrated but only for initial symmetry-broken states corresponding to an interaction strength different from the one in the Hamiltonian, i.e., $\lambda(t=0) \neq \lambda_0$.
Here, we explore other types of initial states, namely, fully polarised states either along the positive $x$ direction, $\{j_x,j_y,j_z\}=\{1/2,0,0\}$, or negative $z$ direction, $\{j_x,j_y,j_z\}=\{0,0,-1/2\}$, which we label as $\lvert\Rightarrow\rangle$  or $\lvert\Downarrow\rangle$, respectively. A symmetry-broken state interpolates between these two limits. For the DM, we include a small fluctuation in the photon mode, such that $a(t=0)=0.01$. 

We present in Figs.~\ref{fig:5}(a) and \ref{fig:5}(b) the evolution of the phase diagrams as a function of the dissipation strength for initial fully polarised states $\lvert\Rightarrow\rangle$ and $\lvert\Downarrow\rangle$, respectively.
Crucially, we find that, for both types of fully polarised initial states, time-crystalline phases are absent in the closed system models $\kappa/\omega_0=\{0,10^3,\infty\}$, and the phase diagrams are dominated by thermal phases, see also Figs.~\ref{fig:1}(d) and \ref{fig:1}(e). The behaviour is strikingly different for dissipative cases as seen in Fig.~\ref{fig:5} for $\kappa/\omega_0=\{1,10\}$. The choice of initial state between $\lvert\Rightarrow\rangle$ and $\lvert\Downarrow\rangle$ does not significantly alter the area in the phase diagram with TCs. This is further emphasised if we include the initial symmetry-broken state in the comparison as evidenced by the results for $\kappa/\omega_0 \in [1,10]$ in Figs.~\ref{fig:1}(c)-\ref{fig:1}(e). This implies that dissipation allows for flexibility in the fidelity of the initial state preparation. 
In Appendix \ref{app:contdrive}, we observe similar results for a smooth sinusoidal or continuous driving protocol, which further corroborates the positive role of controlled dissipation for infinite-range interacting spin systems.

The results for the ADM and LMG model $\kappa/\omega_0=\{10^3,\infty\}$, shown in Fig.~\ref{fig:5}(b), exhibit resonance lobes reminiscent of parametric resonances that appear when the driving frequency satisfies $\omega_d/\omega_0 = 2/n$, where $n \in \mathbb{Z}^+$.  Notice that in Fig.~\ref{fig:5}(b), the shape of the TC area for $\kappa/\omega_0=10$ is like that of the primary resonance lobe ($\omega_d/\omega_0 = 2$) for $\kappa/\omega_0=\{10^3,\infty\}$. This points to a period-doubling instability arising from a parametric resonance as the main mechanism behind the formation of dissipative TCs with binary driving, like the smooth sinusoidal driving in Refs.~\cite{Chitra2015,Kessler2021}. Note that the parametric resonance also applies to initial symmetry broken states as evinced by the shape of the thermal region, including the TC and TQC phases, in Fig.~\ref{fig:2}(d) for $\kappa=0$. There, the absence of dissipation heats up the system, resulting in a more prominent thermal phase except at the special points along the instability line for initial symmetry-broken states, Eq.~\eqref{eq:cond} .

\section{Quantum results}\label{sec:qm}

We now study the TCs in the limit of a small number of spins, wherein quantum effects and many-body correlations become dominant. Platforms for physical implementations of a relatively small number of artificial or effective spins include circuit QED systems based on superconducting qubits \cite{Blais2007,Mlynek2014,Bamba2016,Diaz2017,Yoshihara2017,Gong2018} and ion chains \cite{Korenblit2012,Jurcevic2017,Monroe2021}.
In the following, we obtain the full quantum results using the QuantumOptics.jl library \cite{Kramer2018} and we employ the discrete truncated Wigner approximation (DTWA) \cite{Schachenmayer2015,Huber2022} for a larger number of spins beyond the reach of full quantum mechanical simulations.

We focus on the initial fully polarised state along the positive $x$ direction $\lvert\Rightarrow \rangle = \bigotimes_N \lvert\rightarrow\rangle$, which in the mean-field regime corresponds to $\{j_x,j_y,j_z\}=\{\frac{1}{2},0,0\}$, to gain insights into the features of TCs in the quantum regime. By comparing exact quantum and DTWA results, we will also assess the applicability of DTWA in capturing the time crystalline dynamics for periodically driven infinite-range interacting spins. For the DM, the photon mode is initialised in the vacuum state $|0\rangle$, such that the initial state of the system is $\lvert \psi(t=0)\rangle = \lvert\Rightarrow \rangle \otimes |0\rangle$.

\begin{figure}[!t]
\centering
\includegraphics[width=\columnwidth]{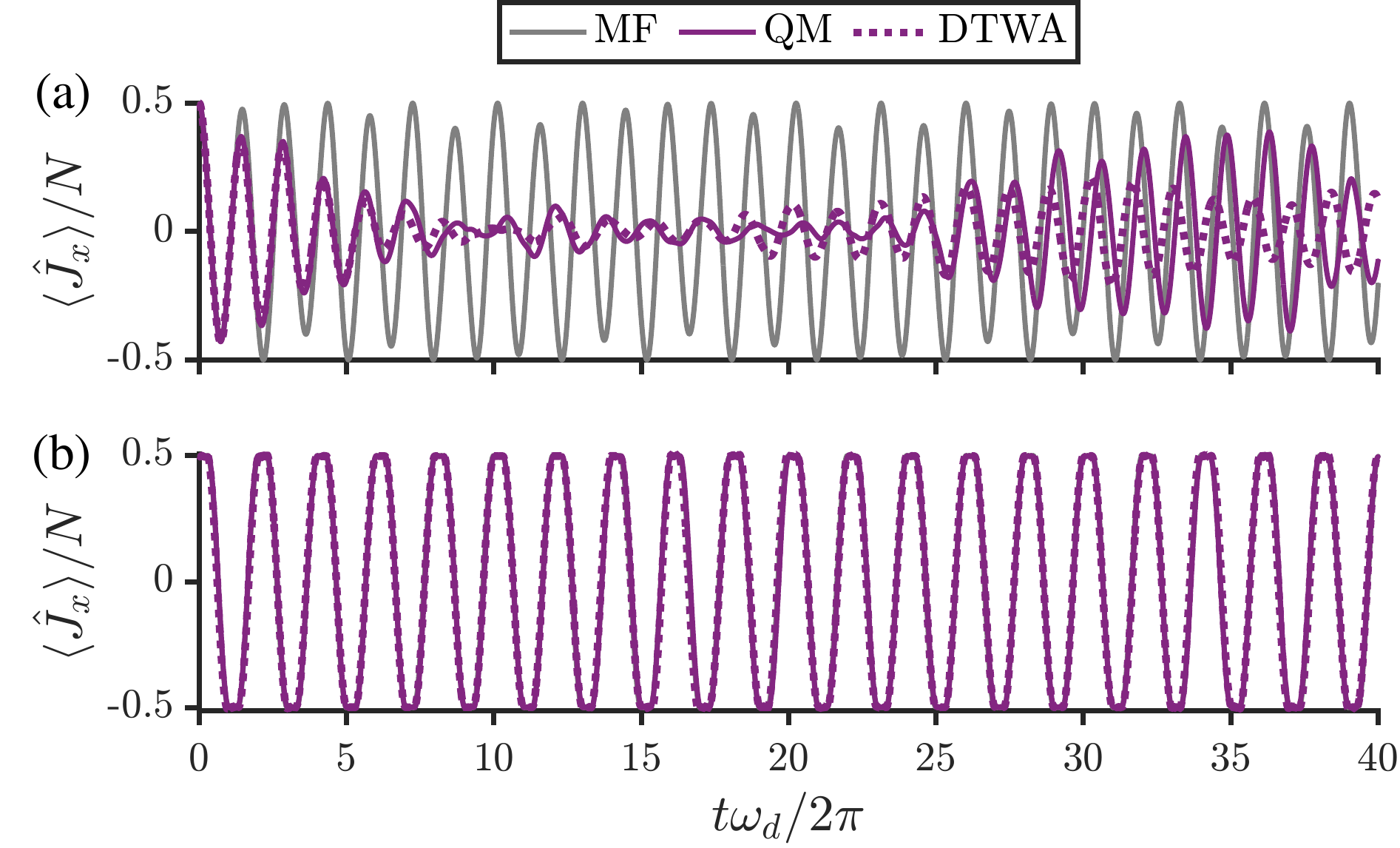}
\caption{Dynamics of the expectation value of the total magnetisation along the $x$-direction for an initial fully polarised state $\lvert\Rightarrow\rangle$ in the Lipkin-Meshkov-Glick (LMG) model. The interaction strengths are (a) $\lambda_0 = 1.1\lambda_\mathrm{cr}$ and (b) $\lambda_0 = 4.0\lambda_\mathrm{cr}$. For the exact quantum mechanical (QM) results and discrete truncated Wigner approximation (DTWA), the number of spins is $N=8$. The driving parameters are $D=0.3$ and $\omega_d=1.4\omega_0$.}
\label{fig:6} 
\end{figure}
The results for the LMG model with $N=8$ spins and driving parameters $\{D,\omega_d\}=\{0.3,1.4\omega_0\}$ are depicted in Fig.~\ref{fig:6}. In Fig.~\ref{fig:6}(a) for $\lambda_0=1.1\lambda_\mathrm{cr}$, the system is in the thermal phase even in the mean-field limit of $N\to \infty$. This again exemplifies the importance of initialising the system in a symmetry broken eigenstate to create a TC in the closed-system limit. In the quantum regime, the irregular mean-field dynamics translate into a beating of the oscillations in the expectation value of the total magnetisation $\langle \hat{J}_x \rangle /N$ like the behaviour found in the kicked LMG model \cite{Russomanno2017}. The full quantum mechanical and DTWA results agree on the overall qualitative behaviour of the dynamics. While we find excellent agreement between the exact and DTWA results for short times, quantitative deviations appear in the long-time dynamics, which is expected in simulations of closed system quantum dynamics using phase-space methods \cite{Polkovnikov2010}. 

For stronger interactions, e.g., $\lambda_0 = 4 \lambda_\mathrm{cr}$ in Fig.~\ref{fig:6}(b), a TC is formed, and interestingly, the mean-field, exact quantum, and DTWA results agree for the entire simulation time of 100 driving cycles, which is noteworthy, considering the relatively small number of spins $N=8$. This also hints at the ability of the DTWA to capture the dynamics of TCs even for long times, provided that the interactions in a fully connected model are sufficiently strong.
We note that the overlap of a symmetry-broken eigenstate with the fully polarised state along the $x$ direction increases with the interaction strength, which can also be inferred from the mean-field steady-state solution in Eq.~\eqref{eq:sr1}. This explains the appearance of a TC in Fig.~\ref{fig:6}(b) despite the initial state not being a symmetry-broken eigenstate for $\lambda_0 = 4 \lambda_\mathrm{cr}$. Thus, we propose utilising large interactions strengths for creating TCs in fully connected systems with few spins if, for a given platform, it is easier to prepare an initial fully polarised state.

\begin{figure}[!hbtp]
\centering
\includegraphics[width=\columnwidth]{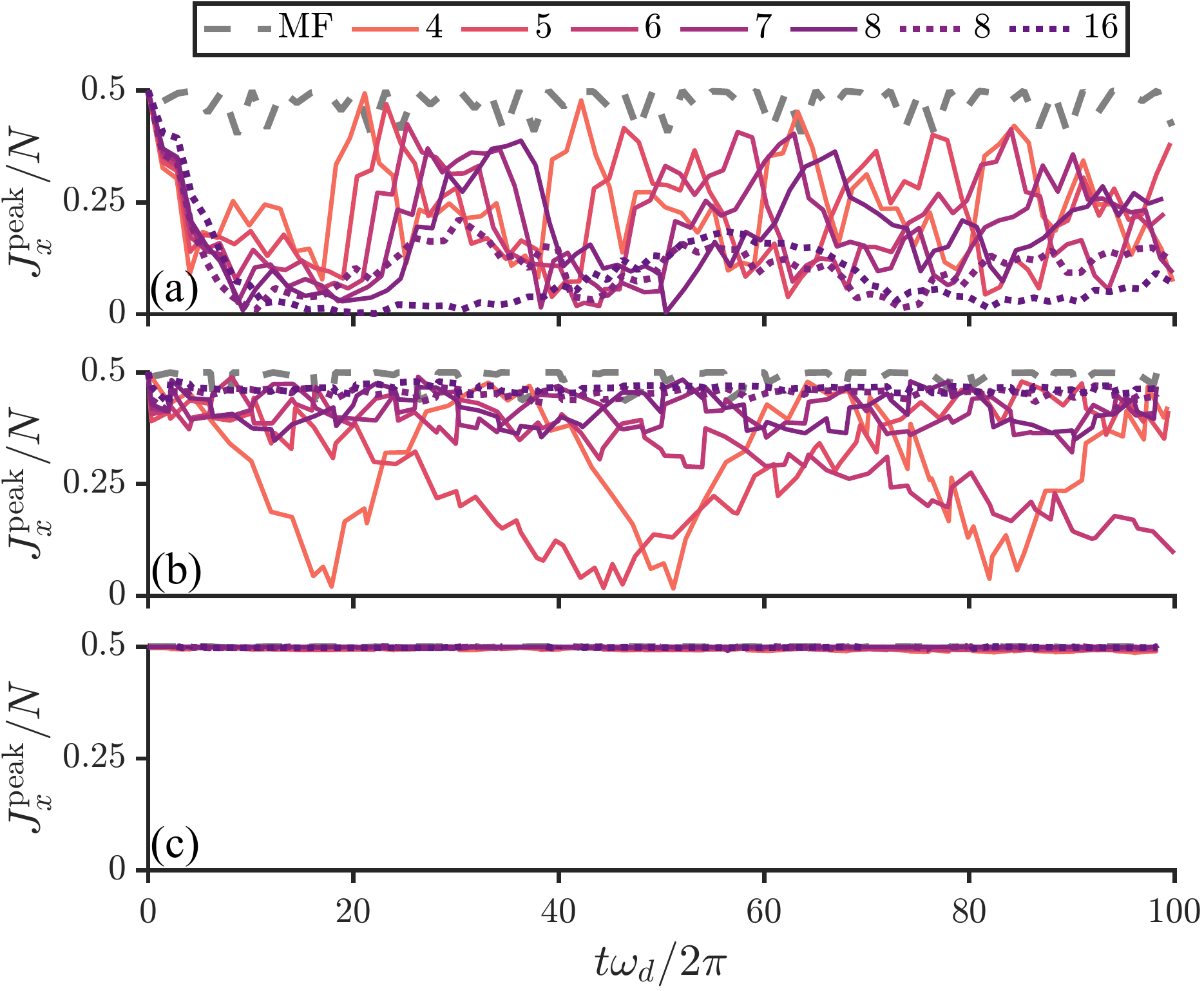}
\caption{Dynamics of the peaks in the total magnetisation in the Lipkin-Meshkov-Glick (LMG) model for an initial state of $\lvert\Rightarrow\rangle$. The solid (dotted) curves denote the full quantum [discrete trucated Wigner approximation (DTWA)] results. The interaction strengths are (a) $\lambda_0 = 1.1\lambda_\mathrm{cr}$, (b) $\lambda_0 = 2.0\lambda_\mathrm{cr}$, and (c) $\lambda_0 = 4.0\lambda_\mathrm{cr}$. The driving parameters are the same as in Fig.~\ref{fig:6}.}
\label{fig:7} 
\end{figure}
Next, we study the dependence of the beating oscillations on the number of spins in the LMG model. To this end, we obtain the peaks in the oscillatory dynamics of the magnetisation $J^\mathrm{peak}_x$, which is directly related to the envelope of the oscillations in $\langle \hat{J}_x \rangle$. In Fig.~\ref{fig:7}, we display the dynamics of $J^\mathrm{peak}_x$ for different $N$ including the mean-field limit. For weak interactions, the chosen driving parameters in Fig.~\ref{fig:7} lead to irregular and therefore non-time-crystalline dynamics. The convergence toward the mean-field limit for increasing $N$ is slow and can only be seen at short times due to the irregularity of the long-time dynamics. The tendency toward the mean-field prediction becomes more clear for stronger interactions, as seen in Fig.~\ref{fig:7}(b) for  $\lambda_0=2\lambda_\mathrm{cr}$. We observe that the beat period increases with $N$, implying that it becomes infinitely large as $N \to \infty$, consistent with the mean-field prediction of an infinitely long-lived TC. This behaviour is more apparent if the system is initialised in a symmetry-broken eigenstate as shown in Appendix \ref{app:qm}. For sufficiently strong interactions represented by $\lambda_0=4\lambda_\mathrm{cr}$ in Fig.~\ref{fig:7}(b), we recover results consistent with Fig.~\ref{fig:6}(b), especially the emergence of long-lived period-doubling response for a relatively small number of spins ($N\sim 4$).

\begin{figure}[!b]
\centering
\includegraphics[width=\columnwidth]{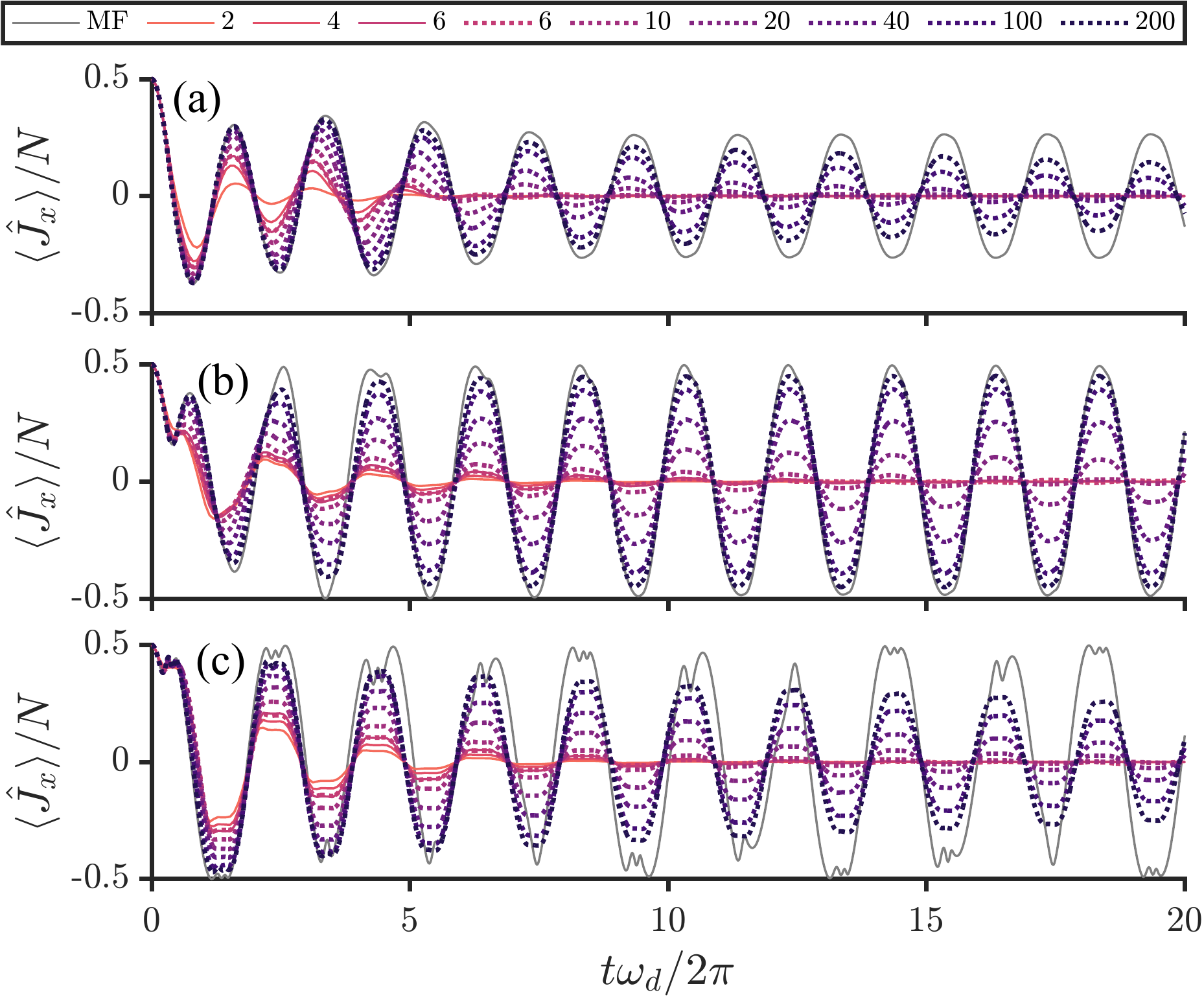}
\caption{Like Fig.~\ref{fig:6} but for the open Dicke model with $\kappa = \omega_0$. The light-matter coupling strengths are (a) $\lambda_0 = 1.1\lambda_\mathrm{cr}$, (b) $\lambda_0 = 2.0\lambda_\mathrm{cr}$, and (c) $\lambda_0 = 4.0\lambda_\mathrm{cr}$. The photon frequency is $\omega_p=\omega_0$. The driving parameters are $D=0.5$ and $\omega_d=1.6\omega_0$.}
\label{fig:8} 
\end{figure}
We present in Fig.~\ref{fig:8} the quantum dynamics in the open DM for $\kappa=\omega_0$. In Fig.~\ref{fig:8}(a), the driving parameters correspond to a TC in the mean-field limit. For few spins, the period-doubling oscillations rapidly decay and for $N<10$, the time-translation symmetry-breaking response only survives for short times, typically around five driving cycles $t \approx  5 T_d$. These exponentially decaying oscillations are analogous to the beating oscillations in the closed-system limit. However, unlike the beat period in the LMG model, the decay constant characterising the exponential suppression of oscillations in the open DM does not monotonously depend on the interaction strength. This is evident from the longer-lived oscillations in Fig.~\ref{fig:8}(b) compared with those in Fig.~\ref{fig:8}(c), even though $\lambda_0$ is larger in Fig.~\ref{fig:8}(c). This means that using the interaction strength to prolong the lifetime of a TC in the open DM is not as efficient as in closed systems, if the driving parameters are fixed. Alternatively, increasing the number of spins could also increase the lifetime of a dissipative TC \cite{Tucker2018,Iemini2018,Gambetta2019}. Indeed, we find in Fig.~\ref{fig:8} that the decay slows down with $N$, irrespective of the interaction strength. In contrast to the LMG model, in which as few as $N=4$ spins generate a period-doubling signal lasting for $t>10T_d$, the number of spins needed for the open DM for the same time scale is $N > 20$.

We point out that, in Fig.~\ref{fig:8}(c), despite the mean-field dynamics showing irregular or chaotic behaviour, both full quantum and DTWA simulations predict periodic albeit decaying oscillations. This apparent inconsistency between mean-field and quantum approaches, regarding the presence or absence of a transition to a chaotic phase, is also reported in a driven-dissipative LMG or fully connected Ising model \cite{Zhihao2023}.
Lastly, we note that, for $N=6$, in Fig.~\ref{fig:8}, DTWA is in good agreement with the numerical data obtained from the full quantum mechanical treatment, thereby suggesting that, in dissipative scenarios, DTWA can capture time-crystalline dynamics even for small $N$. This stabilising effect of dissipation on the performance of DTWA as a method is like that found in the positive-$P$ approach for driven-dissipative bosons \cite{Deuar2021}.

\section{Summary and Discussion}\label{sec:conc}

In this paper, we have extensively studied the influence of dissipation on TCs in a spin system with infinite-range interactions with binary driving. We have employed both mean-field and quantum mechanical treatments of the dynamics in the open DM for different dissipation strengths.
For large dissipation strengths $\kappa > 10^2$, we approximate the system as closed using the ADM and LMG model. In Table \ref{table:1}, we summarise the key properties of TCs, specifically, robustness in the thermodynamic limit and dynamical features in the quantum limit, for the closed-system and dissipative regimes.

\begin{table}[!htp]
\renewcommand{\arraystretch}{1.25}
\begin{tabular}{l c c c}
\hline\hline
 & Closed  & & Open   \\
 & (LMG model) & & (DM) \\
\hline
Mean-field & &  & \\
\quad Robust against: &  & & \\
\quad\quad Random errors in the drive & Weak & & Strong \\
\quad\quad Choice of initial state & Weak & & Strong \\
\quad\quad Choice of driving protocol & Weak & & Strong \\
\quad\quad Variation in system parameters  & Strong* &  & Strong$^\dagger$ \\
\\
\hline
Quantum & &  \\
\quad Oscillations & Beating & & Exponential  \\
\quad & & &  decay \\
\\
\quad Lifetime increases with & Interaction  & & Number of  \\
\quad & strength &  &  spins \\
\hline\hline
\end{tabular}
\normalsize
\caption{Summary of the properties of the period-doubling time crystals in infinite-range interacting spins. In the LMG model, ``Strong*" means that it is strongly robust only for variations and random errors in the interactions strength. In the open Dicke model, ``Strong$^\dagger$" means strong robustness only within the resonance area in the phase diagram.}
\label{table:1}
\end{table}

From our mean-field approach, we have identified a simple but finely tuned set of conditions, involving the driving parameters and initial state, for creating a period-doubling response in the closed-system limit. We have demonstrated that dissipation expands this instability line to include larger areas in parameter space. Thus, we connect the TC phenomenology in the open- and closed-system limits of the infinite-range interacting spins. Moreover, we have observed that the presence and lifetime of TCs do not monotonously depend on the dissipation strength. This implies the existence of an optimal dissipation strength for realising TCs, like dissipative-driven Heisenberg chains \cite{Vu2022}. However, here we show that the optimal dissipation depends strongly on the specific choice of driving parameters, and in certain cases, the absence of dissipation, $\kappa=0$ or $\kappa\to\infty$,  could in fact be the optimal choice, if one is only interested in generating a period-doubling response. If the goal, however, is to create a TC that is also robust against unwanted errors in the drive and imperfect preparation of the initial state, we ascertain that controlled dissipation is helpful. We find that the TC area in the phase diagram becomes relatively large for intermediate dissipation strengths  $\kappa\sim\omega_0$. A large TC area in the phase diagram contributes to the robustness not only against variations in system parameters but also against noise in the drive. Furthermore, we demonstrate that dissipation can form TCs, which are insensitive to the choice of initial state.  We also attribute the formation of dissipative TCs using a binary drive to a period-doubling instability of a parametric resonance, and thus, we generalise the mechanism and conditions proposed in Ref.~\cite{Gong2018}. 

Our quantum results for finite $N$ obtained using numerically exact calculations and the DTWA indicate an exponential decay of the period-doubling oscillations when dissipation is present. On the other hand, in the two extremes $\kappa=0$ or $\kappa \to \infty$, the TCs exhibit beating behaviour, the period of which increases with the number of spins, consistent with Ref.~\cite{Russomanno2017}.
The scaling with the interaction strength of the lifetime of closed-system TCs is more favourable than the scaling with the number of spins for open-system TCs. This suggests a possible advantage of TCs in the closed-system limits if the underlying platform operates with few spins, albeit the driving parameters must be finely tuned according to Eq.~\eqref{eq:cond}.

Finally, we remark on the apparent lack of experimental evidence for TCs in the closed fully connected spin systems. As we have shown in this paper, the period-doubling instability in the LMG model and the closed DM strongly depends on the specific driving protocol. 
For sinusoidal driving, which was utilised for the realisation of dissipative TC in the small-$\kappa$ regime of a cavity-QED system \cite{Kessler2021}, the DM with $\kappa=0$ and $\kappa\to\infty$ does not host any TCs as shown in Appendix \ref{app:contdrive}. Instead, a binary drive according to Eq.~\eqref{eq:lamt} is required to induce a period-doubling response but only in a narrow region in the phase diagram spanned by the driving parameters, i.e., they must follow Eq.~\eqref{eq:cond}. 
It remains to be seen whether alternative schemes that periodically drive the transverse field (as in Refs.~\cite{Russomanno2017,Pizzi2021NC}), instead of the spin-spin interaction strength (as done here), would yield a larger TC area in the relevant phase diagram.
Assuming a binary drive, high-fidelity state preparation is still required, i.e., the initial state should not veer too far from the symmetry-broken state of the Hamiltonian during the bright times.
For the cavity-QED system operating in the regime that emulates the ADM and LMG models, which is realised for dissipation strengths that are several orders of magnitude larger than the atomic transition frequency, $\kappa \gg \omega_0$ \cite{Baumann2010}, the above considerations for the driving protocol and initial state preparation may not be an issue. However, for this system, authors of future studies need to address whether the large bandwidth of the cavity would cause higher momentum modes to participate in the dynamics. If so, this leads to a breakdown of the two-level approximation of the atoms and therefore the mapping onto effective spin-$\frac{1}{2}$ particles.

\begin{acknowledgments}
This paper was funded by the UP System Balik PhD Program (OVPAA-BPhD-2021-04) and the Deutsche Forschungsgemeinschaft SFB-925 Project No. 170620586 and the Cluster of Excellence Advanced Imaging of Matter (EXC 2056), Project No. 390715994. J.S. acknowledges support from the German Academic Scholarship Foundation.  We thank C. Sevilla for helpful discussions.
\end{acknowledgments}

\setcounter{equation}{0}
\setcounter{table}{0}
\appendix

\begin{figure*}[!hptb]
\centering
\includegraphics[width=2\columnwidth]{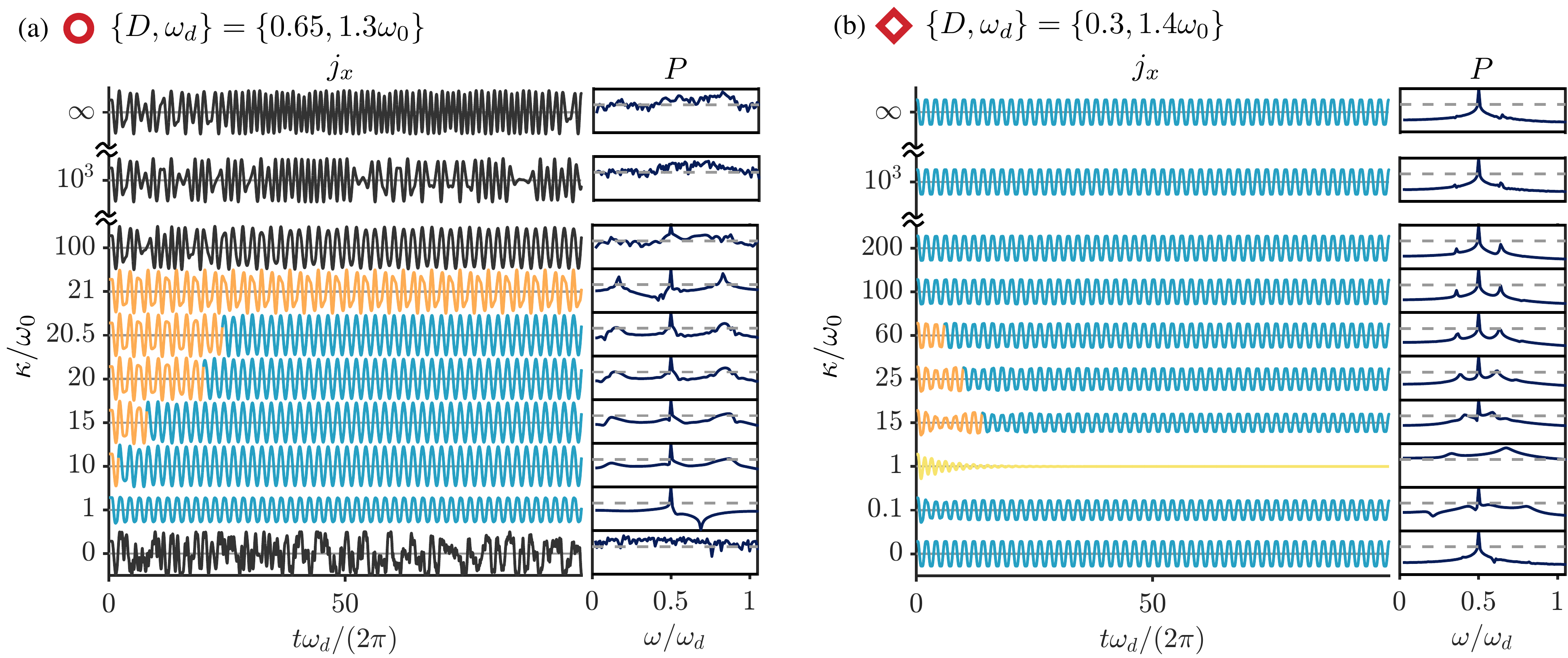}
\caption{Exemplary dynamics for different dissipation strengths $\kappa$ with fixed driving parameters denoted by the circles and diamonds in Figs.~\ref{fig:2}(d)-\ref{fig:2}(j). The left panels depict the dynamics of $j_x$, and the right panels show the corresponding power spectrum $\ln P$. The $y$ axis range of each plot is $[-0.5,0.5]$ for $j_x$ and $[{-22},0]$ for $\ln P$. The horizontal line in the power spectrum plots denote the threshold used for identifying the presence of a time quasicrystal (TQC), which is $\ln P = -8$. The remaining parameters are the same as in Fig.~\ref{fig:2}.}
\label{fig:dyn} 
\end{figure*}
\section{Equations of motion}\label{app:eom}
For a Hamiltonian $\hat{H}$ and the type of dissipator in Eq.~\eqref{eq:rho}, the dynamics of the expectation value of an operator $\hat{O}$ is
\begin{equation}
\partial_t\langle \hat{O}\rangle = \frac{i}{\hbar}\langle[\hat{H},\hat{O}]\rangle + \kappa \biggl\langle \left( 2\hat{a}^{\dagger}\hat{O}\hat{a} -  \hat{a}^{\dagger}\hat{a}\hat{O} - \hat{O}\hat{a}^{\dagger}\hat{a} \right)\biggr \rangle.
\end{equation}
Within mean-field theory, we approximate $\langle \hat{a} \hat{J}_\mu \rangle \approx \langle \hat{a} \rangle \langle \hat{J}_\mu \rangle$. We present the equations of motion for collective spins and individual spins, as the former is used in the mean-field treatment, while the latter is used in DTWA.

\subsection{DM}
The equations of motion for the DM are
\begin{align}
\partial_t a &= -(i \omega_p + \kappa){a}  - i{2\lambda}j_x \\
\partial_t j_x &=-\omega_0 j_y \\
\partial_t j_y &= \omega_{0} j_x - 2\lambda (a+a^*)  j_z \\
\partial_t j_z &=2\lambda (a+a^*) j_y 
\end{align}
If we decompose $j_\mu$ in terms of individual spins, we obtain
\begin{align}
\partial_t a &= -(i \omega_p + \kappa){a}  - i{\lambda}\frac{1}{\sqrt{N}}\sum_i s^x_i \\
\partial_t s^x_i &=-\omega_0 s^y_i \\
\partial_t s^y_i &=\omega_0 s^x_i  - 2\lambda\frac{1}{\sqrt{N}}(a+a^*) s^z_i\\
\partial_t s^z_i &= 2\lambda\frac{1}{\sqrt{N}}(a+a^*) s^y_i
\end{align}
For beyond mean-field approaches, a fluctuation or stochastic term associated with the dissipation must be included in the equations of motion \cite{Ritsch2013,Huber2022}. In our implementation of the equations of motion governing the trajectories in the DTWA, we separate the real and imaginary components of the photon field, $a = a_\mathrm{R} + i a_\mathrm{I}$, which yields 
\begin{align}
d a_\mathrm{R} &=  \left(- \kappa a_\mathrm{R} + \omega_p a_\mathrm{I} \right)dt  + \sqrt{\frac{\kappa}{2}}dW_1 \\
d a_\mathrm{I} &=  \left(-\omega_p a_\mathrm{R}   - \kappa a_\mathrm{I}  - \frac{{\lambda}}{\sqrt{N}}\sum_j s^x_j\right) dt  +  \sqrt{\frac{\kappa}{2}}dW_2 \\
\partial_t s^x_i &=-\omega_0 s^y_i \\
\partial_t s^y_i &=\omega_0 s^x_i  - 4\lambda\frac{1}{\sqrt{N}}a_\mathrm{R} s^z_i\\
\partial_t s^z_i &= 4\lambda\frac{1}{\sqrt{N}}a_\mathrm{R} s^y_i.
\end{align}
The two independent Wiener processes $W_1$ and $W_2$ account for the stochastic noise, and they satisfy $\langle dW_i \rangle = 0$ and $\langle dW_i dW_j \rangle = \delta_{i,j}\, dt$.

\subsection{ADM}
Next, for the ADM in the thermodynamic limit, the equations of motion are \cite{Damanet2019,Jager2022}
\begin{align}
\partial_t j_x &=-\omega_0 j_y \\
\partial_t j_y &= \omega_{0} j_x + \frac{8\lambda^2\omega_p}{(\kappa^2 + \omega^2_p)} j_x j_z + \frac{16\lambda^2\kappa \omega_p\omega_0}{(\kappa^2 + \omega_p^2)^2} j_y j_z \\
\partial_t j_z &=- \frac{8\lambda^2\omega_p}{(\kappa^2 + \omega^2_p)} j_x j_y - \frac{16\lambda^2\kappa \omega_p\omega_0}{(\kappa^2 + \omega_p^2)^2}  (j_y)^2
\end{align}
The corresponding equations for the individual spins are
\begin{align}
\partial_t s^x_j &=-\biggl[\omega_0-  \biggl( \frac{2\lambda^2 \omega_0 (\omega_p^2-\kappa^2)}{N(\kappa^2 + \omega_p^2)} \biggr)\biggr] s^y_j  \\
\partial_t s^y_j &= \omega_{0} s^x_j + \frac{4\lambda^2\omega_p}{N(\kappa^2 + \omega_p^2)}s^z_j\sum_{i=1}^N s^x_i \\ \nonumber
&\qquad + \frac{8\lambda^2\kappa \omega_p\omega_0}{N(\kappa^2 + \omega_p^2)^2}s^z_j\sum_{i=1}^N s^y_i \\
\partial_t s^z_j &=-\frac{4\lambda^2\omega_p}{N(\kappa^2 + \omega_p^2)}  s^y_j \sum_{i=1}^N s^x_i  - \frac{8\lambda^2\kappa \omega_p\omega_0}{N(\kappa^2 + \omega_p^2)^2}s^y_j\sum_{i=1}^N s^y_i .
\end{align}

\subsection{LMG}
Finally, for the LMG model, we have \cite{Vidal2004,Morrison2008}
\begin{align}
\partial_t j_x &=-\omega_0 j_y \\
\partial_t j_y &= \omega_{0} j_x + \frac{8\lambda^2\omega_p}{(\kappa^2 + \omega_p^2)} j_x j_z \\
\partial_t j_z &=- \frac{8\lambda^2\omega_p}{(\kappa^2 + \omega_p^2)} j_x j_y.
\end{align}
For the individual spins, we have
\begin{align}
\partial_t s^x_j &=-\omega_0 s^y_j  \\
\partial_t s^y_j &= \omega_{0} s^x_j + \frac{4\lambda^2\omega_p}{N(\kappa^2 + \omega_p^2)}  s^z_j\sum_{i=1}^N s^x_i  \\
\partial_t s^z_j &=-\frac{4\lambda^2\omega_p}{N(\kappa^2 + \omega_p^2)}  s^y_j \sum_{i=1}^N s^x_i.
\end{align}

\section{Exemplary dynamics for different dissipation strengths}\label{app:ex}

We display in Fig.~\ref{fig:dyn} the exemplary dynamics for different values of dissipation strength as indicated by the labels along the vertical axis. The left panels depict the mean-field results for the time evolution of $j_x$ for driving parameters as indicated in the figure. For the specific choice of driving parameters in Fig.~\ref{fig:dyn}(a), the optimal dissipation strength, identified by a response that is mostly dominated by a clean period doubling, appears to be in the intermediate range $\kappa \sim \omega_0$. As we further increase the dissipation strength, time-quasicrystalline dynamics permeate during the early times, which is signified by the appearance of extra peaks in the power spectrum. The lifetimes of the TQCs increase with the dissipation rate, as seen in Fig.~\ref{fig:dyn} for $\kappa/\omega_0 \in [10,21]$. The system is in a thermal phase for zero- and strong-dissipation limits $\kappa/\omega_0=\{0,10^2,10^3,\infty\}$.

In Fig.~\ref{fig:dyn}(b), we show the dynamics for a set of driving parameters along the instability line defined by Eq.~\eqref{eq:cond}. Here, we find period-doubling response in the nondissipative regimes and a light-induced NP for an intermediate dissipation strength $\kappa= \omega_0$.

\section{Dependence on the light-matter coupling strength and photon frequency}\label{app:param}

\begin{figure}[!htpb]
\centering
\includegraphics[width=1\columnwidth]{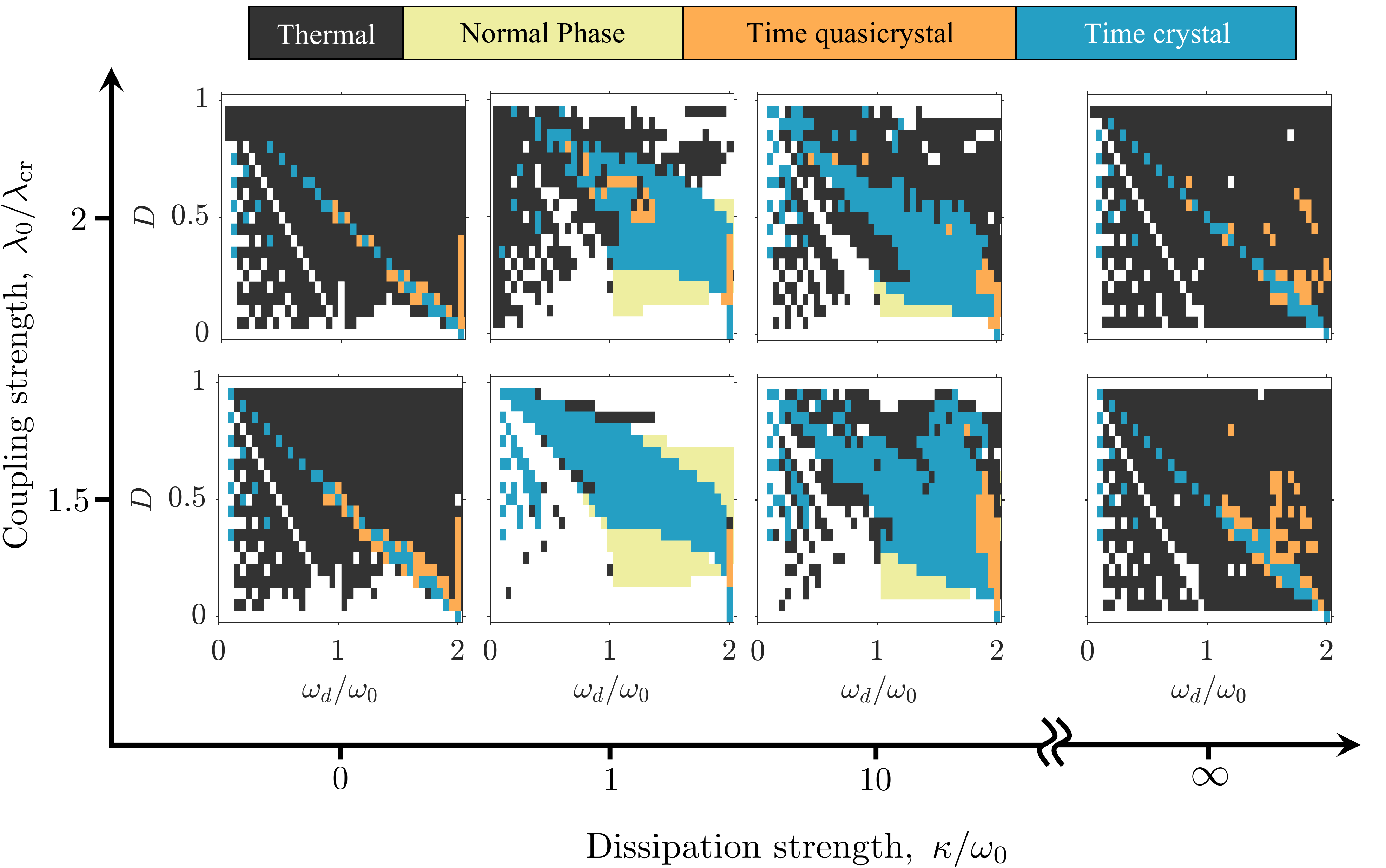}
\caption{Dynamical phase diagrams for different dissipation strengths. Along the vertical axis, we vary the the coupling strength and fix the frequency to $\omega_p=\omega_0$.}
\label{fig:lam} 
\end{figure}

The phase diagrams for varying dissipations strengths and coupling strengths $\lambda$ are depicted in Fig.~\ref{fig:lam}. We find similar results as discussed in the main text. More importantly, we demonstrate in Fig.~\ref{fig:lam} that the TCs persist for larger coupling strengths.

\begin{figure}[!htpb]
\centering
\includegraphics[width=1\columnwidth]{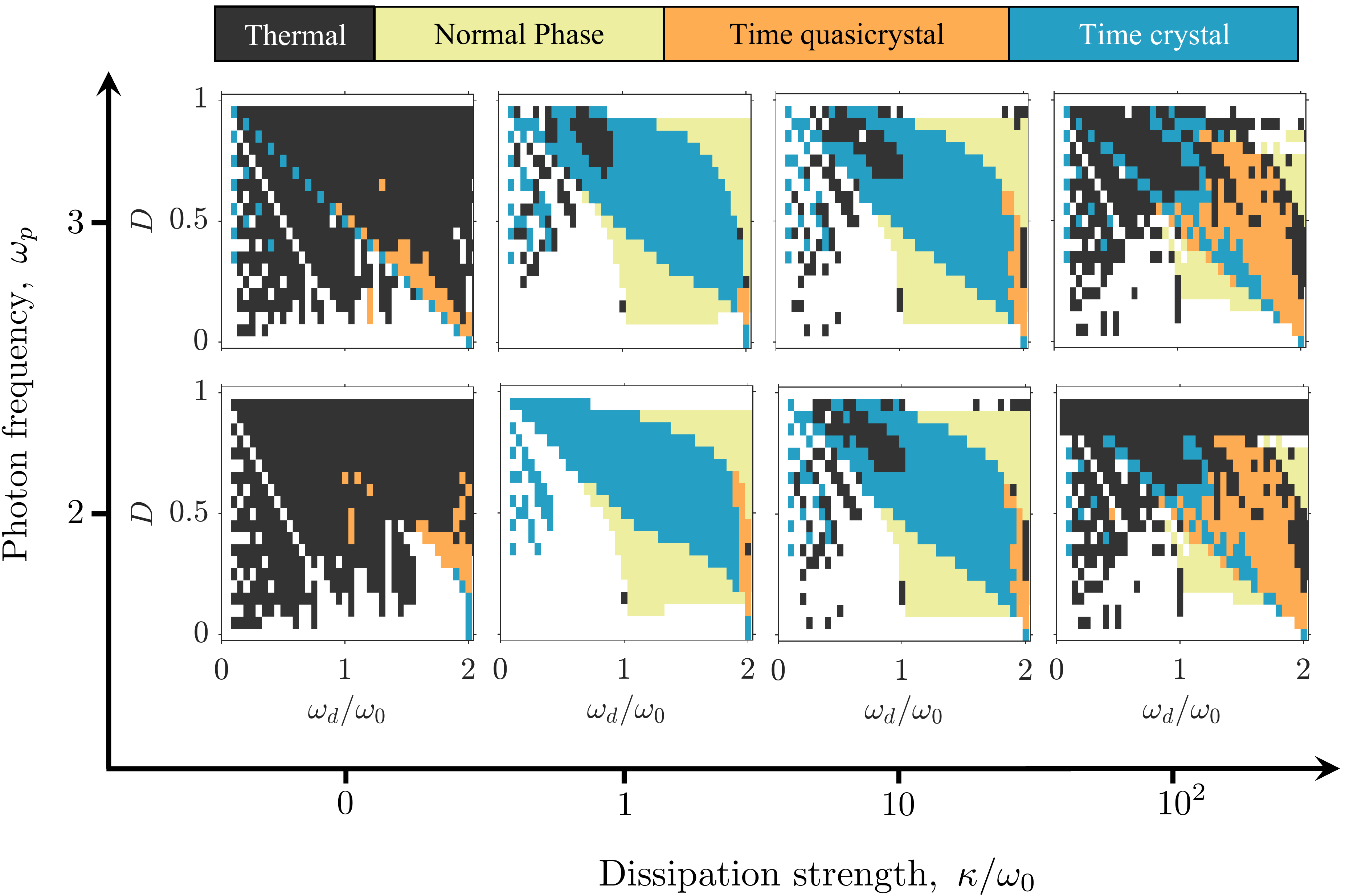}
\caption{Similar to Fig.~\ref{fig:lam} but for varying photon frequency along the vertical axis and fixed coupling strength $\lambda_0 = 1.1\lambda_\mathrm{cr}$.}
\label{fig:wp} 
\end{figure}
\begin{figure}[!htpb]
\centering
\includegraphics[width=1\columnwidth]{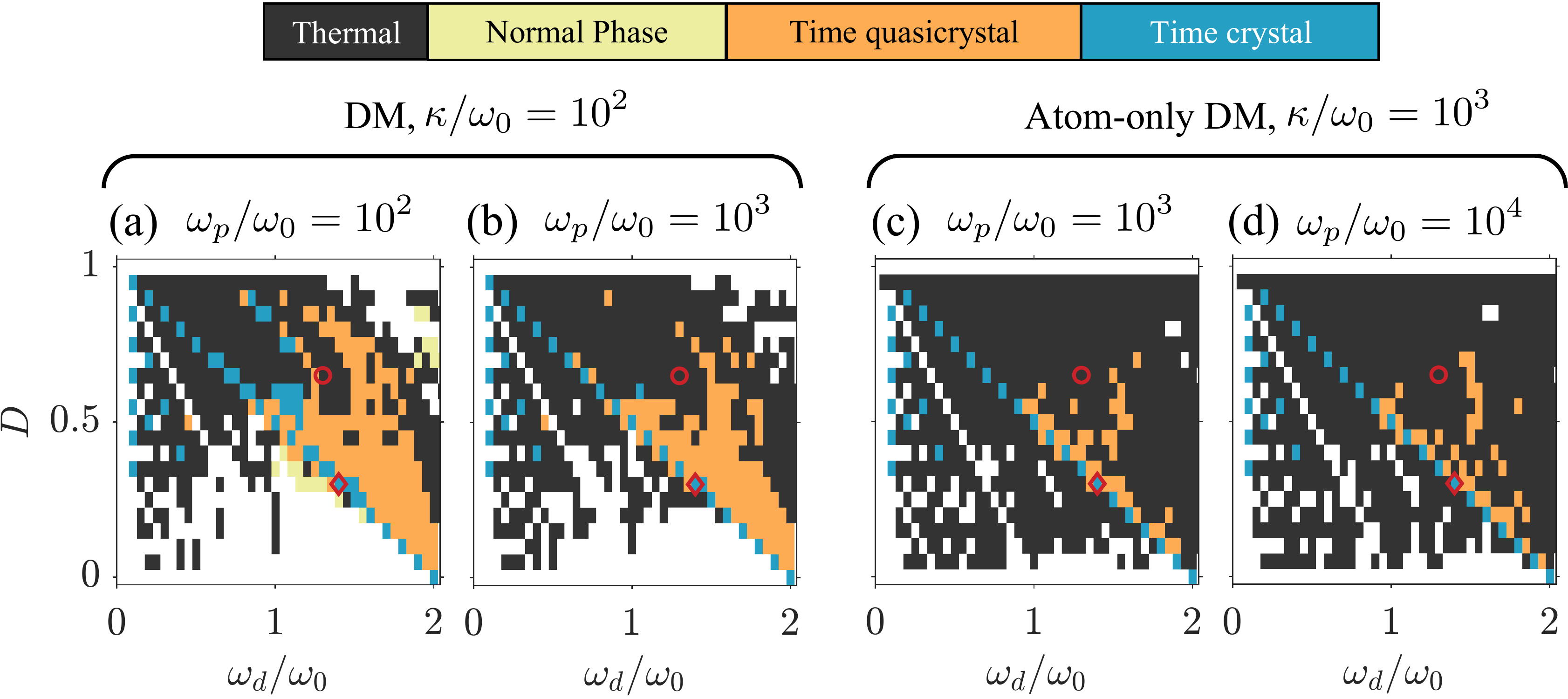}
\caption{Dynamical phase diagrams according to the (a,b) Dicke model and (c,d) ADM for large dissipation strength and photon frequency as indicated. The coupling strength is fixed at $\lambda_0 = 1.1\lambda_\mathrm{cr}$.}
\label{fig:wp_big} 
\end{figure}

The results for other choices of photon frequency $\omega_p$ are shown in Fig.~\ref{fig:wp}. We find that the phase diagrams for the dissipative scenarios are weakly affected by $\omega_p$. Motivated by the typical values of the photon frequency in Ref.~\cite{Baumann2010}, we present in Fig.~\ref{fig:wp_big} the results for photon frequencies $\omega_p$ that are comparable with or larger than the dissipation strength $\kappa$. The phase diagrams for both DM and ADM corroborate our claim that the regions with TCs do not significantly change with $\omega_p$. In fact, the number of thermal phases increases with $\omega_p$.

\section{Continuous sinusoidal driving}\label{app:contdrive}
\begin{figure}[!hbp]
\centering
\includegraphics[width=\columnwidth]{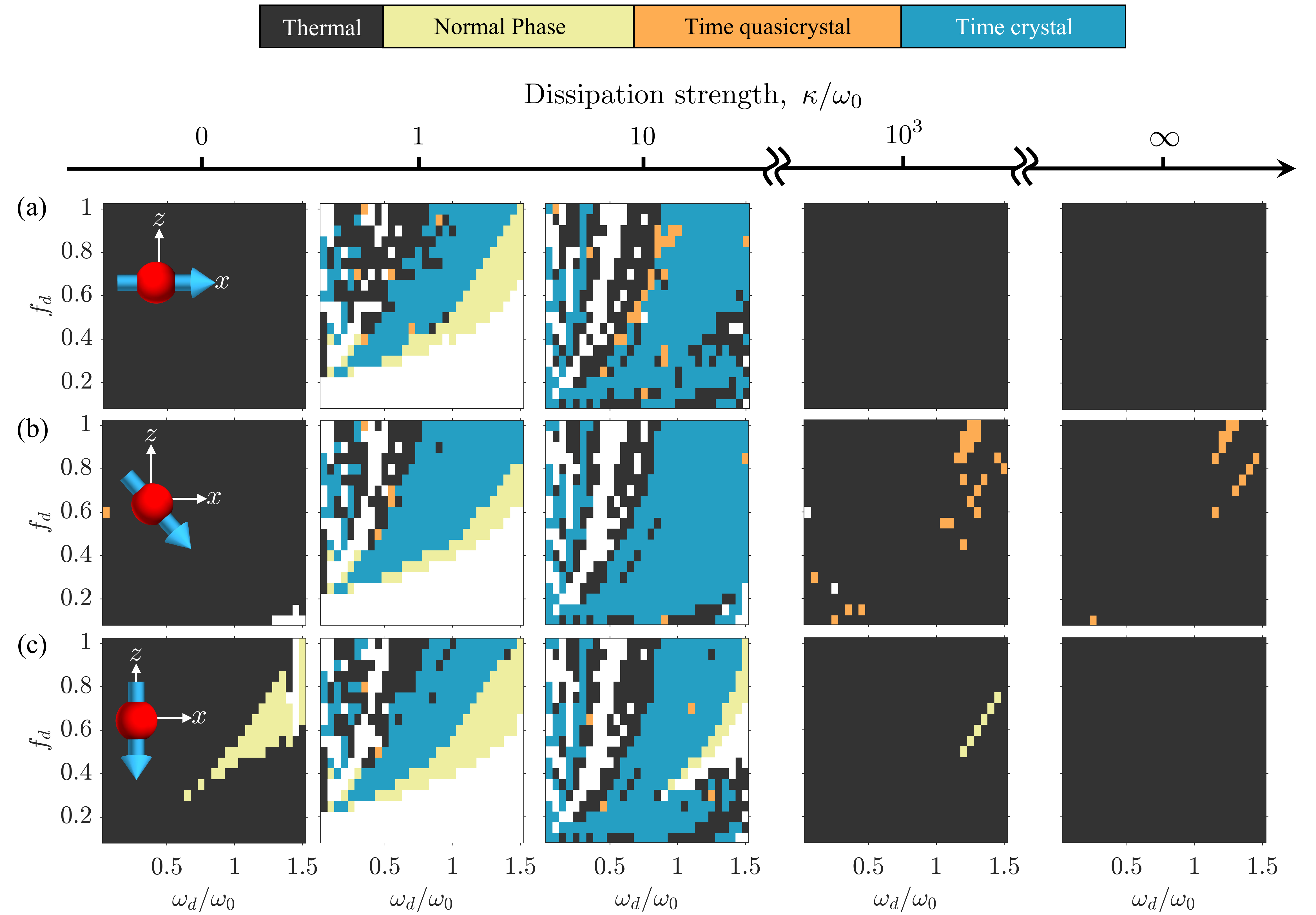}
\caption{Dynamical phase diagrams for a smooth sinusoidal drive and varying dissipation strengths. The remaining parameters are the same as in Fig.~\ref{fig:2}. The insets in the leftmost panels depict the initial state, namely, (a) fully polarised along the positive $x$ direction, (b) one of the symmetry-broken states, and (c) fully polarised along the negative $z$ direction.}
\label{fig:drive} 
\end{figure}
We briefly consider a different driving protocol given by a smooth sinusoidal drive of the light-matter coupling strength,
\begin{equation}
\lambda(t) = \lambda_0\left[1+f_d \sin(\omega_d t) \right],
\end{equation}
where $f_d$ is the modulation or driving strength. This protocol has been implemented to experimentally observe the Dicke time crystal in the cavity-QED platform composed of Bose-Einstein condensates inside a high-finesse optical cavity pumped in the transverse direction by an optical standing wave \cite{Kessler2021}.

\begin{figure}[!b]
\centering
\includegraphics[width=\columnwidth]{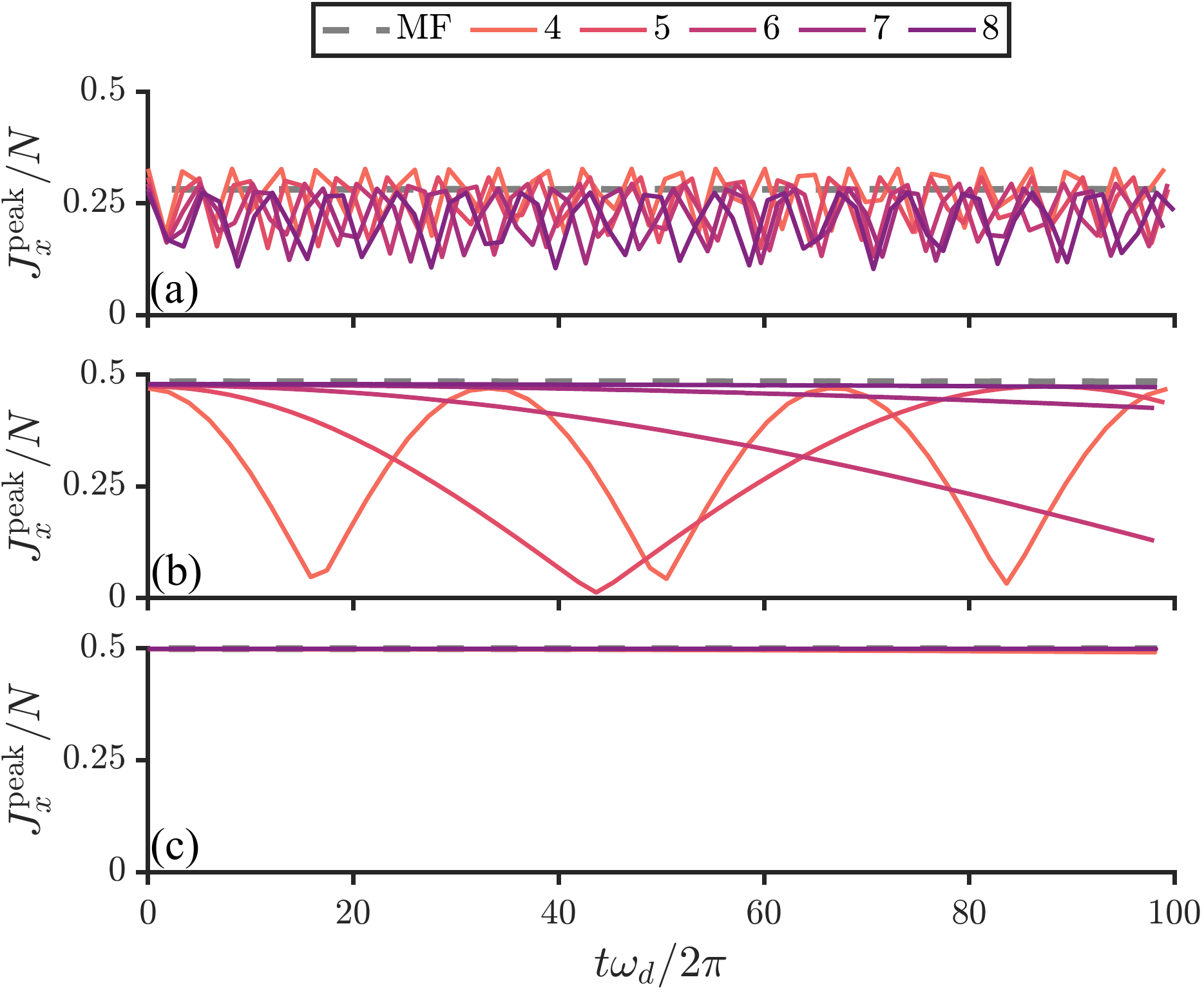}
\caption{Dynamics in the Lipkin-Meshkov-Glick (LMG) model  for an initial symmetry-broken eigenstate. The interaction strengths are (a) $\lambda_0 = 1.1\lambda_\mathrm{cr}$, (b) $\lambda_0 = 2.0\lambda_\mathrm{cr}$, and (c) $\lambda_0 = 4.0\lambda_\mathrm{cr}$.  The driving parameters are $D=0.3$ and $\omega_d=1.4\omega_0$.}
\label{fig:eigs} 
\end{figure}

In Fig.~\ref{fig:drive}, we present the dynamical phase diagrams for such a continuous driving scheme. In addition to varying the dissipation strength, we also consider different initial states as sketched in the insets of Fig.~\ref{fig:drive}. In the dissipative cases $\kappa/\omega_0=\{1,10\}$, time-crystalline phases appear within the resonance lobes, which have similar shape as those found in the cavity-QED simulator for the DM \cite{Kessler2021,Tuquero2022}. Contrary to the binary drive, we do not observe any TCs in the closed-system limits $\kappa/\omega_0 = \{0,10^3,\infty\}$, irrespective of the initial state, for a sinusoidal drive as depicted in Fig.~\ref{fig:drive}.

\section{Quantum results for an initial symmetry-broken state in the LMG model}\label{app:qm}

The results obtained using full quantum simulations for an initial symmetry-broken eigenstate are shown in Fig.~\ref{fig:eigs}. The driving parameters are chosen such that the system is in a time-crystalline phase in the thermodynamic limit for the interaction strengths considered in Fig.~\ref{fig:eigs}. The beat period clearly increases with number of spins $N$, see Fig.~\ref{fig:eigs}(b). Furthermore, the dynamics shown in Fig.~\ref{fig:7}(b) appear to fluctuate around the dynamics in Fig.~\ref{fig:eigs}(b).

\bibliography{biblio}

\providecommand{\noopsort}[1]{}\providecommand{\singleletter}[1]{#1}%
\begin{thebibliography}{79}%
\makeatletter
\providecommand \@ifxundefined [1]{%
 \@ifx{#1\undefined}
}%
\providecommand \@ifnum [1]{%
 \ifnum #1\expandafter \@firstoftwo
 \else \expandafter \@secondoftwo
 \fi
}%
\providecommand \@ifx [1]{%
 \ifx #1\expandafter \@firstoftwo
 \else \expandafter \@secondoftwo
 \fi
}%
\providecommand \natexlab [1]{#1}%
\providecommand \enquote  [1]{``#1''}%
\providecommand \bibnamefont  [1]{#1}%
\providecommand \bibfnamefont [1]{#1}%
\providecommand \citenamefont [1]{#1}%
\providecommand \href@noop [0]{\@secondoftwo}%
\providecommand \href [0]{\begingroup \@sanitize@url \@href}%
\providecommand \@href[1]{\@@startlink{#1}\@@href}%
\providecommand \@@href[1]{\endgroup#1\@@endlink}%
\providecommand \@sanitize@url [0]{\catcode `\\12\catcode `\$12\catcode
  `\&12\catcode `\#12\catcode `\^12\catcode `\_12\catcode `\%12\relax}%
\providecommand \@@startlink[1]{}%
\providecommand \@@endlink[0]{}%
\providecommand \url  [0]{\begingroup\@sanitize@url \@url }%
\providecommand \@url [1]{\endgroup\@href {#1}{\urlprefix }}%
\providecommand \urlprefix  [0]{URL }%
\providecommand \Eprint [0]{\href }%
\providecommand \doibase [0]{https://doi.org/}%
\providecommand \selectlanguage [0]{\@gobble}%
\providecommand \bibinfo  [0]{\@secondoftwo}%
\providecommand \bibfield  [0]{\@secondoftwo}%
\providecommand \translation [1]{[#1]}%
\providecommand \BibitemOpen [0]{}%
\providecommand \bibitemStop [0]{}%
\providecommand \bibitemNoStop [0]{.\EOS\space}%
\providecommand \EOS [0]{\spacefactor3000\relax}%
\providecommand \BibitemShut  [1]{\csname bibitem#1\endcsname}%
\let\auto@bib@innerbib\@empty
\bibitem [{\citenamefont {Wilczek}(2012)}]{Wilczek2012}%
  \BibitemOpen
  \bibfield  {author} {\bibinfo {author} {\bibfnamefont {F.}~\bibnamefont
  {Wilczek}},\ }\bibfield  {title} {\bibinfo {title} {{Quantum Time
  Crystals}},\ }\href {https://doi.org/10.1103/PhysRevLett.109.160401}
  {\bibfield  {journal} {\bibinfo  {journal} {Phys. Rev. Lett.}\ }\textbf
  {\bibinfo {volume} {109}},\ \bibinfo {pages} {160401} (\bibinfo {year}
  {2012})}\BibitemShut {NoStop}%
\bibitem [{\citenamefont {Sacha}(2015)}]{Sacha2015}%
  \BibitemOpen
  \bibfield  {author} {\bibinfo {author} {\bibfnamefont {K.}~\bibnamefont
  {Sacha}},\ }\bibfield  {title} {\bibinfo {title} {Modeling spontaneous
  breaking of time-translation symmetry},\ }\href
  {https://doi.org/10.1103/PhysRevA.91.033617} {\bibfield  {journal} {\bibinfo
  {journal} {Phys. Rev. A}\ }\textbf {\bibinfo {volume} {91}},\ \bibinfo
  {pages} {033617} (\bibinfo {year} {2015})}\BibitemShut {NoStop}%
\bibitem [{\citenamefont {{Khemani}}\ \emph {et~al.}(2019)\citenamefont
  {{Khemani}}, \citenamefont {{Moessner}},\ and\ \citenamefont
  {{Sondhi}}}]{Khemani2019}%
  \BibitemOpen
  \bibfield  {author} {\bibinfo {author} {\bibfnamefont {V.}~\bibnamefont
  {{Khemani}}}, \bibinfo {author} {\bibfnamefont {R.}~\bibnamefont
  {{Moessner}}},\ and\ \bibinfo {author} {\bibfnamefont {S.~L.}\ \bibnamefont
  {{Sondhi}}},\ }\bibfield  {title} {\bibinfo {title} {{A Brief History of Time
  Crystals}},\ }\href@noop {} {\bibfield  {journal} {\bibinfo  {journal} {arXiv
  e-prints}\ ,\ \bibinfo {eid} {arXiv:1910.10745}} (\bibinfo {year} {2019})},\
  \Eprint {https://arxiv.org/abs/1910.10745} {arXiv:1910.10745} \BibitemShut
  {NoStop}%
\bibitem [{\citenamefont {{Else}}\ \emph {et~al.}(2020)\citenamefont {{Else}},
  \citenamefont {{Monroe}}, \citenamefont {{Nayak}},\ and\ \citenamefont
  {{Yao}}}]{Else2020}%
  \BibitemOpen
  \bibfield  {author} {\bibinfo {author} {\bibfnamefont {D.~V.}\ \bibnamefont
  {{Else}}}, \bibinfo {author} {\bibfnamefont {C.}~\bibnamefont {{Monroe}}},
  \bibinfo {author} {\bibfnamefont {C.}~\bibnamefont {{Nayak}}},\ and\ \bibinfo
  {author} {\bibfnamefont {N.~Y.}\ \bibnamefont {{Yao}}},\ }\bibfield  {title}
  {\bibinfo {title} {Discrete time crystals},\ }\href
  {https://doi.org/10.1146/annurev-conmatphys-031119-050658} {\bibfield
  {journal} {\bibinfo  {journal} {{Annu. Rev. Condens. Matter Phys.}}\ }\textbf
  {\bibinfo {volume} {11}},\ \bibinfo {pages} {467} (\bibinfo {year}
  {2020})}\BibitemShut {NoStop}%
\bibitem [{\citenamefont {Sacha}(2020)}]{Sacha2020}%
  \BibitemOpen
  \bibfield  {author} {\bibinfo {author} {\bibfnamefont {K.}~\bibnamefont
  {Sacha}},\ }\href@noop {} {\emph {\bibinfo {title} {{Time Crystals}}}}\
  (\bibinfo  {publisher} {Springer, Cham},\ \bibinfo {year} {2020})\BibitemShut
  {NoStop}%
\bibitem [{\citenamefont {Else}\ \emph {et~al.}(2016)\citenamefont {Else},
  \citenamefont {Bauer},\ and\ \citenamefont {Nayak}}]{Else2016}%
  \BibitemOpen
  \bibfield  {author} {\bibinfo {author} {\bibfnamefont {D.~V.}\ \bibnamefont
  {Else}}, \bibinfo {author} {\bibfnamefont {B.}~\bibnamefont {Bauer}},\ and\
  \bibinfo {author} {\bibfnamefont {C.}~\bibnamefont {Nayak}},\ }\bibfield
  {title} {\bibinfo {title} {{Floquet Time Crystals}},\ }\href
  {https://doi.org/10.1103/PhysRevLett.117.090402} {\bibfield  {journal}
  {\bibinfo  {journal} {Phys. Rev. Lett.}\ }\textbf {\bibinfo {volume} {117}},\
  \bibinfo {pages} {090402} (\bibinfo {year} {2016})}\BibitemShut {NoStop}%
\bibitem [{\citenamefont {Yao}\ \emph {et~al.}(2017)\citenamefont {Yao},
  \citenamefont {Potter}, \citenamefont {Potirniche},\ and\ \citenamefont
  {Vishwanath}}]{Yao2017}%
  \BibitemOpen
  \bibfield  {author} {\bibinfo {author} {\bibfnamefont {N.~Y.}\ \bibnamefont
  {Yao}}, \bibinfo {author} {\bibfnamefont {A.~C.}\ \bibnamefont {Potter}},
  \bibinfo {author} {\bibfnamefont {I.-D.}\ \bibnamefont {Potirniche}},\ and\
  \bibinfo {author} {\bibfnamefont {A.}~\bibnamefont {Vishwanath}},\ }\bibfield
   {title} {\bibinfo {title} {{Discrete Time Crystals: Rigidity, Criticality,
  and Realizations}},\ }\href {https://doi.org/10.1103/PhysRevLett.118.030401}
  {\bibfield  {journal} {\bibinfo  {journal} {Phys. Rev. Lett.}\ }\textbf
  {\bibinfo {volume} {118}},\ \bibinfo {pages} {030401} (\bibinfo {year}
  {2017})}\BibitemShut {NoStop}%
\bibitem [{\citenamefont {Khemani}\ \emph {et~al.}(2016)\citenamefont
  {Khemani}, \citenamefont {Lazarides}, \citenamefont {Moessner},\ and\
  \citenamefont {Sondhi}}]{Khemani2016}%
  \BibitemOpen
  \bibfield  {author} {\bibinfo {author} {\bibfnamefont {V.}~\bibnamefont
  {Khemani}}, \bibinfo {author} {\bibfnamefont {A.}~\bibnamefont {Lazarides}},
  \bibinfo {author} {\bibfnamefont {R.}~\bibnamefont {Moessner}},\ and\
  \bibinfo {author} {\bibfnamefont {S.~L.}\ \bibnamefont {Sondhi}},\ }\bibfield
   {title} {\bibinfo {title} {{Phase Structure of Driven Quantum Systems}},\
  }\href {https://doi.org/10.1103/PhysRevLett.116.250401} {\bibfield  {journal}
  {\bibinfo  {journal} {Phys. Rev. Lett.}\ }\textbf {\bibinfo {volume} {116}},\
  \bibinfo {pages} {250401} (\bibinfo {year} {2016})}\BibitemShut {NoStop}%
\bibitem [{\citenamefont {Russomanno}\ \emph {et~al.}(2017)\citenamefont
  {Russomanno}, \citenamefont {Iemini}, \citenamefont {Dalmonte},\ and\
  \citenamefont {Fazio}}]{Russomanno2017}%
  \BibitemOpen
  \bibfield  {author} {\bibinfo {author} {\bibfnamefont {A.}~\bibnamefont
  {Russomanno}}, \bibinfo {author} {\bibfnamefont {F.}~\bibnamefont {Iemini}},
  \bibinfo {author} {\bibfnamefont {M.}~\bibnamefont {Dalmonte}},\ and\
  \bibinfo {author} {\bibfnamefont {R.}~\bibnamefont {Fazio}},\ }\bibfield
  {title} {\bibinfo {title} {{Floquet time crystal in the Lipkin-Meshkov-Glick
  model}},\ }\href {https://doi.org/10.1103/PhysRevB.95.214307} {\bibfield
  {journal} {\bibinfo  {journal} {Phys. Rev. B}\ }\textbf {\bibinfo {volume}
  {95}},\ \bibinfo {pages} {214307} (\bibinfo {year} {2017})}\BibitemShut
  {NoStop}%
\bibitem [{\citenamefont {{Pizzi}}\ \emph {et~al.}(2021)\citenamefont
  {{Pizzi}}, \citenamefont {{Knolle}},\ and\ \citenamefont
  {{Nunnenkamp}}}]{Pizzi2021NC}%
  \BibitemOpen
  \bibfield  {author} {\bibinfo {author} {\bibfnamefont {A.}~\bibnamefont
  {{Pizzi}}}, \bibinfo {author} {\bibfnamefont {J.}~\bibnamefont {{Knolle}}},\
  and\ \bibinfo {author} {\bibfnamefont {A.}~\bibnamefont {{Nunnenkamp}}},\
  }\bibfield  {title} {\bibinfo {title} {{Higher-order and fractional discrete
  time crystals in clean long-range interacting systems}},\ }\href
  {https://doi.org/10.1038/s41467-021-22583-5} {\bibfield  {journal} {\bibinfo
  {journal} {Nat. Commun.}\ }\textbf {\bibinfo {volume} {12}},\ \bibinfo {eid}
  {2341} (\bibinfo {year} {2021})}\BibitemShut {NoStop}%
\bibitem [{\citenamefont {Barfknecht}\ \emph {et~al.}(2019)\citenamefont
  {Barfknecht}, \citenamefont {Rasmussen}, \citenamefont {Foerster},\ and\
  \citenamefont {Zinner}}]{Barfknecht2019}%
  \BibitemOpen
  \bibfield  {author} {\bibinfo {author} {\bibfnamefont {R.~E.}\ \bibnamefont
  {Barfknecht}}, \bibinfo {author} {\bibfnamefont {S.~E.}\ \bibnamefont
  {Rasmussen}}, \bibinfo {author} {\bibfnamefont {A.}~\bibnamefont
  {Foerster}},\ and\ \bibinfo {author} {\bibfnamefont {N.~T.}\ \bibnamefont
  {Zinner}},\ }\bibfield  {title} {\bibinfo {title} {Realizing time crystals in
  discrete quantum few-body systems},\ }\href
  {https://doi.org/10.1103/PhysRevB.99.144304} {\bibfield  {journal} {\bibinfo
  {journal} {Phys. Rev. B}\ }\textbf {\bibinfo {volume} {99}},\ \bibinfo
  {pages} {144304} (\bibinfo {year} {2019})}\BibitemShut {NoStop}%
\bibitem [{\citenamefont {Estarellas}\ \emph {et~al.}(2020)\citenamefont
  {Estarellas}, \citenamefont {Osada}, \citenamefont {Bastidas}, \citenamefont
  {Renoust}, \citenamefont {Sanaka}, \citenamefont {Munro},\ and\ \citenamefont
  {Nemoto}}]{Estarellas2020}%
  \BibitemOpen
  \bibfield  {author} {\bibinfo {author} {\bibfnamefont {M.~P.}\ \bibnamefont
  {Estarellas}}, \bibinfo {author} {\bibfnamefont {T.}~\bibnamefont {Osada}},
  \bibinfo {author} {\bibfnamefont {V.~M.}\ \bibnamefont {Bastidas}}, \bibinfo
  {author} {\bibfnamefont {B.}~\bibnamefont {Renoust}}, \bibinfo {author}
  {\bibfnamefont {K.}~\bibnamefont {Sanaka}}, \bibinfo {author} {\bibfnamefont
  {W.~J.}\ \bibnamefont {Munro}},\ and\ \bibinfo {author} {\bibfnamefont
  {K.}~\bibnamefont {Nemoto}},\ }\bibfield  {title} {\bibinfo {title}
  {Simulating complex quantum networks with time crystals},\ }\href
  {https://doi.org/10.1126/sciadv.aay8892} {\bibfield  {journal} {\bibinfo
  {journal} {Science Advances}\ }\textbf {\bibinfo {volume} {6}},\ \bibinfo
  {pages} {eaay8892} (\bibinfo {year} {2020})}\BibitemShut {NoStop}%
\bibitem [{\citenamefont {Pizzi}\ \emph
  {et~al.}(2021{\natexlab{a}})\citenamefont {Pizzi}, \citenamefont
  {Nunnenkamp},\ and\ \citenamefont {Knolle}}]{Pizzi2021a}%
  \BibitemOpen
  \bibfield  {author} {\bibinfo {author} {\bibfnamefont {A.}~\bibnamefont
  {Pizzi}}, \bibinfo {author} {\bibfnamefont {A.}~\bibnamefont {Nunnenkamp}},\
  and\ \bibinfo {author} {\bibfnamefont {J.}~\bibnamefont {Knolle}},\
  }\bibfield  {title} {\bibinfo {title} {Classical prethermal phases of
  matter},\ }\href {https://doi.org/10.1103/PhysRevLett.127.140602} {\bibfield
  {journal} {\bibinfo  {journal} {Phys. Rev. Lett.}\ }\textbf {\bibinfo
  {volume} {127}},\ \bibinfo {pages} {140602} (\bibinfo {year}
  {2021}{\natexlab{a}})}\BibitemShut {NoStop}%
\bibitem [{\citenamefont {Ye}\ \emph {et~al.}(2021)\citenamefont {Ye},
  \citenamefont {Machado},\ and\ \citenamefont {Yao}}]{Ye2021}%
  \BibitemOpen
  \bibfield  {author} {\bibinfo {author} {\bibfnamefont {B.}~\bibnamefont
  {Ye}}, \bibinfo {author} {\bibfnamefont {F.}~\bibnamefont {Machado}},\ and\
  \bibinfo {author} {\bibfnamefont {N.~Y.}\ \bibnamefont {Yao}},\ }\bibfield
  {title} {\bibinfo {title} {{Floquet Phases of Matter via Classical
  Prethermalization}},\ }\href {https://doi.org/10.1103/PhysRevLett.127.140603}
  {\bibfield  {journal} {\bibinfo  {journal} {Phys. Rev. Lett.}\ }\textbf
  {\bibinfo {volume} {127}},\ \bibinfo {pages} {140603} (\bibinfo {year}
  {2021})}\BibitemShut {NoStop}%
\bibitem [{\citenamefont {{Zhang}}\ \emph {et~al.}(2017)\citenamefont
  {{Zhang}}, \citenamefont {{Hess}}, \citenamefont {{Kyprianidis}},
  \citenamefont {{Becker}}, \citenamefont {{Lee}}, \citenamefont {{Smith}},
  \citenamefont {{Pagano}}, \citenamefont {{Potirniche}}, \citenamefont
  {{Potter}}, \citenamefont {{Vishwanath}}, \citenamefont {{Yao}},\ and\
  \citenamefont {{Monroe}}}]{Zhang2017}%
  \BibitemOpen
  \bibfield  {author} {\bibinfo {author} {\bibfnamefont {J.}~\bibnamefont
  {{Zhang}}}, \bibinfo {author} {\bibfnamefont {P.~W.}\ \bibnamefont {{Hess}}},
  \bibinfo {author} {\bibfnamefont {A.}~\bibnamefont {{Kyprianidis}}}, \bibinfo
  {author} {\bibfnamefont {P.}~\bibnamefont {{Becker}}}, \bibinfo {author}
  {\bibfnamefont {A.}~\bibnamefont {{Lee}}}, \bibinfo {author} {\bibfnamefont
  {J.}~\bibnamefont {{Smith}}}, \bibinfo {author} {\bibfnamefont
  {G.}~\bibnamefont {{Pagano}}}, \bibinfo {author} {\bibfnamefont {I.-D.}\
  \bibnamefont {{Potirniche}}}, \bibinfo {author} {\bibfnamefont {A.~C.}\
  \bibnamefont {{Potter}}}, \bibinfo {author} {\bibfnamefont {A.}~\bibnamefont
  {{Vishwanath}}}, \bibinfo {author} {\bibfnamefont {N.~Y.}\ \bibnamefont
  {{Yao}}},\ and\ \bibinfo {author} {\bibfnamefont {C.}~\bibnamefont
  {{Monroe}}},\ }\bibfield  {title} {\bibinfo {title} {{Observation of a
  discrete time crystal}},\ }\href {https://doi.org/10.1038/nature21413}
  {\bibfield  {journal} {\bibinfo  {journal} {Nature}\ }\textbf {\bibinfo
  {volume} {543}},\ \bibinfo {pages} {217} (\bibinfo {year}
  {2017})}\BibitemShut {NoStop}%
\bibitem [{\citenamefont {Choi}\ \emph {et~al.}(2017)\citenamefont {Choi},
  \citenamefont {Choi}, \citenamefont {Landig}, \citenamefont {Kucsko},
  \citenamefont {Zhou}, \citenamefont {Isoya}, \citenamefont {Jelezko},
  \citenamefont {Onoda}, \citenamefont {Sumiya}, \citenamefont {Khemani},
  \citenamefont {{Von Keyserlingk}}, \citenamefont {Yao}, \citenamefont
  {Demler},\ and\ \citenamefont {Lukin}}]{Choi2017}%
  \BibitemOpen
  \bibfield  {author} {\bibinfo {author} {\bibfnamefont {S.}~\bibnamefont
  {Choi}}, \bibinfo {author} {\bibfnamefont {J.}~\bibnamefont {Choi}}, \bibinfo
  {author} {\bibfnamefont {R.}~\bibnamefont {Landig}}, \bibinfo {author}
  {\bibfnamefont {G.}~\bibnamefont {Kucsko}}, \bibinfo {author} {\bibfnamefont
  {H.}~\bibnamefont {Zhou}}, \bibinfo {author} {\bibfnamefont {J.}~\bibnamefont
  {Isoya}}, \bibinfo {author} {\bibfnamefont {F.}~\bibnamefont {Jelezko}},
  \bibinfo {author} {\bibfnamefont {S.}~\bibnamefont {Onoda}}, \bibinfo
  {author} {\bibfnamefont {H.}~\bibnamefont {Sumiya}}, \bibinfo {author}
  {\bibfnamefont {V.}~\bibnamefont {Khemani}}, \bibinfo {author} {\bibfnamefont
  {C.}~\bibnamefont {{Von Keyserlingk}}}, \bibinfo {author} {\bibfnamefont
  {N.~Y.}\ \bibnamefont {Yao}}, \bibinfo {author} {\bibfnamefont
  {E.}~\bibnamefont {Demler}},\ and\ \bibinfo {author} {\bibfnamefont {M.~D.}\
  \bibnamefont {Lukin}},\ }\bibfield  {title} {\bibinfo {title} {{Observation
  of discrete time-crystalline order in a disordered dipolar many-body
  system}},\ }\href {https://doi.org/10.1038/nature21426} {\bibfield  {journal}
  {\bibinfo  {journal} {Nature}\ }\textbf {\bibinfo {volume} {543}},\ \bibinfo
  {pages} {221} (\bibinfo {year} {2017})}\BibitemShut {NoStop}%
\bibitem [{\citenamefont {Rovny}\ \emph {et~al.}(2018)\citenamefont {Rovny},
  \citenamefont {Blum},\ and\ \citenamefont {Barrett}}]{Rovny2018}%
  \BibitemOpen
  \bibfield  {author} {\bibinfo {author} {\bibfnamefont {J.}~\bibnamefont
  {Rovny}}, \bibinfo {author} {\bibfnamefont {R.~L.}\ \bibnamefont {Blum}},\
  and\ \bibinfo {author} {\bibfnamefont {S.~E.}\ \bibnamefont {Barrett}},\
  }\bibfield  {title} {\bibinfo {title} {{Observation of Discrete-Time-Crystal
  Signatures in an Ordered Dipolar Many-Body System}},\ }\href
  {https://doi.org/10.1103/PhysRevLett.120.180603} {\bibfield  {journal}
  {\bibinfo  {journal} {Phys. Rev. Lett.}\ }\textbf {\bibinfo {volume} {120}},\
  \bibinfo {pages} {180603} (\bibinfo {year} {2018})}\BibitemShut {NoStop}%
\bibitem [{\citenamefont {{Kyprianidis}}\ \emph {et~al.}(2021)\citenamefont
  {{Kyprianidis}}, \citenamefont {{Machado}}, \citenamefont {{Morong}},
  \citenamefont {{Becker}}, \citenamefont {{Collins}}, \citenamefont {{Else}},
  \citenamefont {{Feng}}, \citenamefont {{Hess}}, \citenamefont {{Nayak}},
  \citenamefont {{Pagano}}, \citenamefont {{Yao}},\ and\ \citenamefont
  {{Monroe}}}]{Kyprianidis2021}%
  \BibitemOpen
  \bibfield  {author} {\bibinfo {author} {\bibfnamefont {A.}~\bibnamefont
  {{Kyprianidis}}}, \bibinfo {author} {\bibfnamefont {F.}~\bibnamefont
  {{Machado}}}, \bibinfo {author} {\bibfnamefont {W.}~\bibnamefont {{Morong}}},
  \bibinfo {author} {\bibfnamefont {P.}~\bibnamefont {{Becker}}}, \bibinfo
  {author} {\bibfnamefont {K.~S.}\ \bibnamefont {{Collins}}}, \bibinfo {author}
  {\bibfnamefont {D.~V.}\ \bibnamefont {{Else}}}, \bibinfo {author}
  {\bibfnamefont {L.}~\bibnamefont {{Feng}}}, \bibinfo {author} {\bibfnamefont
  {P.~W.}\ \bibnamefont {{Hess}}}, \bibinfo {author} {\bibfnamefont
  {C.}~\bibnamefont {{Nayak}}}, \bibinfo {author} {\bibfnamefont
  {G.}~\bibnamefont {{Pagano}}}, \bibinfo {author} {\bibfnamefont {N.~Y.}\
  \bibnamefont {{Yao}}},\ and\ \bibinfo {author} {\bibfnamefont
  {C.}~\bibnamefont {{Monroe}}},\ }\bibfield  {title} {\bibinfo {title}
  {{Observation of a prethermal discrete time crystal}},\ }\href
  {https://doi.org/10.1126/science.abg8102} {\bibfield  {journal} {\bibinfo
  {journal} {Science}\ }\textbf {\bibinfo {volume} {372}},\ \bibinfo {pages}
  {1192} (\bibinfo {year} {2021})}\BibitemShut {NoStop}%
\bibitem [{\citenamefont {{Randall}}\ \emph {et~al.}(2021)\citenamefont
  {{Randall}}, \citenamefont {{Bradley}}, \citenamefont {{van der Gronden}},
  \citenamefont {{Galicia}}, \citenamefont {{Abobeih}}, \citenamefont
  {{Markham}}, \citenamefont {{Twitchen}}, \citenamefont {{Machado}},
  \citenamefont {{Yao}},\ and\ \citenamefont {{Taminiau}}}]{Randall2021}%
  \BibitemOpen
  \bibfield  {author} {\bibinfo {author} {\bibfnamefont {J.}~\bibnamefont
  {{Randall}}}, \bibinfo {author} {\bibfnamefont {C.~E.}\ \bibnamefont
  {{Bradley}}}, \bibinfo {author} {\bibfnamefont {F.~V.}\ \bibnamefont {{van
  der Gronden}}}, \bibinfo {author} {\bibfnamefont {A.}~\bibnamefont
  {{Galicia}}}, \bibinfo {author} {\bibfnamefont {M.~H.}\ \bibnamefont
  {{Abobeih}}}, \bibinfo {author} {\bibfnamefont {M.}~\bibnamefont
  {{Markham}}}, \bibinfo {author} {\bibfnamefont {D.~J.}\ \bibnamefont
  {{Twitchen}}}, \bibinfo {author} {\bibfnamefont {F.}~\bibnamefont
  {{Machado}}}, \bibinfo {author} {\bibfnamefont {N.~Y.}\ \bibnamefont
  {{Yao}}},\ and\ \bibinfo {author} {\bibfnamefont {T.~H.}\ \bibnamefont
  {{Taminiau}}},\ }\bibfield  {title} {\bibinfo {title}
  {{Many-body{\textendash}localized discrete time crystal with a programmable
  spin-based quantum simulator}},\ }\href
  {https://doi.org/10.1126/science.abk0603} {\bibfield  {journal} {\bibinfo
  {journal} {Science}\ }\textbf {\bibinfo {volume} {374}},\ \bibinfo {pages}
  {1474} (\bibinfo {year} {2021})}\BibitemShut {NoStop}%
\bibitem [{\citenamefont {Mu\~noz Arias}\ \emph {et~al.}(2022)\citenamefont
  {Mu\~noz Arias}, \citenamefont {Chinni},\ and\ \citenamefont
  {Poggi}}]{Munoz2022}%
  \BibitemOpen
  \bibfield  {author} {\bibinfo {author} {\bibfnamefont {M.~H.}\ \bibnamefont
  {Mu\~noz Arias}}, \bibinfo {author} {\bibfnamefont {K.}~\bibnamefont
  {Chinni}},\ and\ \bibinfo {author} {\bibfnamefont {P.~M.}\ \bibnamefont
  {Poggi}},\ }\bibfield  {title} {\bibinfo {title} {Floquet time crystals in
  driven spin systems with all-to-all $p$-body interactions},\ }\href
  {https://doi.org/10.1103/PhysRevResearch.4.023018} {\bibfield  {journal}
  {\bibinfo  {journal} {Phys. Rev. Res.}\ }\textbf {\bibinfo {volume} {4}},\
  \bibinfo {pages} {023018} (\bibinfo {year} {2022})}\BibitemShut {NoStop}%
\bibitem [{\citenamefont {{Smits}}\ \emph {et~al.}(2018)\citenamefont
  {{Smits}}, \citenamefont {{Liao}}, \citenamefont {{Stoof}},\ and\
  \citenamefont {{van der Straten}}}]{Smits2018}%
  \BibitemOpen
  \bibfield  {author} {\bibinfo {author} {\bibfnamefont {J.}~\bibnamefont
  {{Smits}}}, \bibinfo {author} {\bibfnamefont {L.}~\bibnamefont {{Liao}}},
  \bibinfo {author} {\bibfnamefont {H.~T.~C.}\ \bibnamefont {{Stoof}}},\ and\
  \bibinfo {author} {\bibfnamefont {P.}~\bibnamefont {{van der Straten}}},\
  }\bibfield  {title} {\bibinfo {title} {{Observation of a Space-Time Crystal
  in a Superfluid Quantum Gas}},\ }\href
  {https://doi.org/10.1103/PhysRevLett.121.185301} {\bibfield  {journal}
  {\bibinfo  {journal} {Phys. Rev. Lett.}\ }\textbf {\bibinfo {volume} {121}},\
  \bibinfo {pages} {185301} (\bibinfo {year} {2018})}\BibitemShut {NoStop}%
\bibitem [{\citenamefont {Autti}\ \emph {et~al.}(2018)\citenamefont {Autti},
  \citenamefont {Eltsov},\ and\ \citenamefont {Volovik}}]{Autti2018}%
  \BibitemOpen
  \bibfield  {author} {\bibinfo {author} {\bibfnamefont {S.}~\bibnamefont
  {Autti}}, \bibinfo {author} {\bibfnamefont {V.~B.}\ \bibnamefont {Eltsov}},\
  and\ \bibinfo {author} {\bibfnamefont {G.~E.}\ \bibnamefont {Volovik}},\
  }\bibfield  {title} {\bibinfo {title} {{Observation of a Time Quasicrystal
  and Its Transition to a Superfluid Time Crystal}},\ }\href
  {https://doi.org/10.1103/PhysRevLett.120.215301} {\bibfield  {journal}
  {\bibinfo  {journal} {Phys. Rev. Lett.}\ }\textbf {\bibinfo {volume} {120}},\
  \bibinfo {pages} {215301} (\bibinfo {year} {2018})}\BibitemShut {NoStop}%
\bibitem [{\citenamefont {Huang}\ \emph {et~al.}(2018)\citenamefont {Huang},
  \citenamefont {Wu},\ and\ \citenamefont {Liu}}]{Huang2018}%
  \BibitemOpen
  \bibfield  {author} {\bibinfo {author} {\bibfnamefont {B.}~\bibnamefont
  {Huang}}, \bibinfo {author} {\bibfnamefont {Y.-H.}\ \bibnamefont {Wu}},\ and\
  \bibinfo {author} {\bibfnamefont {W.~V.}\ \bibnamefont {Liu}},\ }\bibfield
  {title} {\bibinfo {title} {{Clean Floquet Time Crystals: Models and
  Realizations in Cold Atoms}},\ }\href
  {https://doi.org/10.1103/PhysRevLett.120.110603} {\bibfield  {journal}
  {\bibinfo  {journal} {Phys. Rev. Lett.}\ }\textbf {\bibinfo {volume} {120}},\
  \bibinfo {pages} {110603} (\bibinfo {year} {2018})}\BibitemShut {NoStop}%
\bibitem [{\citenamefont {Gong}\ \emph {et~al.}(2018)\citenamefont {Gong},
  \citenamefont {Hamazaki},\ and\ \citenamefont {Ueda}}]{Gong2018}%
  \BibitemOpen
  \bibfield  {author} {\bibinfo {author} {\bibfnamefont {Z.}~\bibnamefont
  {Gong}}, \bibinfo {author} {\bibfnamefont {R.}~\bibnamefont {Hamazaki}},\
  and\ \bibinfo {author} {\bibfnamefont {M.}~\bibnamefont {Ueda}},\ }\bibfield
  {title} {\bibinfo {title} {{Discrete Time-Crystalline Order in Cavity and
  Circuit QED Systems}},\ }\href
  {https://doi.org/10.1103/PhysRevLett.120.040404} {\bibfield  {journal}
  {\bibinfo  {journal} {Phys. Rev. Lett.}\ }\textbf {\bibinfo {volume} {120}},\
  \bibinfo {pages} {040404} (\bibinfo {year} {2018})}\BibitemShut {NoStop}%
\bibitem [{\citenamefont {{Zhu}}\ \emph {et~al.}(2019)\citenamefont {{Zhu}},
  \citenamefont {{Marino}}, \citenamefont {{Yao}}, \citenamefont {{Lukin}},\
  and\ \citenamefont {{Demler}}}]{Zhu2019}%
  \BibitemOpen
  \bibfield  {author} {\bibinfo {author} {\bibfnamefont {B.}~\bibnamefont
  {{Zhu}}}, \bibinfo {author} {\bibfnamefont {J.}~\bibnamefont {{Marino}}},
  \bibinfo {author} {\bibfnamefont {N.~Y.}\ \bibnamefont {{Yao}}}, \bibinfo
  {author} {\bibfnamefont {M.~D.}\ \bibnamefont {{Lukin}}},\ and\ \bibinfo
  {author} {\bibfnamefont {E.~A.}\ \bibnamefont {{Demler}}},\ }\bibfield
  {title} {\bibinfo {title} {{Dicke time crystals in driven-dissipative quantum
  many-body systems}},\ }\href {https://doi.org/10.1088/1367-2630/ab2afe}
  {\bibfield  {journal} {\bibinfo  {journal} {New J. Phys.}\ }\textbf {\bibinfo
  {volume} {21}},\ \bibinfo {pages} {073028} (\bibinfo {year}
  {2019})}\BibitemShut {NoStop}%
\bibitem [{\citenamefont {Iemini}\ \emph {et~al.}(2018)\citenamefont {Iemini},
  \citenamefont {Russomanno}, \citenamefont {Keeling}, \citenamefont
  {Schir\`o}, \citenamefont {Dalmonte},\ and\ \citenamefont
  {Fazio}}]{Iemini2018}%
  \BibitemOpen
  \bibfield  {author} {\bibinfo {author} {\bibfnamefont {F.}~\bibnamefont
  {Iemini}}, \bibinfo {author} {\bibfnamefont {A.}~\bibnamefont {Russomanno}},
  \bibinfo {author} {\bibfnamefont {J.}~\bibnamefont {Keeling}}, \bibinfo
  {author} {\bibfnamefont {M.}~\bibnamefont {Schir\`o}}, \bibinfo {author}
  {\bibfnamefont {M.}~\bibnamefont {Dalmonte}},\ and\ \bibinfo {author}
  {\bibfnamefont {R.}~\bibnamefont {Fazio}},\ }\bibfield  {title} {\bibinfo
  {title} {{Boundary Time Crystals}},\ }\href
  {https://doi.org/10.1103/PhysRevLett.121.035301} {\bibfield  {journal}
  {\bibinfo  {journal} {Phys. Rev. Lett.}\ }\textbf {\bibinfo {volume} {121}},\
  \bibinfo {pages} {035301} (\bibinfo {year} {2018})}\BibitemShut {NoStop}%
\bibitem [{\citenamefont {{Bu{\v{c}}a}}\ \emph {et~al.}(2019)\citenamefont
  {{Bu{\v{c}}a}}, \citenamefont {{Tindall}},\ and\ \citenamefont
  {{Jaksch}}}]{Buca2019}%
  \BibitemOpen
  \bibfield  {author} {\bibinfo {author} {\bibfnamefont {B.}~\bibnamefont
  {{Bu{\v{c}}a}}}, \bibinfo {author} {\bibfnamefont {J.}~\bibnamefont
  {{Tindall}}},\ and\ \bibinfo {author} {\bibfnamefont {D.}~\bibnamefont
  {{Jaksch}}},\ }\bibfield  {title} {\bibinfo {title} {{Non-stationary coherent
  quantum many-body dynamics through dissipation}},\ }\href
  {https://doi.org/10.1038/s41467-019-09757-y} {\bibfield  {journal} {\bibinfo
  {journal} {Nat. Commun.}\ }\textbf {\bibinfo {volume} {10}},\ \bibinfo
  {pages} {1730} (\bibinfo {year} {2019})}\BibitemShut {NoStop}%
\bibitem [{\citenamefont {Gambetta}\ \emph {et~al.}(2019)\citenamefont
  {Gambetta}, \citenamefont {Carollo}, \citenamefont {Marcuzzi}, \citenamefont
  {Garrahan},\ and\ \citenamefont {Lesanovsky}}]{Gambetta2019}%
  \BibitemOpen
  \bibfield  {author} {\bibinfo {author} {\bibfnamefont {F.~M.}\ \bibnamefont
  {Gambetta}}, \bibinfo {author} {\bibfnamefont {F.}~\bibnamefont {Carollo}},
  \bibinfo {author} {\bibfnamefont {M.}~\bibnamefont {Marcuzzi}}, \bibinfo
  {author} {\bibfnamefont {J.~P.}\ \bibnamefont {Garrahan}},\ and\ \bibinfo
  {author} {\bibfnamefont {I.}~\bibnamefont {Lesanovsky}},\ }\bibfield  {title}
  {\bibinfo {title} {{Discrete Time Crystals in the Absence of Manifest
  Symmetries or Disorder in Open Quantum Systems}},\ }\href
  {https://doi.org/10.1103/PhysRevLett.122.015701} {\bibfield  {journal}
  {\bibinfo  {journal} {Phys. Rev. Lett.}\ }\textbf {\bibinfo {volume} {122}},\
  \bibinfo {pages} {015701} (\bibinfo {year} {2019})}\BibitemShut {NoStop}%
\bibitem [{\citenamefont {{O'Sullivan}}\ \emph {et~al.}(2020)\citenamefont
  {{O'Sullivan}}, \citenamefont {{Lunt}}, \citenamefont {{Zollitsch}},
  \citenamefont {{Thewalt}}, \citenamefont {{Morton}},\ and\ \citenamefont
  {{Pal}}}]{Sullivan2020}%
  \BibitemOpen
  \bibfield  {author} {\bibinfo {author} {\bibfnamefont {J.}~\bibnamefont
  {{O'Sullivan}}}, \bibinfo {author} {\bibfnamefont {O.}~\bibnamefont
  {{Lunt}}}, \bibinfo {author} {\bibfnamefont {C.~W.}\ \bibnamefont
  {{Zollitsch}}}, \bibinfo {author} {\bibfnamefont {M.~L.~W.}\ \bibnamefont
  {{Thewalt}}}, \bibinfo {author} {\bibfnamefont {J.~J.~L.}\ \bibnamefont
  {{Morton}}},\ and\ \bibinfo {author} {\bibfnamefont {A.}~\bibnamefont
  {{Pal}}},\ }\bibfield  {title} {\bibinfo {title} {{Signatures of discrete
  time crystalline order in dissipative spin ensembles}},\ }\href
  {https://doi.org/10.1088/1367-2630/ab9fbe} {\bibfield  {journal} {\bibinfo
  {journal} {New J. Phys.}\ }\textbf {\bibinfo {volume} {22}},\ \bibinfo {eid}
  {085001} (\bibinfo {year} {2020})}\BibitemShut {NoStop}%
\bibitem [{\citenamefont {Skulte}\ \emph {et~al.}(2021)\citenamefont {Skulte},
  \citenamefont {Kongkhambut}, \citenamefont {Ke\ss{}ler}, \citenamefont
  {Hemmerich}, \citenamefont {Mathey},\ and\ \citenamefont
  {Cosme}}]{Skulte2021}%
  \BibitemOpen
  \bibfield  {author} {\bibinfo {author} {\bibfnamefont {J.}~\bibnamefont
  {Skulte}}, \bibinfo {author} {\bibfnamefont {P.}~\bibnamefont {Kongkhambut}},
  \bibinfo {author} {\bibfnamefont {H.}~\bibnamefont {Ke\ss{}ler}}, \bibinfo
  {author} {\bibfnamefont {A.}~\bibnamefont {Hemmerich}}, \bibinfo {author}
  {\bibfnamefont {L.}~\bibnamefont {Mathey}},\ and\ \bibinfo {author}
  {\bibfnamefont {J.~G.}\ \bibnamefont {Cosme}},\ }\bibfield  {title} {\bibinfo
  {title} {{Parametrically driven dissipative three-level Dicke model}},\
  }\href {https://doi.org/10.1103/PhysRevA.104.063705} {\bibfield  {journal}
  {\bibinfo  {journal} {Phys. Rev. A}\ }\textbf {\bibinfo {volume} {104}},\
  \bibinfo {pages} {063705} (\bibinfo {year} {2021})}\BibitemShut {NoStop}%
\bibitem [{\citenamefont {Hajdu\ifmmode~\check{s}\else \v{s}\fi{}ek}\ \emph
  {et~al.}(2022)\citenamefont {Hajdu\ifmmode~\check{s}\else \v{s}\fi{}ek},
  \citenamefont {Solanki}, \citenamefont {Fazio},\ and\ \citenamefont
  {Vinjanampathy}}]{Michal2022}%
  \BibitemOpen
  \bibfield  {author} {\bibinfo {author} {\bibfnamefont {M.}~\bibnamefont
  {Hajdu\ifmmode~\check{s}\else \v{s}\fi{}ek}}, \bibinfo {author}
  {\bibfnamefont {P.}~\bibnamefont {Solanki}}, \bibinfo {author} {\bibfnamefont
  {R.}~\bibnamefont {Fazio}},\ and\ \bibinfo {author} {\bibfnamefont
  {S.}~\bibnamefont {Vinjanampathy}},\ }\bibfield  {title} {\bibinfo {title}
  {{Seeding Crystallization in Time}},\ }\href
  {https://doi.org/10.1103/PhysRevLett.128.080603} {\bibfield  {journal}
  {\bibinfo  {journal} {Phys. Rev. Lett.}\ }\textbf {\bibinfo {volume} {128}},\
  \bibinfo {pages} {080603} (\bibinfo {year} {2022})}\BibitemShut {NoStop}%
\bibitem [{\citenamefont {Cabot}\ \emph {et~al.}(2022)\citenamefont {Cabot},
  \citenamefont {Carollo},\ and\ \citenamefont {Lesanovsky}}]{Cabot2022}%
  \BibitemOpen
  \bibfield  {author} {\bibinfo {author} {\bibfnamefont {A.}~\bibnamefont
  {Cabot}}, \bibinfo {author} {\bibfnamefont {F.}~\bibnamefont {Carollo}},\
  and\ \bibinfo {author} {\bibfnamefont {I.}~\bibnamefont {Lesanovsky}},\
  }\bibfield  {title} {\bibinfo {title} {Metastable discrete time-crystal
  resonances in a dissipative central spin system},\ }\href
  {https://doi.org/10.1103/PhysRevB.106.134311} {\bibfield  {journal} {\bibinfo
   {journal} {Phys. Rev. B}\ }\textbf {\bibinfo {volume} {106}},\ \bibinfo
  {pages} {134311} (\bibinfo {year} {2022})}\BibitemShut {NoStop}%
\bibitem [{\citenamefont {Vu}\ and\ \citenamefont {Das~Sarma}(2023)}]{Vu2022}%
  \BibitemOpen
  \bibfield  {author} {\bibinfo {author} {\bibfnamefont {D.}~\bibnamefont
  {Vu}}\ and\ \bibinfo {author} {\bibfnamefont {S.}~\bibnamefont {Das~Sarma}},\
  }\bibfield  {title} {\bibinfo {title} {{Dissipative Prethermal Discrete Time
  Crystal}},\ }\href {https://doi.org/10.1103/PhysRevLett.130.130401}
  {\bibfield  {journal} {\bibinfo  {journal} {Phys. Rev. Lett.}\ }\textbf
  {\bibinfo {volume} {130}},\ \bibinfo {pages} {130401} (\bibinfo {year}
  {2023})}\BibitemShut {NoStop}%
\bibitem [{\citenamefont {Ke\ss{}ler}\ \emph {et~al.}(2021)\citenamefont
  {Ke\ss{}ler}, \citenamefont {Kongkhambut}, \citenamefont {Georges},
  \citenamefont {Mathey}, \citenamefont {Cosme},\ and\ \citenamefont
  {Hemmerich}}]{Kessler2021}%
  \BibitemOpen
  \bibfield  {author} {\bibinfo {author} {\bibfnamefont {H.}~\bibnamefont
  {Ke\ss{}ler}}, \bibinfo {author} {\bibfnamefont {P.}~\bibnamefont
  {Kongkhambut}}, \bibinfo {author} {\bibfnamefont {C.}~\bibnamefont
  {Georges}}, \bibinfo {author} {\bibfnamefont {L.}~\bibnamefont {Mathey}},
  \bibinfo {author} {\bibfnamefont {J.~G.}\ \bibnamefont {Cosme}},\ and\
  \bibinfo {author} {\bibfnamefont {A.}~\bibnamefont {Hemmerich}},\ }\bibfield
  {title} {\bibinfo {title} {{Observation of a Dissipative Time Crystal}},\
  }\href {https://doi.org/10.1103/PhysRevLett.127.043602} {\bibfield  {journal}
  {\bibinfo  {journal} {Phys. Rev. Lett.}\ }\textbf {\bibinfo {volume} {127}},\
  \bibinfo {pages} {043602} (\bibinfo {year} {2021})}\BibitemShut {NoStop}%
\bibitem [{\citenamefont {Nie}\ and\ \citenamefont {Zheng}(2023)}]{Nie2023}%
  \BibitemOpen
  \bibfield  {author} {\bibinfo {author} {\bibfnamefont {X.}~\bibnamefont
  {Nie}}\ and\ \bibinfo {author} {\bibfnamefont {W.}~\bibnamefont {Zheng}},\
  }\bibfield  {title} {\bibinfo {title} {Mode softening in time-crystalline
  transitions of open quantum systems},\ }\href
  {https://doi.org/10.1103/PhysRevA.107.033311} {\bibfield  {journal} {\bibinfo
   {journal} {Phys. Rev. A}\ }\textbf {\bibinfo {volume} {107}},\ \bibinfo
  {pages} {033311} (\bibinfo {year} {2023})}\BibitemShut {NoStop}%
\bibitem [{\citenamefont {{Ke{\ss}ler}}\ \emph {et~al.}(2020)\citenamefont
  {{Ke{\ss}ler}}, \citenamefont {{Cosme}}, \citenamefont {{Georges}},
  \citenamefont {{Mathey}},\ and\ \citenamefont {{Hemmerich}}}]{Kessler2020}%
  \BibitemOpen
  \bibfield  {author} {\bibinfo {author} {\bibfnamefont {H.}~\bibnamefont
  {{Ke{\ss}ler}}}, \bibinfo {author} {\bibfnamefont {J.~G.}\ \bibnamefont
  {{Cosme}}}, \bibinfo {author} {\bibfnamefont {C.}~\bibnamefont {{Georges}}},
  \bibinfo {author} {\bibfnamefont {L.}~\bibnamefont {{Mathey}}},\ and\
  \bibinfo {author} {\bibfnamefont {A.}~\bibnamefont {{Hemmerich}}},\
  }\bibfield  {title} {\bibinfo {title} {{From a continuous to a discrete time
  crystal in a dissipative atom-cavity system}},\ }\href
  {https://doi.org/10.1088/1367-2630/ab9fc0} {\bibfield  {journal} {\bibinfo
  {journal} {New J. Phys.}\ }\textbf {\bibinfo {volume} {22}},\ \bibinfo {eid}
  {085002} (\bibinfo {year} {2020})}\BibitemShut {NoStop}%
\bibitem [{\citenamefont {{Alaeian}}\ and\ \citenamefont
  {{Bu{\v{c}}a}}(2022)}]{Alaeian2022}%
  \BibitemOpen
  \bibfield  {author} {\bibinfo {author} {\bibfnamefont {H.}~\bibnamefont
  {{Alaeian}}}\ and\ \bibinfo {author} {\bibfnamefont {B.}~\bibnamefont
  {{Bu{\v{c}}a}}},\ }\bibfield  {title} {\bibinfo {title} {{Exact
  multistability and dissipative time crystals in interacting fermionic
  lattices}},\ }\href {https://doi.org/10.1038/s42005-022-01090-z} {\bibfield
  {journal} {\bibinfo  {journal} {Commun. Phys.}\ }\textbf {\bibinfo {volume}
  {5}},\ \bibinfo {eid} {318} (\bibinfo {year} {2022})}\BibitemShut {NoStop}%
\bibitem [{\citenamefont {Heugel}\ \emph {et~al.}(2019)\citenamefont {Heugel},
  \citenamefont {Oscity}, \citenamefont {Eichler}, \citenamefont {Zilberberg},\
  and\ \citenamefont {Chitra}}]{Heugel2019}%
  \BibitemOpen
  \bibfield  {author} {\bibinfo {author} {\bibfnamefont {T.~L.}\ \bibnamefont
  {Heugel}}, \bibinfo {author} {\bibfnamefont {M.}~\bibnamefont {Oscity}},
  \bibinfo {author} {\bibfnamefont {A.}~\bibnamefont {Eichler}}, \bibinfo
  {author} {\bibfnamefont {O.}~\bibnamefont {Zilberberg}},\ and\ \bibinfo
  {author} {\bibfnamefont {R.}~\bibnamefont {Chitra}},\ }\bibfield  {title}
  {\bibinfo {title} {{Classical Many-Body Time Crystals}},\ }\href
  {https://doi.org/10.1103/PhysRevLett.123.124301} {\bibfield  {journal}
  {\bibinfo  {journal} {Phys. Rev. Lett.}\ }\textbf {\bibinfo {volume} {123}},\
  \bibinfo {pages} {124301} (\bibinfo {year} {2019})}\BibitemShut {NoStop}%
\bibitem [{\citenamefont {Kongkhambut}\ \emph {et~al.}(2022)\citenamefont
  {Kongkhambut}, \citenamefont {Skulte}, \citenamefont {Mathey}, \citenamefont
  {Cosme}, \citenamefont {Hemmerich},\ and\ \citenamefont
  {Ke{\ss}ler}}]{Kongkhambut2022}%
  \BibitemOpen
  \bibfield  {author} {\bibinfo {author} {\bibfnamefont {P.}~\bibnamefont
  {Kongkhambut}}, \bibinfo {author} {\bibfnamefont {J.}~\bibnamefont {Skulte}},
  \bibinfo {author} {\bibfnamefont {L.}~\bibnamefont {Mathey}}, \bibinfo
  {author} {\bibfnamefont {J.~G.}\ \bibnamefont {Cosme}}, \bibinfo {author}
  {\bibfnamefont {A.}~\bibnamefont {Hemmerich}},\ and\ \bibinfo {author}
  {\bibfnamefont {H.}~\bibnamefont {Ke{\ss}ler}},\ }\bibfield  {title}
  {\bibinfo {title} {Observation of a continuous time crystal},\ }\href
  {https://doi.org/10.1126/science.abo3382} {\bibfield  {journal} {\bibinfo
  {journal} {Science}\ }\textbf {\bibinfo {volume} {377}},\ \bibinfo {pages}
  {670} (\bibinfo {year} {2022})}\BibitemShut {NoStop}%
\bibitem [{\citenamefont {{Taheri}}\ \emph {et~al.}(2022)\citenamefont
  {{Taheri}}, \citenamefont {{Matsko}}, \citenamefont {{Maleki}},\ and\
  \citenamefont {{Sacha}}}]{Taheri2022}%
  \BibitemOpen
  \bibfield  {author} {\bibinfo {author} {\bibfnamefont {H.}~\bibnamefont
  {{Taheri}}}, \bibinfo {author} {\bibfnamefont {A.~B.}\ \bibnamefont
  {{Matsko}}}, \bibinfo {author} {\bibfnamefont {L.}~\bibnamefont {{Maleki}}},\
  and\ \bibinfo {author} {\bibfnamefont {K.}~\bibnamefont {{Sacha}}},\
  }\bibfield  {title} {\bibinfo {title} {{All-optical dissipative discrete time
  crystals}},\ }\href
  {https://doi.org/https://doi.org/10.1038/s41467-022-28462-x} {\bibfield
  {journal} {\bibinfo  {journal} {Nat. Commun.}\ }\textbf {\bibinfo {volume}
  {13}},\ \bibinfo {eid} {848} (\bibinfo {year} {2022})}\BibitemShut {NoStop}%
\bibitem [{\citenamefont {{Lipkin}}\ \emph {et~al.}(1965)\citenamefont
  {{Lipkin}}, \citenamefont {{Meshkov}},\ and\ \citenamefont
  {{Glick}}}]{Lipkin1965}%
  \BibitemOpen
  \bibfield  {author} {\bibinfo {author} {\bibfnamefont {H.~J.}\ \bibnamefont
  {{Lipkin}}}, \bibinfo {author} {\bibfnamefont {N.}~\bibnamefont
  {{Meshkov}}},\ and\ \bibinfo {author} {\bibfnamefont {A.~J.}\ \bibnamefont
  {{Glick}}},\ }\bibfield  {title} {\bibinfo {title} {{Validity of many-body
  approximation methods for a solvable model. (I). Exact solutions and
  perturbation theory}},\ }\href {https://doi.org/10.1016/0029-5582(65)90862-X}
  {\bibfield  {journal} {\bibinfo  {journal} {Nucl. Phys.}\ }\textbf {\bibinfo
  {volume} {62}},\ \bibinfo {pages} {188} (\bibinfo {year} {1965})}\BibitemShut
  {NoStop}%
\bibitem [{\citenamefont {{Meshkov}}\ \emph {et~al.}(1965)\citenamefont
  {{Meshkov}}, \citenamefont {{Glick}},\ and\ \citenamefont
  {{Lipkin}}}]{Meshkov1965}%
  \BibitemOpen
  \bibfield  {author} {\bibinfo {author} {\bibfnamefont {N.}~\bibnamefont
  {{Meshkov}}}, \bibinfo {author} {\bibfnamefont {A.~J.}\ \bibnamefont
  {{Glick}}},\ and\ \bibinfo {author} {\bibfnamefont {H.~J.}\ \bibnamefont
  {{Lipkin}}},\ }\bibfield  {title} {\bibinfo {title} {{Validity of many-body
  approximation methods for a solvable model. (II). Linearization
  procedures}},\ }\href {https://doi.org/10.1016/0029-5582(65)90863-1}
  {\bibfield  {journal} {\bibinfo  {journal} {Nucl. Phys.}\ }\textbf {\bibinfo
  {volume} {62}},\ \bibinfo {pages} {199} (\bibinfo {year} {1965})}\BibitemShut
  {NoStop}%
\bibitem [{\citenamefont {{Glick}}\ \emph {et~al.}(1965)\citenamefont
  {{Glick}}, \citenamefont {{Lipkin}},\ and\ \citenamefont
  {{Meshkov}}}]{Glick1965}%
  \BibitemOpen
  \bibfield  {author} {\bibinfo {author} {\bibfnamefont {A.~J.}\ \bibnamefont
  {{Glick}}}, \bibinfo {author} {\bibfnamefont {H.~J.}\ \bibnamefont
  {{Lipkin}}},\ and\ \bibinfo {author} {\bibfnamefont {N.}~\bibnamefont
  {{Meshkov}}},\ }\bibfield  {title} {\bibinfo {title} {{Validity of many-body
  approximation methods for a solvable model. (III). Diagram summations}},\
  }\href {https://doi.org/10.1016/0029-5582(65)90864-3} {\bibfield  {journal}
  {\bibinfo  {journal} {Nucl. Phys.}\ }\textbf {\bibinfo {volume} {62}},\
  \bibinfo {pages} {211} (\bibinfo {year} {1965})}\BibitemShut {NoStop}%
\bibitem [{\citenamefont {Vidal}\ \emph {et~al.}(2004)\citenamefont {Vidal},
  \citenamefont {Palacios},\ and\ \citenamefont {Aslangul}}]{Vidal2004}%
  \BibitemOpen
  \bibfield  {author} {\bibinfo {author} {\bibfnamefont {J.}~\bibnamefont
  {Vidal}}, \bibinfo {author} {\bibfnamefont {G.}~\bibnamefont {Palacios}},\
  and\ \bibinfo {author} {\bibfnamefont {C.}~\bibnamefont {Aslangul}},\
  }\bibfield  {title} {\bibinfo {title} {{Entanglement dynamics in the
  Lipkin-Meshkov-Glick model}},\ }\href
  {https://doi.org/10.1103/PhysRevA.70.062304} {\bibfield  {journal} {\bibinfo
  {journal} {Phys. Rev. A}\ }\textbf {\bibinfo {volume} {70}},\ \bibinfo
  {pages} {062304} (\bibinfo {year} {2004})}\BibitemShut {NoStop}%
\bibitem [{\citenamefont {Morrison}\ and\ \citenamefont
  {Parkins}(2008)}]{Morrison2008}%
  \BibitemOpen
  \bibfield  {author} {\bibinfo {author} {\bibfnamefont {S.}~\bibnamefont
  {Morrison}}\ and\ \bibinfo {author} {\bibfnamefont {A.~S.}\ \bibnamefont
  {Parkins}},\ }\bibfield  {title} {\bibinfo {title} {{Dynamical Quantum Phase
  Transitions in the Dissipative Lipkin-Meshkov-Glick Model with Proposed
  Realization in Optical Cavity QED}},\ }\href
  {https://doi.org/10.1103/PhysRevLett.100.040403} {\bibfield  {journal}
  {\bibinfo  {journal} {Phys. Rev. Lett.}\ }\textbf {\bibinfo {volume} {100}},\
  \bibinfo {pages} {040403} (\bibinfo {year} {2008})}\BibitemShut {NoStop}%
\bibitem [{\citenamefont {Dicke}(1954)}]{Dicke1954}%
  \BibitemOpen
  \bibfield  {author} {\bibinfo {author} {\bibfnamefont {R.~H.}\ \bibnamefont
  {Dicke}},\ }\bibfield  {title} {\bibinfo {title} {{Coherence in Spontaneous
  Radiation Processes}},\ }\href {https://doi.org/10.1103/PhysRev.93.99}
  {\bibfield  {journal} {\bibinfo  {journal} {Phys. Rev.}\ }\textbf {\bibinfo
  {volume} {93}},\ \bibinfo {pages} {99} (\bibinfo {year} {1954})}\BibitemShut
  {NoStop}%
\bibitem [{\citenamefont {Kirton}\ \emph {et~al.}(2019)\citenamefont {Kirton},
  \citenamefont {Roses}, \citenamefont {Keeling},\ and\ \citenamefont
  {Dalla~Torre}}]{Kirton2019}%
  \BibitemOpen
  \bibfield  {author} {\bibinfo {author} {\bibfnamefont {P.}~\bibnamefont
  {Kirton}}, \bibinfo {author} {\bibfnamefont {M.~M.}\ \bibnamefont {Roses}},
  \bibinfo {author} {\bibfnamefont {J.}~\bibnamefont {Keeling}},\ and\ \bibinfo
  {author} {\bibfnamefont {E.~G.}\ \bibnamefont {Dalla~Torre}},\ }\bibfield
  {title} {\bibinfo {title} {{Introduction to the Dicke Model: From Equilibrium
  to Nonequilibrium, and Vice Versa}},\ }\href
  {https://doi.org/10.1002/qute.201800043} {\bibfield  {journal} {\bibinfo
  {journal} {Adv. Quantum Technol.}\ }\textbf {\bibinfo {volume} {2}},\
  \bibinfo {pages} {1800043} (\bibinfo {year} {2019})}\BibitemShut {NoStop}%
\bibitem [{\citenamefont {Larson}\ and\ \citenamefont
  {Mavrogordatos}(2021)}]{Larson2021}%
  \BibitemOpen
  \bibfield  {author} {\bibinfo {author} {\bibfnamefont {J.}~\bibnamefont
  {Larson}}\ and\ \bibinfo {author} {\bibfnamefont {T.}~\bibnamefont
  {Mavrogordatos}},\ }\href@noop {} {\emph {\bibinfo {title} {{The
  Jaynes–Cummings Model and Its Descendants}}}}\ (\bibinfo  {publisher} {IOP
  Publishing, Bristol},\ \bibinfo {year} {2021})\BibitemShut {NoStop}%
\bibitem [{\citenamefont {Kelly}\ \emph {et~al.}(2021)\citenamefont {Kelly},
  \citenamefont {Timmermans}, \citenamefont {Marino},\ and\ \citenamefont
  {Tsai}}]{Kelly2021}%
  \BibitemOpen
  \bibfield  {author} {\bibinfo {author} {\bibfnamefont {S.~P.}\ \bibnamefont
  {Kelly}}, \bibinfo {author} {\bibfnamefont {E.}~\bibnamefont {Timmermans}},
  \bibinfo {author} {\bibfnamefont {J.}~\bibnamefont {Marino}},\ and\ \bibinfo
  {author} {\bibfnamefont {S.-W.}\ \bibnamefont {Tsai}},\ }\bibfield  {title}
  {\bibinfo {title} {{Stroboscopic aliasing in long-range interacting quantum
  systems}},\ }\href {https://doi.org/10.21468/SciPostPhysCore.4.3.021}
  {\bibfield  {journal} {\bibinfo  {journal} {SciPost Phys. Core}\ }\textbf
  {\bibinfo {volume} {4}},\ \bibinfo {pages} {021} (\bibinfo {year}
  {2021})}\BibitemShut {NoStop}%
\bibitem [{\citenamefont {{Baumann}}\ \emph {et~al.}(2010)\citenamefont
  {{Baumann}}, \citenamefont {{Guerlin}}, \citenamefont {{Brennecke}},\ and\
  \citenamefont {{Esslinger}}}]{Baumann2010}%
  \BibitemOpen
  \bibfield  {author} {\bibinfo {author} {\bibfnamefont {K.}~\bibnamefont
  {{Baumann}}}, \bibinfo {author} {\bibfnamefont {C.}~\bibnamefont
  {{Guerlin}}}, \bibinfo {author} {\bibfnamefont {F.}~\bibnamefont
  {{Brennecke}}},\ and\ \bibinfo {author} {\bibfnamefont {T.}~\bibnamefont
  {{Esslinger}}},\ }\bibfield  {title} {\bibinfo {title} {{Dicke quantum phase
  transition with a superfluid gas in an optical cavity}},\ }\href
  {https://doi.org/10.1038/nature09009} {\bibfield  {journal} {\bibinfo
  {journal} {Nature}\ }\textbf {\bibinfo {volume} {464}},\ \bibinfo {pages}
  {1301} (\bibinfo {year} {2010})}\BibitemShut {NoStop}%
\bibitem [{\citenamefont {{Korenblit}}\ \emph {et~al.}(2012)\citenamefont
  {{Korenblit}}, \citenamefont {{Kafri}}, \citenamefont {{Campbell}},
  \citenamefont {{Islam}}, \citenamefont {{Edwards}}, \citenamefont {{Gong}},
  \citenamefont {{Lin}}, \citenamefont {{Duan}}, \citenamefont {{Kim}},
  \citenamefont {{Kim}},\ and\ \citenamefont {{Monroe}}}]{Korenblit2012}%
  \BibitemOpen
  \bibfield  {author} {\bibinfo {author} {\bibfnamefont {S.}~\bibnamefont
  {{Korenblit}}}, \bibinfo {author} {\bibfnamefont {D.}~\bibnamefont
  {{Kafri}}}, \bibinfo {author} {\bibfnamefont {W.~C.}\ \bibnamefont
  {{Campbell}}}, \bibinfo {author} {\bibfnamefont {R.}~\bibnamefont {{Islam}}},
  \bibinfo {author} {\bibfnamefont {E.~E.}\ \bibnamefont {{Edwards}}}, \bibinfo
  {author} {\bibfnamefont {Z.~X.}\ \bibnamefont {{Gong}}}, \bibinfo {author}
  {\bibfnamefont {G.~D.}\ \bibnamefont {{Lin}}}, \bibinfo {author}
  {\bibfnamefont {L.~M.}\ \bibnamefont {{Duan}}}, \bibinfo {author}
  {\bibfnamefont {J.}~\bibnamefont {{Kim}}}, \bibinfo {author} {\bibfnamefont
  {K.}~\bibnamefont {{Kim}}},\ and\ \bibinfo {author} {\bibfnamefont
  {C.}~\bibnamefont {{Monroe}}},\ }\bibfield  {title} {\bibinfo {title}
  {{Quantum simulation of spin models on an arbitrary lattice with trapped
  ions}},\ }\href {https://doi.org/10.1088/1367-2630/14/9/095024} {\bibfield
  {journal} {\bibinfo  {journal} {New J. Phys.}\ }\textbf {\bibinfo {volume}
  {14}},\ \bibinfo {eid} {095024} (\bibinfo {year} {2012})}\BibitemShut
  {NoStop}%
\bibitem [{\citenamefont {Jurcevic}\ \emph {et~al.}(2017)\citenamefont
  {Jurcevic}, \citenamefont {Shen}, \citenamefont {Hauke}, \citenamefont
  {Maier}, \citenamefont {Brydges}, \citenamefont {Hempel}, \citenamefont
  {Lanyon}, \citenamefont {Heyl}, \citenamefont {Blatt},\ and\ \citenamefont
  {Roos}}]{Jurcevic2017}%
  \BibitemOpen
  \bibfield  {author} {\bibinfo {author} {\bibfnamefont {P.}~\bibnamefont
  {Jurcevic}}, \bibinfo {author} {\bibfnamefont {H.}~\bibnamefont {Shen}},
  \bibinfo {author} {\bibfnamefont {P.}~\bibnamefont {Hauke}}, \bibinfo
  {author} {\bibfnamefont {C.}~\bibnamefont {Maier}}, \bibinfo {author}
  {\bibfnamefont {T.}~\bibnamefont {Brydges}}, \bibinfo {author} {\bibfnamefont
  {C.}~\bibnamefont {Hempel}}, \bibinfo {author} {\bibfnamefont {B.~P.}\
  \bibnamefont {Lanyon}}, \bibinfo {author} {\bibfnamefont {M.}~\bibnamefont
  {Heyl}}, \bibinfo {author} {\bibfnamefont {R.}~\bibnamefont {Blatt}},\ and\
  \bibinfo {author} {\bibfnamefont {C.~F.}\ \bibnamefont {Roos}},\ }\bibfield
  {title} {\bibinfo {title} {Direct observation of dynamical quantum phase
  transitions in an interacting many-body system},\ }\href
  {https://doi.org/10.1103/PhysRevLett.119.080501} {\bibfield  {journal}
  {\bibinfo  {journal} {Phys. Rev. Lett.}\ }\textbf {\bibinfo {volume} {119}},\
  \bibinfo {pages} {080501} (\bibinfo {year} {2017})}\BibitemShut {NoStop}%
\bibitem [{\citenamefont {Monroe}\ \emph {et~al.}(2021)\citenamefont {Monroe},
  \citenamefont {Campbell}, \citenamefont {Duan}, \citenamefont {Gong},
  \citenamefont {Gorshkov}, \citenamefont {Hess}, \citenamefont {Islam},
  \citenamefont {Kim}, \citenamefont {Linke}, \citenamefont {Pagano},
  \citenamefont {Richerme}, \citenamefont {Senko},\ and\ \citenamefont
  {Yao}}]{Monroe2021}%
  \BibitemOpen
  \bibfield  {author} {\bibinfo {author} {\bibfnamefont {C.}~\bibnamefont
  {Monroe}}, \bibinfo {author} {\bibfnamefont {W.~C.}\ \bibnamefont
  {Campbell}}, \bibinfo {author} {\bibfnamefont {L.-M.}\ \bibnamefont {Duan}},
  \bibinfo {author} {\bibfnamefont {Z.-X.}\ \bibnamefont {Gong}}, \bibinfo
  {author} {\bibfnamefont {A.~V.}\ \bibnamefont {Gorshkov}}, \bibinfo {author}
  {\bibfnamefont {P.~W.}\ \bibnamefont {Hess}}, \bibinfo {author}
  {\bibfnamefont {R.}~\bibnamefont {Islam}}, \bibinfo {author} {\bibfnamefont
  {K.}~\bibnamefont {Kim}}, \bibinfo {author} {\bibfnamefont {N.~M.}\
  \bibnamefont {Linke}}, \bibinfo {author} {\bibfnamefont {G.}~\bibnamefont
  {Pagano}}, \bibinfo {author} {\bibfnamefont {P.}~\bibnamefont {Richerme}},
  \bibinfo {author} {\bibfnamefont {C.}~\bibnamefont {Senko}},\ and\ \bibinfo
  {author} {\bibfnamefont {N.~Y.}\ \bibnamefont {Yao}},\ }\bibfield  {title}
  {\bibinfo {title} {Programmable quantum simulations of spin systems with
  trapped ions},\ }\href {https://doi.org/10.1103/RevModPhys.93.025001}
  {\bibfield  {journal} {\bibinfo  {journal} {Rev. Mod. Phys.}\ }\textbf
  {\bibinfo {volume} {93}},\ \bibinfo {pages} {025001} (\bibinfo {year}
  {2021})}\BibitemShut {NoStop}%
\bibitem [{\citenamefont {Tuquero}\ \emph {et~al.}(2022)\citenamefont
  {Tuquero}, \citenamefont {Skulte}, \citenamefont {Mathey},\ and\
  \citenamefont {Cosme}}]{Tuquero2022}%
  \BibitemOpen
  \bibfield  {author} {\bibinfo {author} {\bibfnamefont {R.~J.~L.}\
  \bibnamefont {Tuquero}}, \bibinfo {author} {\bibfnamefont {J.}~\bibnamefont
  {Skulte}}, \bibinfo {author} {\bibfnamefont {L.}~\bibnamefont {Mathey}},\
  and\ \bibinfo {author} {\bibfnamefont {J.~G.}\ \bibnamefont {Cosme}},\
  }\bibfield  {title} {\bibinfo {title} {{Dissipative time crystal in an
  atom-cavity system: Influence of trap and competing interactions}},\ }\href
  {https://doi.org/10.1103/PhysRevA.105.043311} {\bibfield  {journal} {\bibinfo
   {journal} {Phys. Rev. A}\ }\textbf {\bibinfo {volume} {105}},\ \bibinfo
  {pages} {043311} (\bibinfo {year} {2022})}\BibitemShut {NoStop}%
\bibitem [{\citenamefont {Keeling}\ \emph {et~al.}(2010)\citenamefont
  {Keeling}, \citenamefont {Bhaseen},\ and\ \citenamefont
  {Simons}}]{Keeling2010}%
  \BibitemOpen
  \bibfield  {author} {\bibinfo {author} {\bibfnamefont {J.}~\bibnamefont
  {Keeling}}, \bibinfo {author} {\bibfnamefont {M.~J.}\ \bibnamefont
  {Bhaseen}},\ and\ \bibinfo {author} {\bibfnamefont {B.~D.}\ \bibnamefont
  {Simons}},\ }\bibfield  {title} {\bibinfo {title} {{Collective Dynamics of
  Bose-Einstein Condensates in Optical Cavities}},\ }\href
  {https://doi.org/10.1103/PhysRevLett.105.043001} {\bibfield  {journal}
  {\bibinfo  {journal} {Phys. Rev. Lett.}\ }\textbf {\bibinfo {volume} {105}},\
  \bibinfo {pages} {043001} (\bibinfo {year} {2010})}\BibitemShut {NoStop}%
\bibitem [{\citenamefont {Dimer}\ \emph {et~al.}(2007)\citenamefont {Dimer},
  \citenamefont {Estienne}, \citenamefont {Parkins},\ and\ \citenamefont
  {Carmichael}}]{Dimer2007}%
  \BibitemOpen
  \bibfield  {author} {\bibinfo {author} {\bibfnamefont {F.}~\bibnamefont
  {Dimer}}, \bibinfo {author} {\bibfnamefont {B.}~\bibnamefont {Estienne}},
  \bibinfo {author} {\bibfnamefont {A.~S.}\ \bibnamefont {Parkins}},\ and\
  \bibinfo {author} {\bibfnamefont {H.~J.}\ \bibnamefont {Carmichael}},\
  }\bibfield  {title} {\bibinfo {title} {{Proposed realization of the
  Dicke-model quantum phase transition in an optical cavity QED system}},\
  }\href {https://doi.org/10.1103/PhysRevA.75.013804} {\bibfield  {journal}
  {\bibinfo  {journal} {Phys. Rev. A}\ }\textbf {\bibinfo {volume} {75}},\
  \bibinfo {pages} {013804} (\bibinfo {year} {2007})}\BibitemShut {NoStop}%
\bibitem [{\citenamefont {Damanet}\ \emph {et~al.}(2019)\citenamefont
  {Damanet}, \citenamefont {Daley},\ and\ \citenamefont
  {Keeling}}]{Damanet2019}%
  \BibitemOpen
  \bibfield  {author} {\bibinfo {author} {\bibfnamefont {F.}~\bibnamefont
  {Damanet}}, \bibinfo {author} {\bibfnamefont {A.~J.}\ \bibnamefont {Daley}},\
  and\ \bibinfo {author} {\bibfnamefont {J.}~\bibnamefont {Keeling}},\
  }\bibfield  {title} {\bibinfo {title} {{Atom-only descriptions of the
  driven-dissipative Dicke model}},\ }\href
  {https://doi.org/10.1103/PhysRevA.99.033845} {\bibfield  {journal} {\bibinfo
  {journal} {Phys. Rev. A}\ }\textbf {\bibinfo {volume} {99}},\ \bibinfo
  {pages} {033845} (\bibinfo {year} {2019})}\BibitemShut {NoStop}%
\bibitem [{\citenamefont {J\"ager}\ \emph {et~al.}(2022)\citenamefont
  {J\"ager}, \citenamefont {Schmit}, \citenamefont {Morigi}, \citenamefont
  {Holland},\ and\ \citenamefont {Betzholz}}]{Jager2022}%
  \BibitemOpen
  \bibfield  {author} {\bibinfo {author} {\bibfnamefont {S.~B.}\ \bibnamefont
  {J\"ager}}, \bibinfo {author} {\bibfnamefont {T.}~\bibnamefont {Schmit}},
  \bibinfo {author} {\bibfnamefont {G.}~\bibnamefont {Morigi}}, \bibinfo
  {author} {\bibfnamefont {M.~J.}\ \bibnamefont {Holland}},\ and\ \bibinfo
  {author} {\bibfnamefont {R.}~\bibnamefont {Betzholz}},\ }\bibfield  {title}
  {\bibinfo {title} {{Lindblad Master Equations for Quantum Systems Coupled to
  Dissipative Bosonic Modes}},\ }\href
  {https://doi.org/10.1103/PhysRevLett.129.063601} {\bibfield  {journal}
  {\bibinfo  {journal} {Phys. Rev. Lett.}\ }\textbf {\bibinfo {volume} {129}},\
  \bibinfo {pages} {063601} (\bibinfo {year} {2022})}\BibitemShut {NoStop}%
\bibitem [{\citenamefont {Engelhardt}\ \emph {et~al.}(2013)\citenamefont
  {Engelhardt}, \citenamefont {Bastidas}, \citenamefont {Emary},\ and\
  \citenamefont {Brandes}}]{Engelhardt2013}%
  \BibitemOpen
  \bibfield  {author} {\bibinfo {author} {\bibfnamefont {G.}~\bibnamefont
  {Engelhardt}}, \bibinfo {author} {\bibfnamefont {V.~M.}\ \bibnamefont
  {Bastidas}}, \bibinfo {author} {\bibfnamefont {C.}~\bibnamefont {Emary}},\
  and\ \bibinfo {author} {\bibfnamefont {T.}~\bibnamefont {Brandes}},\
  }\bibfield  {title} {\bibinfo {title} {ac-driven quantum phase transition in
  the lipkin-meshkov-glick model},\ }\href
  {https://doi.org/10.1103/PhysRevE.87.052110} {\bibfield  {journal} {\bibinfo
  {journal} {Phys. Rev. E}\ }\textbf {\bibinfo {volume} {87}},\ \bibinfo
  {pages} {052110} (\bibinfo {year} {2013})}\BibitemShut {NoStop}%
\bibitem [{\citenamefont {Ritsch}\ \emph {et~al.}(2013)\citenamefont {Ritsch},
  \citenamefont {Domokos}, \citenamefont {Brennecke},\ and\ \citenamefont
  {Esslinger}}]{Ritsch2013}%
  \BibitemOpen
  \bibfield  {author} {\bibinfo {author} {\bibfnamefont {H.}~\bibnamefont
  {Ritsch}}, \bibinfo {author} {\bibfnamefont {P.}~\bibnamefont {Domokos}},
  \bibinfo {author} {\bibfnamefont {F.}~\bibnamefont {Brennecke}},\ and\
  \bibinfo {author} {\bibfnamefont {T.}~\bibnamefont {Esslinger}},\ }\bibfield
  {title} {\bibinfo {title} {Cold atoms in cavity-generated dynamical optical
  potentials},\ }\href {https://doi.org/10.1103/RevModPhys.85.553} {\bibfield
  {journal} {\bibinfo  {journal} {Rev. Mod. Phys.}\ }\textbf {\bibinfo {volume}
  {85}},\ \bibinfo {pages} {553} (\bibinfo {year} {2013})}\BibitemShut
  {NoStop}%
\bibitem [{\citenamefont {{Mivehvar}}\ \emph {et~al.}(2021)\citenamefont
  {{Mivehvar}}, \citenamefont {{Piazza}}, \citenamefont {{Donner}},\ and\
  \citenamefont {{Ritsch}}}]{Mivehvar2021}%
  \BibitemOpen
  \bibfield  {author} {\bibinfo {author} {\bibfnamefont {F.}~\bibnamefont
  {{Mivehvar}}}, \bibinfo {author} {\bibfnamefont {F.}~\bibnamefont
  {{Piazza}}}, \bibinfo {author} {\bibfnamefont {T.}~\bibnamefont {{Donner}}},\
  and\ \bibinfo {author} {\bibfnamefont {H.}~\bibnamefont {{Ritsch}}},\
  }\bibfield  {title} {\bibinfo {title} {{Cavity QED with quantum gases: new
  paradigms in many-body physics}},\ }\href
  {https://doi.org/10.1080/00018732.2021.1969727} {\bibfield  {journal}
  {\bibinfo  {journal} {Adv. Phys.}\ }\textbf {\bibinfo {volume} {70}},\
  \bibinfo {pages} {1} (\bibinfo {year} {2021})}\BibitemShut {NoStop}%
\bibitem [{\citenamefont {{Klinder}}\ \emph {et~al.}(2015)\citenamefont
  {{Klinder}}, \citenamefont {{Ke{\ss}ler}}, \citenamefont {{Wolke}},
  \citenamefont {{Mathey}},\ and\ \citenamefont {{Hemmerich}}}]{Klinder2015}%
  \BibitemOpen
  \bibfield  {author} {\bibinfo {author} {\bibfnamefont {J.}~\bibnamefont
  {{Klinder}}}, \bibinfo {author} {\bibfnamefont {H.}~\bibnamefont
  {{Ke{\ss}ler}}}, \bibinfo {author} {\bibfnamefont {M.}~\bibnamefont
  {{Wolke}}}, \bibinfo {author} {\bibfnamefont {L.}~\bibnamefont {{Mathey}}},\
  and\ \bibinfo {author} {\bibfnamefont {A.}~\bibnamefont {{Hemmerich}}},\
  }\bibfield  {title} {\bibinfo {title} {{Dynamical phase transition in the
  open Dicke model}},\ }\href {https://doi.org/10.1073/pnas.1417132112}
  {\bibfield  {journal} {\bibinfo  {journal} {Proc. Natl. Acad. Sci. USA}\
  }\textbf {\bibinfo {volume} {112}},\ \bibinfo {pages} {3290} (\bibinfo {year}
  {2015})}\BibitemShut {NoStop}%
\bibitem [{\citenamefont {D'Alessio}\ and\ \citenamefont
  {Rigol}(2014)}]{Alessio2014}%
  \BibitemOpen
  \bibfield  {author} {\bibinfo {author} {\bibfnamefont {L.}~\bibnamefont
  {D'Alessio}}\ and\ \bibinfo {author} {\bibfnamefont {M.}~\bibnamefont
  {Rigol}},\ }\bibfield  {title} {\bibinfo {title} {{Long-time Behavior of
  Isolated Periodically Driven Interacting Lattice Systems}},\ }\href
  {https://doi.org/10.1103/PhysRevX.4.041048} {\bibfield  {journal} {\bibinfo
  {journal} {Phys. Rev. X}\ }\textbf {\bibinfo {volume} {4}},\ \bibinfo {pages}
  {041048} (\bibinfo {year} {2014})}\BibitemShut {NoStop}%
\bibitem [{\citenamefont {{Bukov}}\ \emph {et~al.}(2015)\citenamefont
  {{Bukov}}, \citenamefont {{D'Alessio}},\ and\ \citenamefont
  {{Polkovnikov}}}]{Bukov2015}%
  \BibitemOpen
  \bibfield  {author} {\bibinfo {author} {\bibfnamefont {M.}~\bibnamefont
  {{Bukov}}}, \bibinfo {author} {\bibfnamefont {L.}~\bibnamefont
  {{D'Alessio}}},\ and\ \bibinfo {author} {\bibfnamefont {A.}~\bibnamefont
  {{Polkovnikov}}},\ }\bibfield  {title} {\bibinfo {title} {{Universal
  high-frequency behavior of periodically driven systems: from dynamical
  stabilization to Floquet engineering}},\ }\href
  {https://doi.org/10.1080/00018732.2015.1055918} {\bibfield  {journal}
  {\bibinfo  {journal} {Adv. Phys.}\ }\textbf {\bibinfo {volume} {64}},\
  \bibinfo {pages} {139} (\bibinfo {year} {2015})}\BibitemShut {NoStop}%
\bibitem [{\citenamefont {{Pizzi}}\ \emph {et~al.}(2019)\citenamefont
  {{Pizzi}}, \citenamefont {{Knolle}},\ and\ \citenamefont
  {{Nunnenkamp}}}]{Pizzi2019}%
  \BibitemOpen
  \bibfield  {author} {\bibinfo {author} {\bibfnamefont {A.}~\bibnamefont
  {{Pizzi}}}, \bibinfo {author} {\bibfnamefont {J.}~\bibnamefont {{Knolle}}},\
  and\ \bibinfo {author} {\bibfnamefont {A.}~\bibnamefont {{Nunnenkamp}}},\
  }\bibfield  {title} {\bibinfo {title} {{Period-$n$ discrete time crystals and
  quasicrystals with ultracold bosons}},\ }\href
  {https://doi.org/10.1103/PhysRevLett.123.150601} {\bibfield  {journal}
  {\bibinfo  {journal} {Phys. Rev. Lett.}\ }\textbf {\bibinfo {volume} {123}},\
  \bibinfo {pages} {150601} (\bibinfo {year} {2019})}\BibitemShut {NoStop}%
\bibitem [{\citenamefont {Pizzi}\ \emph
  {et~al.}(2021{\natexlab{b}})\citenamefont {Pizzi}, \citenamefont
  {Nunnenkamp},\ and\ \citenamefont {Knolle}}]{Pizzi2021}%
  \BibitemOpen
  \bibfield  {author} {\bibinfo {author} {\bibfnamefont {A.}~\bibnamefont
  {Pizzi}}, \bibinfo {author} {\bibfnamefont {A.}~\bibnamefont {Nunnenkamp}},\
  and\ \bibinfo {author} {\bibfnamefont {J.}~\bibnamefont {Knolle}},\
  }\bibfield  {title} {\bibinfo {title} {Classical approaches to prethermal
  discrete time crystals in one, two, and three dimensions},\ }\href
  {https://doi.org/10.1103/PhysRevB.104.094308} {\bibfield  {journal} {\bibinfo
   {journal} {Phys. Rev. B}\ }\textbf {\bibinfo {volume} {104}},\ \bibinfo
  {pages} {094308} (\bibinfo {year} {2021}{\natexlab{b}})}\BibitemShut
  {NoStop}%
\bibitem [{\citenamefont {Chitra}\ and\ \citenamefont
  {Zilberberg}(2015)}]{Chitra2015}%
  \BibitemOpen
  \bibfield  {author} {\bibinfo {author} {\bibfnamefont {R.}~\bibnamefont
  {Chitra}}\ and\ \bibinfo {author} {\bibfnamefont {O.}~\bibnamefont
  {Zilberberg}},\ }\bibfield  {title} {\bibinfo {title} {{Dynamical many-body
  phases of the parametrically driven, dissipative Dicke model}},\ }\href
  {https://doi.org/10.1103/PhysRevA.92.023815} {\bibfield  {journal} {\bibinfo
  {journal} {Phys. Rev. A}\ }\textbf {\bibinfo {volume} {92}},\ \bibinfo
  {pages} {023815} (\bibinfo {year} {2015})}\BibitemShut {NoStop}%
\bibitem [{\citenamefont {Blais}\ \emph {et~al.}(2007)\citenamefont {Blais},
  \citenamefont {Gambetta}, \citenamefont {Wallraff}, \citenamefont {Schuster},
  \citenamefont {Girvin}, \citenamefont {Devoret},\ and\ \citenamefont
  {Schoelkopf}}]{Blais2007}%
  \BibitemOpen
  \bibfield  {author} {\bibinfo {author} {\bibfnamefont {A.}~\bibnamefont
  {Blais}}, \bibinfo {author} {\bibfnamefont {J.}~\bibnamefont {Gambetta}},
  \bibinfo {author} {\bibfnamefont {A.}~\bibnamefont {Wallraff}}, \bibinfo
  {author} {\bibfnamefont {D.~I.}\ \bibnamefont {Schuster}}, \bibinfo {author}
  {\bibfnamefont {S.~M.}\ \bibnamefont {Girvin}}, \bibinfo {author}
  {\bibfnamefont {M.~H.}\ \bibnamefont {Devoret}},\ and\ \bibinfo {author}
  {\bibfnamefont {R.~J.}\ \bibnamefont {Schoelkopf}},\ }\bibfield  {title}
  {\bibinfo {title} {Quantum-information processing with circuit quantum
  electrodynamics},\ }\href {https://doi.org/10.1103/PhysRevA.75.032329}
  {\bibfield  {journal} {\bibinfo  {journal} {Phys. Rev. A}\ }\textbf {\bibinfo
  {volume} {75}},\ \bibinfo {pages} {032329} (\bibinfo {year}
  {2007})}\BibitemShut {NoStop}%
\bibitem [{\citenamefont {{Mlynek}}\ \emph {et~al.}(2014)\citenamefont
  {{Mlynek}}, \citenamefont {{Abdumalikov}}, \citenamefont {{Eichler}},\ and\
  \citenamefont {{Wallraff}}}]{Mlynek2014}%
  \BibitemOpen
  \bibfield  {author} {\bibinfo {author} {\bibfnamefont {J.~A.}\ \bibnamefont
  {{Mlynek}}}, \bibinfo {author} {\bibfnamefont {A.~A.}\ \bibnamefont
  {{Abdumalikov}}}, \bibinfo {author} {\bibfnamefont {C.}~\bibnamefont
  {{Eichler}}},\ and\ \bibinfo {author} {\bibfnamefont {A.}~\bibnamefont
  {{Wallraff}}},\ }\bibfield  {title} {\bibinfo {title} {{Observation of Dicke
  superradiance for two artificial atoms in a cavity with high decay rate}},\
  }\href {https://doi.org/10.1038/ncomms6186} {\bibfield  {journal} {\bibinfo
  {journal} {Nat. Commun.}\ }\textbf {\bibinfo {volume} {5}},\ \bibinfo {eid}
  {5186} (\bibinfo {year} {2014})}\BibitemShut {NoStop}%
\bibitem [{\citenamefont {Bamba}\ \emph {et~al.}(2016)\citenamefont {Bamba},
  \citenamefont {Inomata},\ and\ \citenamefont {Nakamura}}]{Bamba2016}%
  \BibitemOpen
  \bibfield  {author} {\bibinfo {author} {\bibfnamefont {M.}~\bibnamefont
  {Bamba}}, \bibinfo {author} {\bibfnamefont {K.}~\bibnamefont {Inomata}},\
  and\ \bibinfo {author} {\bibfnamefont {Y.}~\bibnamefont {Nakamura}},\
  }\bibfield  {title} {\bibinfo {title} {Superradiant phase transition in a
  superconducting circuit in thermal equilibrium},\ }\href
  {https://doi.org/10.1103/PhysRevLett.117.173601} {\bibfield  {journal}
  {\bibinfo  {journal} {Phys. Rev. Lett.}\ }\textbf {\bibinfo {volume} {117}},\
  \bibinfo {pages} {173601} (\bibinfo {year} {2016})}\BibitemShut {NoStop}%
\bibitem [{\citenamefont {{Forn-D{\'\i}az}}\ \emph {et~al.}(2017)\citenamefont
  {{Forn-D{\'\i}az}}, \citenamefont {{Garc{\'\i}a-Ripoll}}, \citenamefont
  {{Peropadre}}, \citenamefont {{Orgiazzi}}, \citenamefont {{Yurtalan}},
  \citenamefont {{Belyansky}}, \citenamefont {{Wilson}},\ and\ \citenamefont
  {{Lupascu}}}]{Diaz2017}%
  \BibitemOpen
  \bibfield  {author} {\bibinfo {author} {\bibfnamefont {P.}~\bibnamefont
  {{Forn-D{\'\i}az}}}, \bibinfo {author} {\bibfnamefont {J.~J.}\ \bibnamefont
  {{Garc{\'\i}a-Ripoll}}}, \bibinfo {author} {\bibfnamefont {B.}~\bibnamefont
  {{Peropadre}}}, \bibinfo {author} {\bibfnamefont {J.~L.}\ \bibnamefont
  {{Orgiazzi}}}, \bibinfo {author} {\bibfnamefont {M.~A.}\ \bibnamefont
  {{Yurtalan}}}, \bibinfo {author} {\bibfnamefont {R.}~\bibnamefont
  {{Belyansky}}}, \bibinfo {author} {\bibfnamefont {C.~M.}\ \bibnamefont
  {{Wilson}}},\ and\ \bibinfo {author} {\bibfnamefont {A.}~\bibnamefont
  {{Lupascu}}},\ }\bibfield  {title} {\bibinfo {title} {{Ultrastrong coupling
  of a single artificial atom to an electromagnetic continuum in the
  nonperturbative regime}},\ }\href {https://doi.org/10.1038/nphys3905}
  {\bibfield  {journal} {\bibinfo  {journal} {Nat. Phys.}\ }\textbf {\bibinfo
  {volume} {13}},\ \bibinfo {pages} {39} (\bibinfo {year} {2017})}\BibitemShut
  {NoStop}%
\bibitem [{\citenamefont {{Yoshihara}}\ \emph {et~al.}(2017)\citenamefont
  {{Yoshihara}}, \citenamefont {{Fuse}}, \citenamefont {{Ashhab}},
  \citenamefont {{Kakuyanagi}}, \citenamefont {{Saito}},\ and\ \citenamefont
  {{Semba}}}]{Yoshihara2017}%
  \BibitemOpen
  \bibfield  {author} {\bibinfo {author} {\bibfnamefont {F.}~\bibnamefont
  {{Yoshihara}}}, \bibinfo {author} {\bibfnamefont {T.}~\bibnamefont {{Fuse}}},
  \bibinfo {author} {\bibfnamefont {S.}~\bibnamefont {{Ashhab}}}, \bibinfo
  {author} {\bibfnamefont {K.}~\bibnamefont {{Kakuyanagi}}}, \bibinfo {author}
  {\bibfnamefont {S.}~\bibnamefont {{Saito}}},\ and\ \bibinfo {author}
  {\bibfnamefont {K.}~\bibnamefont {{Semba}}},\ }\bibfield  {title} {\bibinfo
  {title} {{Superconducting qubit-oscillator circuit beyond the
  ultrastrong-coupling regime}},\ }\href {https://doi.org/10.1038/nphys3906}
  {\bibfield  {journal} {\bibinfo  {journal} {Nat. Phys.}\ }\textbf {\bibinfo
  {volume} {13}},\ \bibinfo {pages} {44} (\bibinfo {year} {2017})}\BibitemShut
  {NoStop}%
\bibitem [{\citenamefont {Kr{\"a}mer}\ \emph {et~al.}(2018)\citenamefont
  {Kr{\"a}mer}, \citenamefont {Plankensteiner}, \citenamefont {Ostermann},\
  and\ \citenamefont {Ritsch}}]{Kramer2018}%
  \BibitemOpen
  \bibfield  {author} {\bibinfo {author} {\bibfnamefont {S.}~\bibnamefont
  {Kr{\"a}mer}}, \bibinfo {author} {\bibfnamefont {D.}~\bibnamefont
  {Plankensteiner}}, \bibinfo {author} {\bibfnamefont {L.}~\bibnamefont
  {Ostermann}},\ and\ \bibinfo {author} {\bibfnamefont {H.}~\bibnamefont
  {Ritsch}},\ }\bibfield  {title} {\bibinfo {title} {{QuantumOptics. jl: A
  Julia framework for simulating open quantum systems}},\ }\href@noop {}
  {\bibfield  {journal} {\bibinfo  {journal} {Computer Physics Communications}\
  }\textbf {\bibinfo {volume} {227}},\ \bibinfo {pages} {109} (\bibinfo {year}
  {2018})}\BibitemShut {NoStop}%
\bibitem [{\citenamefont {Schachenmayer}\ \emph {et~al.}(2015)\citenamefont
  {Schachenmayer}, \citenamefont {Pikovski},\ and\ \citenamefont
  {Rey}}]{Schachenmayer2015}%
  \BibitemOpen
  \bibfield  {author} {\bibinfo {author} {\bibfnamefont {J.}~\bibnamefont
  {Schachenmayer}}, \bibinfo {author} {\bibfnamefont {A.}~\bibnamefont
  {Pikovski}},\ and\ \bibinfo {author} {\bibfnamefont {A.~M.}\ \bibnamefont
  {Rey}},\ }\bibfield  {title} {\bibinfo {title} {{Many-Body Quantum Spin
  Dynamics with Monte Carlo Trajectories on a Discrete Phase Space}},\ }\href
  {https://doi.org/10.1103/PhysRevX.5.011022} {\bibfield  {journal} {\bibinfo
  {journal} {Phys. Rev. X}\ }\textbf {\bibinfo {volume} {5}},\ \bibinfo {pages}
  {011022} (\bibinfo {year} {2015})}\BibitemShut {NoStop}%
\bibitem [{\citenamefont {Huber}\ \emph {et~al.}(2022)\citenamefont {Huber},
  \citenamefont {Rey},\ and\ \citenamefont {Rabl}}]{Huber2022}%
  \BibitemOpen
  \bibfield  {author} {\bibinfo {author} {\bibfnamefont {J.}~\bibnamefont
  {Huber}}, \bibinfo {author} {\bibfnamefont {A.~M.}\ \bibnamefont {Rey}},\
  and\ \bibinfo {author} {\bibfnamefont {P.}~\bibnamefont {Rabl}},\ }\bibfield
  {title} {\bibinfo {title} {{Realistic simulations of spin squeezing and
  cooperative coupling effects in large ensembles of interacting two-level
  systems}},\ }\href {https://doi.org/10.1103/PhysRevA.105.013716} {\bibfield
  {journal} {\bibinfo  {journal} {Phys. Rev. A}\ }\textbf {\bibinfo {volume}
  {105}},\ \bibinfo {pages} {013716} (\bibinfo {year} {2022})}\BibitemShut
  {NoStop}%
\bibitem [{\citenamefont {Polkovnikov}(2010)}]{Polkovnikov2010}%
  \BibitemOpen
  \bibfield  {author} {\bibinfo {author} {\bibfnamefont {A.}~\bibnamefont
  {Polkovnikov}},\ }\bibfield  {title} {\bibinfo {title} {Phase space
  representation of quantum dynamics},\ }\href
  {https://doi.org/http://dx.doi.org/10.1016/j.aop.2010.02.006} {\bibfield
  {journal} {\bibinfo  {journal} {Ann. Phys.}\ }\textbf {\bibinfo {volume}
  {325}},\ \bibinfo {pages} {1790} (\bibinfo {year} {2010})}\BibitemShut
  {NoStop}%
\bibitem [{\citenamefont {{Tucker}}\ \emph {et~al.}(2018)\citenamefont
  {{Tucker}}, \citenamefont {{Zhu}}, \citenamefont {{Lewis-Swan}},
  \citenamefont {{Marino}}, \citenamefont {{Jimenez}}, \citenamefont
  {{Restrepo}},\ and\ \citenamefont {{Rey}}}]{Tucker2018}%
  \BibitemOpen
  \bibfield  {author} {\bibinfo {author} {\bibfnamefont {K.}~\bibnamefont
  {{Tucker}}}, \bibinfo {author} {\bibfnamefont {B.}~\bibnamefont {{Zhu}}},
  \bibinfo {author} {\bibfnamefont {R.~J.}\ \bibnamefont {{Lewis-Swan}}},
  \bibinfo {author} {\bibfnamefont {J.}~\bibnamefont {{Marino}}}, \bibinfo
  {author} {\bibfnamefont {F.}~\bibnamefont {{Jimenez}}}, \bibinfo {author}
  {\bibfnamefont {J.~G.}\ \bibnamefont {{Restrepo}}},\ and\ \bibinfo {author}
  {\bibfnamefont {A.~M.}\ \bibnamefont {{Rey}}},\ }\bibfield  {title} {\bibinfo
  {title} {{Shattered time: can a dissipative time crystal survive many-body
  correlations?}},\ }\href {https://doi.org/10.1088/1367-2630/aaf18b}
  {\bibfield  {journal} {\bibinfo  {journal} {New J. Phys.}\ }\textbf {\bibinfo
  {volume} {20}},\ \bibinfo {eid} {123003} (\bibinfo {year}
  {2018})}\BibitemShut {NoStop}%
\bibitem [{\citenamefont {{Zhihao}}\ \emph {et~al.}(2023)\citenamefont
  {{Zhihao}}, \citenamefont {{Wu}}, \citenamefont {{Wang}}, \citenamefont
  {{Xianlong}},\ and\ \citenamefont {{Wang}}}]{Zhihao2023}%
  \BibitemOpen
  \bibfield  {author} {\bibinfo {author} {\bibfnamefont {N.}~\bibnamefont
  {{Zhihao}}}, \bibinfo {author} {\bibfnamefont {Q.}~\bibnamefont {{Wu}}},
  \bibinfo {author} {\bibfnamefont {Q.}~\bibnamefont {{Wang}}}, \bibinfo
  {author} {\bibfnamefont {G.}~\bibnamefont {{Xianlong}}},\ and\ \bibinfo
  {author} {\bibfnamefont {P.}~\bibnamefont {{Wang}}},\ }\bibfield  {title}
  {\bibinfo {title} {{The failure of semiclassical approach in the dissipative
  fully-connected Ising model}},\ }\href
  {https://doi.org/10.48550/arXiv.2302.04381} {\bibfield  {journal} {\bibinfo
  {journal} {arXiv e-prints}\ ,\ \bibinfo {eid} {arXiv:2302.04381}} (\bibinfo
  {year} {2023})},\ \Eprint {https://arxiv.org/abs/2302.04381} {2302.04381}
  \BibitemShut {NoStop}%
\bibitem [{\citenamefont {Deuar}\ \emph {et~al.}(2021)\citenamefont {Deuar},
  \citenamefont {Ferrier}, \citenamefont {Matuszewski}, \citenamefont {Orso},\
  and\ \citenamefont {Szyma\ifmmode~\acute{n}\else \'{n}\fi{}ska}}]{Deuar2021}%
  \BibitemOpen
  \bibfield  {author} {\bibinfo {author} {\bibfnamefont {P.}~\bibnamefont
  {Deuar}}, \bibinfo {author} {\bibfnamefont {A.}~\bibnamefont {Ferrier}},
  \bibinfo {author} {\bibfnamefont {M.}~\bibnamefont {Matuszewski}}, \bibinfo
  {author} {\bibfnamefont {G.}~\bibnamefont {Orso}},\ and\ \bibinfo {author}
  {\bibfnamefont {M.~H.}\ \bibnamefont {Szyma\ifmmode~\acute{n}\else
  \'{n}\fi{}ska}},\ }\bibfield  {title} {\bibinfo {title} {Fully quantum
  scalable description of driven-dissipative lattice models},\ }\href
  {https://doi.org/10.1103/PRXQuantum.2.010319} {\bibfield  {journal} {\bibinfo
   {journal} {PRX Quantum}\ }\textbf {\bibinfo {volume} {2}},\ \bibinfo {pages}
  {010319} (\bibinfo {year} {2021})}\BibitemShut {NoStop}%
\end{thebibliography}%

\end{document}